\documentclass[aps,prx,reprint,superscriptaddress,longbibliography,nofootinbib]{revtex4-2}
\usepackage[T1]{fontenc}
\usepackage{lmodern}
\usepackage{amsmath,amssymb,bm,mathrsfs}
\usepackage{tikz}
\usetikzlibrary{calc,arrows.meta,decorations.pathreplacing}
\usepackage{pgfplots}
\pgfplotsset{compat=1.18}
\definecolor{xcol}{RGB}{31,119,180}
\definecolor{ycol}{RGB}{44,160,44}
\definecolor{zcol}{RGB}{214,39,40}
\definecolor{Acol}{RGB}{230,120,0}
\definecolor{Bcol}{RGB}{120,60,170}
\usepackage[colorlinks=true,linkcolor=blue,citecolor=blue,urlcolor=blue]{hyperref}
\hypersetup{
 pdftitle={Optimal Local Preparation of Two-Dimensional Topological States across the Tenfold Way},
 pdfsubject={Exact local preparation of topological free-fermion states},
 pdfkeywords={topological states, tenfold way, quantum circuits, Chern number, quantum signal processing, Kitaev honeycomb model}
}

\newtheorem{theorem}{Theorem}
\newtheorem{lemma}{Lemma}
\newtheorem{proposition}{Proposition}
\newtheorem{corollary}{Corollary}
\newenvironment{proof}[1][Proof]{\par\noindent\textit{#1.}\ }{\hfill$\square$\par}
\newcommand{\Id}{\mathbf 1}
\newcommand{\iu}{\mathrm i}
\newcommand{\Torus}{\mathbb T}
\newcommand{\Z}{\mathbb Z}

\newcommand{\bk}{\bm k}
\newcommand{\br}{\bm r}
\newcommand{\bn}{\bm n}
\newcommand{\btau}{\bm\tau}
\newcommand{\Tr}{\operatorname{Tr}}

\newcommand{\diag}{\operatorname{diag}}
\newcommand{\spec}{\operatorname{spec}}

\begin{document}

\title{Optimal Local Preparation of Two-Dimensional Topological States across the Tenfold Way}

\author{Xie-Hang Yu}
\affiliation{Department of Computing and Mathematical Sciences, California Institute of Technology, Pasadena, CA 91125, USA}
\author{Xun Gao}
\email{xun.gao@colorado.edu}
\affiliation{JILA and Department of Physics, University of Colorado Boulder, Boulder, Colorado 80309, USA}

\begin{abstract}
How does topology constrain the circuit complexity of quantum-state preparation?
We establish optimal depth bounds for the exact preparation of two-dimensional topological insulators and superconductors.
Using quantum signal processing implemented through local Kitaev-honeycomb bond pulses, we construct explicit, spatially uniform, ancilla-free circuits for representative states of all five strong topological symmetry classes (A, D, C, AII, DIII).
On odd $L\times L$ tori, these circuits have depth $O(L)$.
Conversely, we prove a matching $\Omega(L)$ lower bound, valid for every $L$, for arbitrary local circuits, including interacting and spatially nonuniform ones.
This bound extends to states with a nonzero strong topological index and an $L$-independent momentum variation of their Bloch projectors.
Our results establish linear depth as both sufficient and necessary for exact preparation of these topological states.
While quantum circuits can bypass the obstruction to gapped symmetric deformations, locality imposes a fundamental limit on how quickly strong topology can be generated.
\end{abstract}

\maketitle

\section{Introduction}
\label{sec:introduction}

Topological insulators and superconductors exhibit quantized responses and robust boundary modes protected by bulk energy gaps and symmetries~\cite{HasanKane2010,QiZhang2011,FuKaneMele2007,MooreBalents2007}.
The tenfold way provides a unified classification of free-fermion topological phases based on time-reversal, particle-hole, and chiral symmetries~\cite{AltlandZirnbauer1997,Schnyder2008,Kitaev2009,Ryu2010,Chiu2016}.
In two dimensions, five symmetry classes (A, D, C, AII, and DIII) support strong topology, encompassing distinct physical phenomena such as quantum Hall effects~\cite{Klitzing1980,TKNN1982,Haldane1988,Chang2013}, quantum spin Hall effects~\cite{KaneMele2005QSH,KaneMele2005Z2,BHZ2006,Konig2007}, and topological superconductivity~\cite{Senthil1999,ReadGreen2000,Kitaev2001,FuKane2008,QiHughesRaghuZhang2009,Alicea2012,SatoAndo2017}.
This classification provides a natural framework for investigating whether different forms of strong topology share common constraints on quantum-state preparation.
Recent experiments have realized topological band structures in ultracold atoms~\cite{Aidelsburger2013,Miyake2013,Jotzu2014,Aidelsburger2015,Goldman2016,Asteria2019,Cooper2019,Wintersperger2020}, prepared topologically ordered states on programmable quantum platforms~\cite{Satzinger2021,Semeghini2021,Andersen2023,Iqbal2024}, and simulated the Kitaev honeycomb model on a neutral-atom quantum computer~\cite{Evered2025}.
These advances further motivate the search for efficient quantum circuits that prepare topological states.

Adiabatic preparation faces a fundamental topological obstruction: a symmetry-preserving path connecting a nontrivial topological phase to a trivial product state must close the bulk energy gap~\cite{HasanKane2010,Chiu2016}.
Quantum circuits can bypass this restriction because their intermediate states need not remain gapped ground states or preserve the symmetries of the target.
Their preparation depth is instead constrained by locality~\cite{HastingsWen2005,ChenGuWen2010}.
We focus on two-dimensional strong topology, as weak indices can be represented by stacks of one-dimensional phases (Sec.~\ref{sec:topology-background}).

Existing methods do not establish whether these states can be prepared exactly using spatially uniform local circuits of linear depth.
Adiabatic and Floquet ramps~\cite{OkaAoki2009,KitagawaQW2010,KitagawaFloquet2010,Lindner2011,Rudner2013,DAlessioRigol2015,Caio2015,GeRigol2017} may become slow near gap closings, while numerically optimized pulse sequences lack general guarantees on accuracy and depth~\cite{Kalinowski2023,Evered2025}.
Exact constructions based on fermionic Fourier or Bogoliubov transformations and matchgate circuits typically require position-dependent gates, and sometimes ancillas~\cite{Valiant2002,TerhalDiVincenzo2002,VerstraeteCiracLatorre2009,Ferris2014,Wecker2015,SchmollOrus2017,Jiang2018,Kivlichan2018,Babbush2018,Langer2026}.
Other approaches rely on additional resources, including quasi-local unitaries, measurement and feedforward, engineered cooling, or continuum entanglement renormalization~\cite{ChenHermeleStephen2025,RaussendorfBravyiHarrington2005,Piroli2021,Tantivasadakarn2023,Diehl2011,Kishony2025,Haegeman2013,Chu2019}.
This raises our first question: Can representative states of all five two-dimensional strong topological classes be prepared exactly on an $L\times L$ torus using spatially uniform, ancilla-free local circuits of depth $O(L)$?

A separate fundamental question is whether locality alone
imposes a linear-depth lower bound for the exact preparation
of two-dimensional strong topological states.
Lieb--Robinson arguments~\cite{LiebRobinson1972,HastingsKoma2006,NachtergaeleSims2006} establish linear-time obstructions
to preparing intrinsically topologically ordered states
~\cite{BravyiHastingsVerstraete2006},
but do not directly apply to free-fermion phases with unique
ground states.
Related bounds concern one-dimensional topological phases
~\cite{HuangChen2015,McGinleyCooper2018},
single-particle dynamics changing the Bott index
~\cite{LoringHastings2010,Toniolo2018},
Chern-state preparation under translation-invariant Gaussian dynamics
~\cite{Wang2026},
or sequential circuits for chiral states
~\cite{FanWuBaoDai2025}.
Separately, no-go results for strictly finite-range flat-band
Hamiltonians rule out nontrivial strong topology
~\cite{ChenMazaheriSeidelTang2013,DubailRead2013,Read2017},
but do not directly establish finite-size circuit-depth bounds.
None of these results establishes a linear-depth lower bound
for strong topological classes under
arbitrary interacting and spatially nonuniform local circuits,
including classes with vanishing total Chern number.

We answer both questions affirmatively. For the construction, we exploit uniform bond pulses of the Kitaev honeycomb model~\cite{Kitaev2006}, which can be implemented as parallel layers of identical local gates and have been explored in neutral-atom platforms~\cite{DuanDemlerLukin2003,Kalinowski2023,Evered2025}. In a fixed bond sector, these pulses act on free matter Majorana fermions. We show that they naturally implement quantum signal processing (QSP)~\cite{LowYoderChuang2016,LowChuang2017,Gilyen2019,Martyn2021,Dong2021}, with momentum-dependent signal rotations and tunable control rotations (Fig.~\ref{fig:pulses}). This yields explicit pulse sequences that reproduce the target states exactly at every allowed momentum on a finite torus (Fig.~\ref{fig:sampling}). Using one, two, or four uncoupled copies of the honeycomb lattice, our spatially uniform, ancilla-free circuits prepare representative states of all five two-dimensional strong topological classes exactly, with depth $O(L_x+L_y)$.

For the lower bound, we establish a fundamental limitation imposed solely by locality, without assuming Gaussian dynamics, translation invariance, or symmetry-preserving gates. Starting from a product state, a depth-$D$ local circuit can generate two-point correlations only within distance $2D$, even with arbitrary interacting and spatially nonuniform gates (Fig.~\ref{fig:lightcone}). This finite correlation range constrains the momentum dependence of the resulting Bloch projector. For a topologically nontrivial target whose Bloch projector varies smoothly with momentum at an $L$-independent rate, exact preparation with $D\ll L$ would imply a topologically equivalent strictly finite-range projector, contradicting the no-go theorem for such projectors~\cite{Read2017}. We therefore obtain a general lower bound $D=\Omega(L)$, covering all five strong classes, including those with vanishing total Chern number. 

Together, our construction and lower bound establish the optimal preparation depth $\Theta(L)$, indicating that neither interactions nor spatially varying controls can improve its asymptotic scaling. Beyond this optimality result, our work demonstrates the power of quantum signal processing for physical state preparation by realizing it directly through a system's native local couplings rather than relying on an abstract black-box oracle. This approach provides explicit pulse angles for exact finite-size preparation, without variational optimization, adiabatic schedules, or ancillas.

The manuscript is structured as follows. Section~\ref{sec:background}
introduces the free-fermion states and their topology, the honeycomb
pulses that prepare them, the way the pulses realize QSP, and the
general local circuit model to which the lower bound applies.
Section~\ref{sec:results} states the results and explains why the
optimal depth is linear. Section~\ref{sec:seed} constructs the five
target states, Sec.~\ref{sec:preparation} proves that the pulses
prepare them, Sec.~\ref{sec:lower} proves the lower bound, and
Sec.~\ref{app:regularity} evaluates its constants for gapped
Hamiltonians. Section~\ref{sec:discussion} discusses implications.
The appendices treat the discrete Chern numbers of Fukui, Hatsugai,
and Suzuki (FHS)~\cite{FHS2005}.

\section{Topological states, honeycomb pulses, quantum signal processing, and local circuits}
\label{sec:background}

\subsection{Free-fermion topological states on a finite torus}
\label{sec:topology-background}

The states we prepare are Slater determinants of insulators and paired
ground states of superconductors, which are free-fermion states. Such a
state, also called a fermionic Gaussian state, is fixed by its
two-point correlations: higher even-order correlations follow from
Wick's theorem, and odd-order correlations vanish. A Gaussian unitary
is generated by a quadratic Hamiltonian and maps Gaussian states to
Gaussian states.

We place the fermions on the periodic lattice of cells
$\Lambda_{L_x,L_y}=\Z_{L_x}\times\Z_{L_y}$, with a fixed number of
modes per cell and lattice constant one. The integers $L_x,L_y$ count
cells, and the allowed momenta form the grid
\begin{equation}
 \mathcal K_{L_x,L_y}=
 \left\{\left(\frac{2\pi p}{L_x},\frac{2\pi q}{L_y}\right):
 p\in\Z_{L_x},\ q\in\Z_{L_y}\right\}.
 \label{eq:grid}
\end{equation}
Each momentum coordinate is defined modulo $2\pi$, so the Brillouin
zone is the torus $\Torus^2$. We write $\bk=(k_x,k_y)$ for a point on
it and $\bar k_\mu\in[-\pi,\pi)$ for the value of $k_\mu$ reduced to
this interval.

In a translation-invariant state, the two-point correlations are
diagonal in the momentum basis. At each momentum they form a matrix over the
modes of a cell. For a number-conserving state, this matrix is the
occupied-state projector, with entries
$P_{ab}(\bk)=\langle c_{\bk,b}^\dagger c_{\bk,a}\rangle$, where
$c_{\bk,a}$ annihilates mode $a$ at momentum $\bk$. It is an
orthogonal projector because, in a pure Gaussian state, each Bloch
state is either filled or empty. Superconducting states have an
analogous projector built from Majorana correlations
(Sec.~\ref{app:projectors}). We call $P(\bk)$ the Bloch projector of
the state. We specify each target by a Bloch projector that is
continuous and periodic over the whole Brillouin zone. On the finite
torus, only its values at the momenta in Eq.~\eqref{eq:grid} enter, so
the target state is fixed by these $L_xL_y$ matrices. Exact preparation
means producing the pure Gaussian state with these values, or
equivalently all of its normal and pairing correlations. The
topological index instead depends on the projector over the full
Brillouin zone.

Within a symmetry class, two gapped states belong to the same phase if
one can be deformed into the other without closing the gap or breaking
the symmetries. A topological index labels the phases. The trivial
phase is the one that contains the atomic limit: a translation-invariant
product state in which every unit cell of the lattice, in real space,
is in the same state and different cells are uncorrelated. Its
correlations are confined to single cells, so its Bloch projector is
the same at every momentum; conversely, a pure Gaussian state whose
Bloch projector does not depend on momentum is of this form. The
initial states of our construction are atomic-limit states
(Sec.~\ref{sec:classes}). Indices come in two
kinds. \emph{Strong} indices are the genuinely two-dimensional part of
the classification. \emph{Weak} indices come from stacking
one-dimensional phases and require lattice translations for their
protection. A nonzero strong index forbids a gapped, symmetric
deformation to the atomic limit. To compute indices, we keep the
occupied subspace and replace the band energies by constants. Define a
flat-band Hamiltonian $Q$ as
\begin{equation}
 Q(\bk)=2P(\bk)-\Id,\qquad Q^\dagger=Q,\qquad Q^2=\Id,
 \label{eq:flattening}
\end{equation}
where $\Id$ is the identity on the one-particle space. Thus, $-Q$ has
energy $-1$ on the occupied states and $+1$ on the empty ones. The
Gaussian state is the unique ground state of the quadratic Hamiltonian
whose one-particle matrix is $-Q$, so this replacement of the
energies, called spectral flattening, leaves the state
unchanged~\cite{Ryu2010}. We keep the sign of
Eq.~\eqref{eq:flattening}, in which $Q$ is $+1$ on the occupied states,
and call $Q$ the flat-band Hamiltonian, although the physical one is
$-Q$.

For an occupied projector $P(\bk)$ on the two-dimensional Brillouin
zone, the Chern number is
\begin{equation}
 c_1(P)=\frac{-\iu}{2\pi}
 \int_{\Torus^2}\Tr\!\left(P[\partial_{k_x}P,\partial_{k_y}P]\right)
 \,d^2k.
 \label{eq:chern}
\end{equation}
In our convention, the Berry curvature of the occupied bands is
$-\iu\Tr(P[\partial_{k_x}P,\partial_{k_y}P])$. It is the geometric
(Berry) phase per unit area of a small loop in momentum space. The
total Berry flux through the closed Brillouin torus is an integer
multiple of $2\pi$, so $c_1$ is an
integer~\cite{HasanKane2010,QiZhang2011}. The Chern number also decides
whether the occupied states can be described by smooth periodic
vectors. Suppose a single occupied band has an occupied Bloch state
$\psi(\bk)$, a unit vector with $P(\bk)=\psi(\bk)\psi(\bk)^\dagger$,
that is smooth and periodic over the whole zone. Then the Berry
curvature is the curl of the Berry connection
$\mathcal A=-\iu\psi^\dagger\nabla_{\bk}\psi$, and Stokes' theorem on
the closed torus gives $c_1=0$. A band with nonzero Chern number
therefore has no such occupied Bloch state, although its projector
$P(\bk)$ is smooth and periodic~\cite{Thouless1984,Kohmoto1985,Brouder2007,Panati2007}. Physically, in an insulator $c_1$
gives the Hall conductance in units of $e^2/h$~\cite{TKNN1982} and the net number of
chiral edge modes; in a superconductor it counts chiral edge modes and
fixes the quantized thermal Hall response~\cite{ReadGreen2000}.

The symmetry class of a state records its time-reversal, particle-hole,
and chiral symmetries, following the Altland--Zirnbauer (AZ) tenfold
way~\cite{Schnyder2008,Kitaev2009,Ryu2010,Chiu2016}. Time reversal
reverses momentum and keeps occupied states occupied. Particle-hole
symmetry maps occupied states at $\bk$ to empty states at $-\bk$.
Every superconductor has it: Cooper pairing mixes particles and holes,
and the Bogoliubov--de Gennes (BdG) description contains each fermion
both as a particle and as a hole. When present, time reversal
$\mathcal T=U_TK$ and particle-hole symmetry $\mathcal C=U_CK$ require
\begin{align}
 U_TP(\bk)^*U_T^\dagger&=P(-\bk),\notag\\
 U_CP(\bk)^*U_C^\dagger&=\Id-P(-\bk),
 \label{eq:az-projectors}
\end{align}
where $K$ denotes complex conjugation and $U_T,U_C$ are fixed unitary
matrices acting within each cell. The ten AZ classes are fixed by
which of these relations hold and by the signs of
$\mathcal T^2=U_TU_T^*$ and $\mathcal C^2=U_CU_C^*$. When neither
relation holds, they are further distinguished by chiral symmetry, a
unitary matrix $U_S$ with $U_SQ(\bk)U_S^\dagger=-Q(\bk)$ at every
momentum. When time reversal and particle-hole symmetry are both
present, their product is a chiral symmetry.
Class A imposes none of these symmetries; a Chern insulator is an
example. Class D imposes only particle-hole symmetry, with
$\mathcal C^2=+1$, as in a chiral $p+\iu p$ superconductor of spinless
fermions. Class C describes superconductors whose Hamiltonian is
invariant under $SU(2)$ rotations of the electron spin (spin-rotation
symmetry), such as a spin-singlet $d+\iu d$ superconductor.
Spin-rotation symmetry lets one keep only the spin-up particles and the
spin-down holes; in this spin-reduced description
(Sec.~\ref{app:projectors}) the remaining particle-hole operation
obeys $\mathcal C^2=-1$. Class AII imposes only time reversal, with
$\mathcal T^2=-1$, as in a quantum spin Hall insulator. Class DIII
imposes both, with $\mathcal T^2=-1$ and $\mathcal C^2=+1$, as in a
helical superconductor in which spin-up fermions pair as $p+\iu p$ and
spin-down fermions as $p-\iu p$. Table~\ref{tab:targets} gives the
matrices for our target states.

We use the stable classification, in which trivial filled or empty
bands may be added when phases are compared. In two dimensions, the
five strong classes have the nonzero strong groups $\Z$ in A and D,
$2\Z$ in C (in the spin-reduced description), and $\Z_2$ in AII and
DIII~\cite{Kitaev2009,Ryu2010}. The groups $\Z$ and $2\Z$ mean that
the Chern number is any integer or any even integer, respectively.
Time reversal makes the total Chern number zero in AII and DIII, but
it can still obstruct a deformation to the atomic limit that preserves
time reversal. This remaining strong topology is described by a $\Z_2$
index $\nu\in\{0,1\}$. A state with $\nu=1$ has an odd number of
Kramers pairs of counterpropagating (helical) edge modes, which cannot
be gapped without breaking time reversal. In general, $\nu$ is defined
from the whole occupied subspace and the action of time
reversal~\cite{Kitaev2009,Ryu2010}. In our construction the occupied
projector separates into two blocks related by time reversal, with
Chern numbers $c_1$ and $-c_1$, and then $\nu=c_1\bmod 2$.

Weak indices of decoupled stacks pose no new two-dimensional
preparation problem. A decoupled stack of identical
chains has a projector that depends on only one momentum coordinate;
in real space it is a product of identical chain states. If a local
circuit prepares one chain on a periodic row of the torus, running it
on all rows in parallel prepares the stack,
because the row circuits act on disjoint modes. The stack therefore
needs no more depth than one chain, and a chain of $L$ cells can be
prepared exactly in depth $O(L)$, for example by a network of
nearest-neighbor Givens rotations~\cite{Jiang2018}. We therefore study
the preparation cost imposed by the strong index.

\subsection{Kitaev honeycomb model and bond pulses}
\label{sec:honeycomb-pulses}

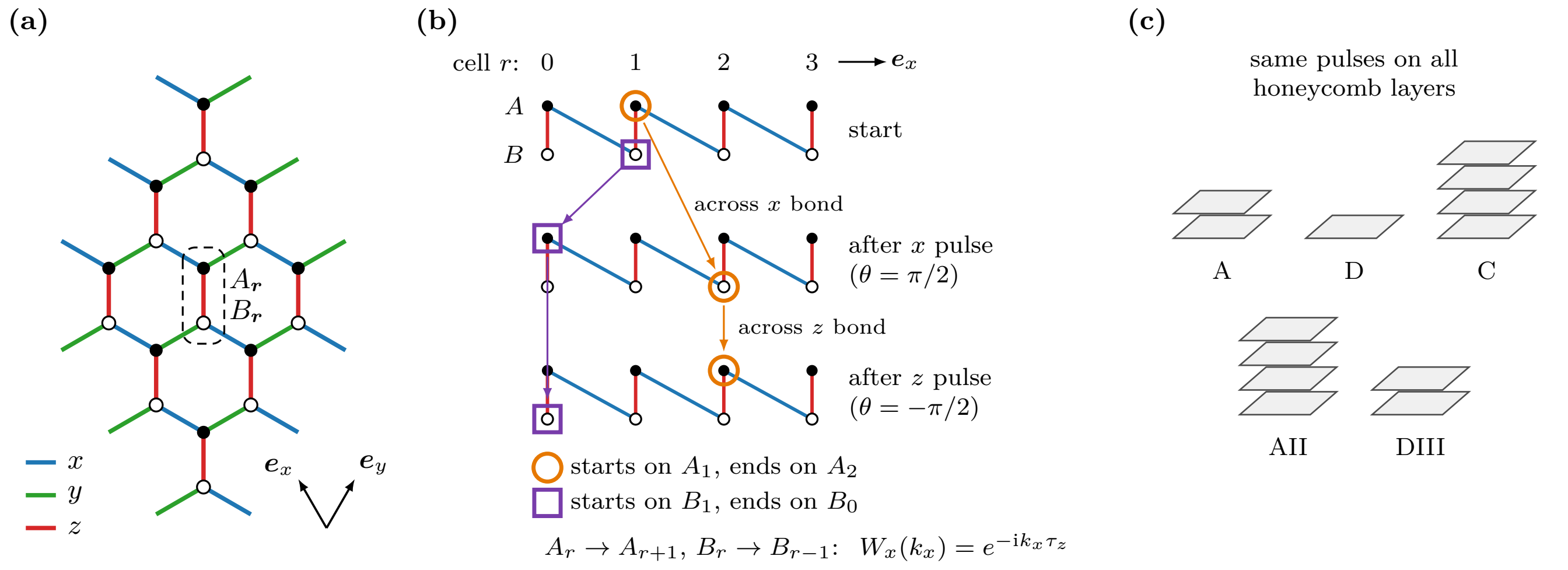
\begin{figure*}[t]
\centering
\begin{tikzpicture}[
  x=1cm,y=1cm,
  siteA/.style={circle,fill=black,inner sep=0pt,minimum size=4.2pt},
  siteB/.style={circle,draw=black,fill=white,inner sep=0pt,minimum size=4.2pt,line width=0.6pt},
  xb/.style={line width=1.4pt,color=xcol},
  yb/.style={line width=1.4pt,color=ycol},
  zb/.style={line width=1.4pt,color=zcol},
  lab/.style={font=\small}]
\begin{scope}[scale=0.62]
  \def\hx{0.866}
  \foreach \m in {0,1,2}{\foreach \n in {0,1,2}{
    \coordinate (A\m\n) at ({(\n-\m)*\hx},{(\m+\n)*1.5});
    \coordinate (B\m\n) at ({(\n-\m)*\hx},{(\m+\n)*1.5-1});
  }}
  \foreach \m in {0,1,2}{\foreach \n in {0,1,2}{
    \draw[zb] (A\m\n) -- (B\m\n);
  }}
  \foreach \m/\mm in {0/1,1/2}{\foreach \n in {0,1,2}{
    \draw[xb] (A\m\n) -- (B\mm\n);
  }}
  \foreach \m in {0,1,2}{\foreach \n/\nn in {0/1,1/2}{
    \draw[yb] (A\m\n) -- (B\m\nn);
  }}
  \foreach \n in {0,1,2}{\draw[xb] (A2\n) -- ++({-\hx},0.5);}
  \foreach \m in {0,1,2}{\draw[yb] (A\m2) -- ++({\hx},0.5);}
  \foreach \n in {0,1,2}{\draw[xb] (B0\n) -- ++({\hx},-0.5);}
  \foreach \m in {0,1,2}{\draw[yb] (B\m0) -- ++({-\hx},-0.5);}
  \foreach \m in {0,1,2}{\foreach \n in {0,1,2}{
    \node[siteA] at (A\m\n) {};
    \node[siteB] at (B\m\n) {};
  }}
  \draw[densely dashed,rounded corners=4pt,line width=0.6pt]
    ($(A11)+(-0.38,0.38)$) rectangle ($(B11)+(0.38,-0.38)$);
  \node[lab,anchor=west,inner sep=1pt] at ($(A11)+(0.4,-0.22)$) {$A_{\br}$};
  \node[lab,anchor=west,inner sep=1pt] at ($(B11)+(0.4,0.22)$) {$B_{\br}$};
  \coordinate (o) at (2.25,-1.75);
  \draw[-{Latex[length=4pt]},line width=0.7pt] (o) -- ++({-0.6*\hx},0.9) node[lab,above left,inner sep=1pt] {$\bm e_x$};
  \draw[-{Latex[length=4pt]},line width=0.7pt] (o) -- ++({0.6*\hx},0.9) node[lab,above right,inner sep=1pt] {$\bm e_y$};
  \begin{scope}[shift={(-3.25,-0.55)}]
    \draw[xb] (0,0) -- (0.55,0) node[lab,right,black] {$x$};
    \draw[yb] (0,-0.6) -- (0.55,-0.6) node[lab,right,black] {$y$};
    \draw[zb] (0,-1.2) -- (0.55,-1.2) node[lab,right,black] {$z$};
  \end{scope}
  \node[font=\bfseries\small] at (-3.2,7.5) {(a)};
\end{scope}
\begin{scope}[shift={(3.9,0.15)}]
  \def\dx{1.0}
  \def\rowgap{1.5}
  \node[font=\bfseries\small] at (-1.25,4.5) {(b)};
  \node[font=\footnotesize,anchor=east] at (-0.2,4.05) {cell $r$:};
  \foreach \c in {0,1,2,3}{\node[font=\footnotesize] at ({\c*\dx},4.05) {$\c$};}
  \draw[-{Latex[length=4pt]},line width=0.6pt] ({3*\dx+0.3},4.05) -- ({3*\dx+0.85},4.05)
    node[font=\footnotesize,right,inner sep=1pt] {$\bm e_x$};
  \foreach \row/\lbl in {0/{start},1/{after $x$ pulse\\ ($\theta=\pi/2$)},2/{after $z$ pulse\\ ($\theta=-\pi/2$)}}{
    \begin{scope}[shift={(0,{(2-\row)*\rowgap})}]
      \foreach \c in {0,1,2,3}{
        \draw[zb,line width=1.1pt] ({\c*\dx},0.55) -- ({\c*\dx},0);
      }
      \foreach \c/\cc in {0/1,1/2,2/3}{
        \draw[xb,line width=1.1pt] ({\c*\dx},0.55) -- ({\cc*\dx},0);
      }
      \foreach \c in {0,1,2,3}{
        \node[siteA,minimum size=3.6pt] at ({\c*\dx},0.55) {};
        \node[siteB,minimum size=3.6pt] at ({\c*\dx},0) {};
      }
      \node[anchor=west,font=\footnotesize,align=left] at ({3*\dx+0.3},0.28) {\lbl};
    \end{scope}
  }
  \node[font=\footnotesize,anchor=east] at (-0.15,{2*\rowgap+0.55}) {$A$};
  \node[font=\footnotesize,anchor=east] at (-0.15,{2*\rowgap}) {$B$};
  \def\markA#1#2{\node[circle,draw=Acol,line width=1.3pt,minimum size=9pt,inner sep=0pt] at (#1,#2) {};}
  \def\markB#1#2{\node[rectangle,draw=Bcol,line width=1.3pt,minimum size=8.5pt,inner sep=0pt] at (#1,#2) {};}
  \markA{\dx}{2*\rowgap+0.55}
  \markB{\dx}{2*\rowgap}
  \markA{2*\dx}{\rowgap}
  \markB{0}{\rowgap+0.55}
  \markA{2*\dx}{0.55}
  \markB{0}{0}
  \draw[-{Latex[length=4pt]},Acol,line width=0.6pt,shorten >=6pt,shorten <=6pt] (\dx,{2*\rowgap+0.55}) -- (2*\dx,\rowgap);
  \draw[-{Latex[length=4pt]},Bcol,line width=0.6pt,shorten >=6pt,shorten <=6pt] (\dx,{2*\rowgap}) -- (0,{\rowgap+0.55});
  \draw[-{Latex[length=4pt]},Acol,line width=0.6pt,shorten >=6pt,shorten <=6pt] (2*\dx,\rowgap) -- (2*\dx,0.55);
  \draw[-{Latex[length=4pt]},Bcol,line width=0.6pt,shorten >=6pt,shorten <=6pt] (0,{\rowgap+0.55}) -- (0,0);
  \node[font=\scriptsize,anchor=west,inner sep=1pt] at ({1.62*\dx},{\rowgap+0.95}) {across $x$ bond};
  \node[font=\scriptsize,anchor=west,inner sep=1pt] at ({2*\dx+0.12},{0.5*\rowgap+0.3}) {across $z$ bond};
  \markA{0}{-0.55}
  \node[font=\footnotesize,anchor=west] at (0.15,-0.55) {starts on $A_1$, ends on $A_2$};
  \markB{0}{-0.95}
  \node[font=\footnotesize,anchor=west] at (0.15,-0.95) {starts on $B_1$, ends on $B_0$};
  \node[font=\footnotesize,anchor=west] at (-0.15,-1.45)
    {$A_r\to A_{r+1}$, $B_r\to B_{r-1}$: \ $W_x(k_x)=e^{-\iu k_x\tau_z}$};
\end{scope}
\begin{scope}[shift={(11.0,0.2)}]
  \node[font=\bfseries\small] at (-0.3,4.45) {(c)};
  \def\layer#1#2#3{
    \filldraw[fill=gray!12,draw=black!70,line width=0.5pt]
      (#1,#2) -- ++(0.8,0) -- ++(0.3,0.26) -- ++(-0.8,0) -- cycle;
    \node[font=\scriptsize] at ($(#1,#2)+(0.55,0.13)$) {$#3$};}
  \layer{0}{2.0}{}\layer{0}{2.28}{}
  \node[font=\footnotesize,align=center,anchor=north] at (0.55,1.85) {A};
  \layer{1.5}{2.0}{}
  \node[font=\footnotesize,align=center,anchor=north] at (2.05,1.85) {D};
  \layer{3.0}{2.0}{}\layer{3.0}{2.28}{}\layer{3.0}{2.56}{}\layer{3.0}{2.84}{}
  \node[font=\footnotesize,align=center,anchor=north] at (3.55,1.85) {C};
  \layer{0.75}{0}{}\layer{0.75}{0.28}{}\layer{0.75}{0.56}{}\layer{0.75}{0.84}{}
  \node[font=\footnotesize,align=center,anchor=north] at (1.3,-0.15) {AII};
  \layer{2.25}{0}{}\layer{2.25}{0.28}{}
  \node[font=\footnotesize,align=center,anchor=north] at (2.8,-0.15) {DIII};
  \node[font=\footnotesize,align=center] at (2.05,3.85) {same pulses on all\\ honeycomb layers};
\end{scope}
\end{tikzpicture}
\caption{Bond pulses of the Kitaev honeycomb model.
(a)~Honeycomb lattice with $x$ (blue), $y$ (green), and $z$ (red)
bonds; $A$ sites are filled and $B$ sites open. The $z$ bond in the
dashed box joins $A_{\br}$ and $B_{\br}$ of the unit cell $\br$; the
$x$ and $y$ bonds join $A_{\br}$ to $B_{\br+\bm e_x}$ and
$B_{\br+\bm e_y}$.
(b)~The signal on one row of cells $r=0,\ldots,3$ along $\bm e_x$;
snapshots read from top to bottom. An $x$ pulse ($\theta=\pi/2$)
followed by a $z$ pulse ($\theta=-\pi/2$) moves the amplitude starting
on $A_1$ (orange circle) to $A_2$ and that starting on $B_1$ (purple
square) to $B_0$; signs are omitted [Eq.~\eqref{eq:signal-proof}]. At
momentum $k_x$ these opposite shifts give
$W_x(k_x)=e^{-\iu k_x\tau_z}$.
(c)~The five strong classes use one, two, or four identical, uncoupled
honeycomb layers, all driven by the same pulses
(Sec.~\ref{sec:classes} and Fig.~\ref{fig:classes}).}
\label{fig:pulses}
\end{figure*}

Our construction uses the controls of the Kitaev honeycomb
model~\cite{Kitaev2006}. Spins $1/2$ sit on the sites of a honeycomb
lattice. Each site has three bonds, one of each type $x$, $y$, and
$z$, distinguished by their orientation
[Fig.~\ref{fig:pulses}(a)]. We group the two sites of each $z$ bond
into a unit cell $\br$, with sites $A_{\br}$ and $B_{\br}$. The $x$
bond then joins $A_{\br}$ to $B_{\br+\bm e_x}$, and the $y$ bond joins
$A_{\br}$ to $B_{\br+\bm e_y}$, where $\bm e_x,\bm e_y$ are the two
cell translations. The cells form the lattice $\Lambda_{L_x,L_y}$ of
Sec.~\ref{sec:topology-background}.

Let $\sigma_i^\alpha$ be the Pauli operator at site $i$, with
$\alpha=x,y,z$. With Kitaev couplings $J_\alpha^{\rm K}$, the
Hamiltonian is the sum of three bond Hamiltonians,
\begin{equation}
 H_K=H_x+H_y+H_z,\qquad
 H_\alpha=-J_\alpha^{\rm K}\sum_{\langle ij\rangle_\alpha}
 \sigma_i^\alpha\sigma_j^\alpha,
 \label{eq:kitaev-pulse-hamiltonian}
\end{equation}
where $\langle ij\rangle_\alpha$ runs over the $\alpha$ bonds. A bond
pulse turns on one $H_\alpha$ for a chosen time. The bonds of one type
are disjoint, so their terms commute. A pulse is therefore exactly one
circuit layer of identical two-site gates applied in parallel over the
whole lattice, with no Trotter error. It needs no spatial addressing,
only the ability to switch each bond type on and off.

Each spin in this model can be written in terms of Majorana
fermions. Then, each bond term becomes a coupling of two matter
Majoranas $\gamma_i,\gamma_j$~\cite{Kitaev2006}:
\begin{equation}
 -J_\alpha^{\rm K}\sigma_i^\alpha\sigma_j^\alpha
 =\iu J_\alpha^{\rm K} u_{ij}\gamma_i\gamma_j,\qquad u_{ij}=\pm1.
 \label{eq:kitaev-bond-matter}
\end{equation}
The bond variable $u_{ij}$ is built from the other Majoranas of the
two sites. It commutes with every bond Hamiltonian, so bond pulses
conserve it. We fix $u_{ij}=1$ on every bond, with $i$ on sublattice
$A$ and $j$ on sublattice $B$. Then no hexagon carries a flux, the
matter Majoranas see translation-invariant couplings, and our results
concern these matter fermions. With the coupling and the pulse
duration $t$ absorbed into the rotation angle
$\theta=-2J_\alpha^{\rm K}t$, the three pulses are
\begin{align}
 \widehat V_\mu(\theta)
 &=\exp\!\Bigl[-\frac\theta2\sum_{\br}
 \gamma_{A,\br}\gamma_{B,\br+\bm e_\mu}\Bigr],
 \qquad \mu=x,y,\notag\\
 \widehat V_z(\theta)
 &=\exp\!\Bigl[-\frac\theta2\sum_{\br}
 \gamma_{A,\br}\gamma_{B,\br}\Bigr].
 \label{eq:real-space-majorana}
\end{align}
On every bond of its type, a pulse rotates the two Majoranas into each
other by the angle $\theta$.

Locality, used later, is already visible in
Eq.~\eqref{eq:real-space-majorana}: each term couples Majoranas on two
neighboring sites, so one pulse moves a fermion operator by at most
one cell.

The same pulses can be applied in parallel to several identical,
uncoupled copies of the honeycomb lattice. We call these copies
honeycomb layers, to distinguish them from circuit layers. Two
honeycomb layers give two Majoranas $\gamma_{a,1,\br}$ and
$\gamma_{a,2,\br}$ on each site $a=A,B$ of a cell, which combine into
a complex fermion $c_{a,\br}=(\gamma_{a,1,\br}+\iu\gamma_{a,2,\br})/2$.
On each bond, with the cell labels suppressed, the anticommutation
relations give $c_B^\dagger c_A-c_A^\dagger c_B
=-(\gamma_{A,1}\gamma_{B,1}+\gamma_{A,2}\gamma_{B,2})/2$, so that
\begin{equation}
 \exp\!\Bigl[\theta\sum_{\br}
 \bigl(c_{B,\br+\bm e_\mu}^\dagger c_{A,\br}
 -c_{A,\br}^\dagger c_{B,\br+\bm e_\mu}\bigr)\Bigr]
 =\widehat V_{\mu,1}(\theta)\,\widehat V_{\mu,2}(\theta),
 \label{eq:hopping-majorana-copies}
\end{equation}
where $\widehat V_{\mu,\ell}$ is the pulse of
Eq.~\eqref{eq:real-space-majorana} on honeycomb layer $\ell=1,2$; the
same holds for the $z$ pulse, with $\bm e_\mu$ replaced by zero.
Applying the same pulse to both honeycomb layers is thus exactly a
nearest-neighbor hopping pulse of the complex fermions. One honeycomb
layer provides the Majorana fermions of a superconductor, and a pair of
honeycomb layers provides number-conserving complex fermions.

Because the pulses act identically in every cell, they do not mix
different momenta. After Fourier transformation,
$\gamma_{\bk,a}=(L_xL_y)^{-1/2}\sum_{\br}e^{-\iu\bk\cdot\br}\gamma_{a,\br}$,
a pulse acts at each momentum on the two components $A$ and $B$ as a
$2\times2$ matrix. For a pair of honeycomb layers, the complex fermions
$c_{\bk,a}$, Fourier transformed in the same way, transform with the
same matrix, because $c_{a,\br}$ is a fixed combination of the
Majoranas of the two layers, which receive the same pulse. Let $\tau_x,\tau_y,\tau_z$ denote the
Pauli matrices on this space and $\btau=(\tau_x,\tau_y,\tau_z)$. The
generators of the three pulses are
\begin{equation}
 g_z=\tau_y,\qquad
 g_\mu(k_\mu)=\sin k_\mu\tau_x+\cos k_\mu\tau_y.
 \label{eq:bond-generators}
\end{equation}
The $z$ bond lies inside a cell and gives the momentum-independent
$\tau_y$, while the phase $e^{\iu k_\mu}$ of the neighboring cell
produces the sine and cosine in $g_\mu$. We write the pulse matrices
as $\mathscr R_z(\theta)=e^{-\iu\theta g_z}$ and
$\mathscr R_\mu(\theta;k_\mu)=e^{-\iu\theta g_\mu(k_\mu)}$. Since
$g_z=\tau_y$, a $z$ pulse rotates the two components about the $y$
axis of their Bloch sphere.

\subsection{Quantum signal processing with bond pulses}
\label{sec:qsp-background}

At each momentum, a pulse sequence acts on the $A$ and $B$ components
as a $2\times2$ unitary $V(\bk)$ and transforms the Bloch projector as
$P\mapsto VPV^\dagger$. The atomic limit has the same projector at
every momentum (Sec.~\ref{sec:topology-background}), whereas a
topological target must depend on momentum. Preparing a target therefore requires a
rotation with a prescribed momentum dependence.

Start with the real-space picture of Fig.~\ref{fig:pulses}(b). An $x$
pulse with $\theta=\pi/2$ moves the amplitude on every site across its
$x$ bond: $A_{\br}\to B_{\br+\bm e_x}$ and $B_{\br}\to-A_{\br-\bm e_x}$.
A $z$ pulse with $\theta=-\pi/2$ then moves it across the $z$ bond of
its new cell: $B_{\br}\to A_{\br}$ and $A_{\br}\to-B_{\br}$. The
net effect is that the $A$ component moves one cell along $+\bm e_x$
and the $B$ component one cell along $-\bm e_x$. In momentum space, a
shift by one cell along $\pm\bm e_\mu$ multiplies an amplitude by
$e^{\mp\iu k_\mu}$, and Eq.~\eqref{eq:bond-generators} gives
\begin{align}
 \mathscr R_z(-\pi/2)\mathscr R_\mu(\pi/2;k_\mu)
 &=(\iu\tau_y)(-\iu g_\mu(k_\mu))\notag\\
 &=\cos k_\mu\,\Id-\iu\sin k_\mu\tau_z
 \label{eq:signal-proof}\\
 &=e^{-\iu k_\mu\tau_z}
 \label{eq:two-layer-signal}
\end{align}
for $\mu=x$ and, in the same way, for $\mu=y$. The relative phase
$2k_\mu$ between the two components is the QSP signal
$W_\mu(k_\mu)=e^{-\iu k_\mu\tau_z}$. The control is a $z$ pulse of
tunable angle $\varphi$, $\mathscr R_z(\varphi)=e^{-\iu\varphi\tau_y}$:
the same pulse type that ends each signal, now with a free angle. It
acts within each cell and is the same at every momentum.

Alternating signal and control make the two components travel along
different paths and interfere. Each signal moves the components one
cell in opposite directions, and each control pulse mixes them. After $d$ signals, the amplitude to arrive at a given
displacement is a sum over all paths with that displacement, and the
control angles set how these paths interfere. In momentum space, the
sequence with $d$ signals is
\begin{equation}
 F(k_\mu)=\mathscr R_z(\varphi_0)
 \prod_{s=1}^{d}\left[W_\mu(k_\mu)\mathscr R_z(\varphi_s)\right],
 \label{eq:qsp-background}
\end{equation}
where $\varphi_0,\ldots,\varphi_d$ are real control angles and the
factors are ordered from left to right in increasing $s$; the
rightmost factor acts first.

Equation~\eqref{eq:qsp-background} has the structure of QSP. QSP began
as a method to design composite pulses for a single
spin~\cite{LowYoderChuang2016} and became a general tool for quantum
algorithms~\cite{LowChuang2017,LowChuang2019,Martyn2021}. In its usual
form, a signal rotation encodes a number between $-1$ and $1$,
interleaved control rotations process it, and one matrix element of the
product becomes a chosen polynomial of that number. In our setting,
each momentum is an independent sector, and the two-component space is
that of the physical components $A$ and $B$. Let
$|\pm y\rangle=(1,\pm\iu)^T/\sqrt2$ be the eigenvectors of $\tau_y$,
which form the eigenbasis of the control pulses. Then
\begin{equation}
 \langle+y|W_\mu(k_\mu)|+y\rangle=\cos k_\mu,
 \label{eq:qsp-scalar-block}
\end{equation}
so $\cos k_\mu$ plays the role of the signal value. In this basis the
control pulses only change phases, and the diagonal entries of $F$ are
polynomials of degree at most $d$ in $\cos k_\mu$, which is the
standard QSP response. Both components belong to the target state, so
the preparation uses the whole $2\times2$ rotation rather than one
matrix element.

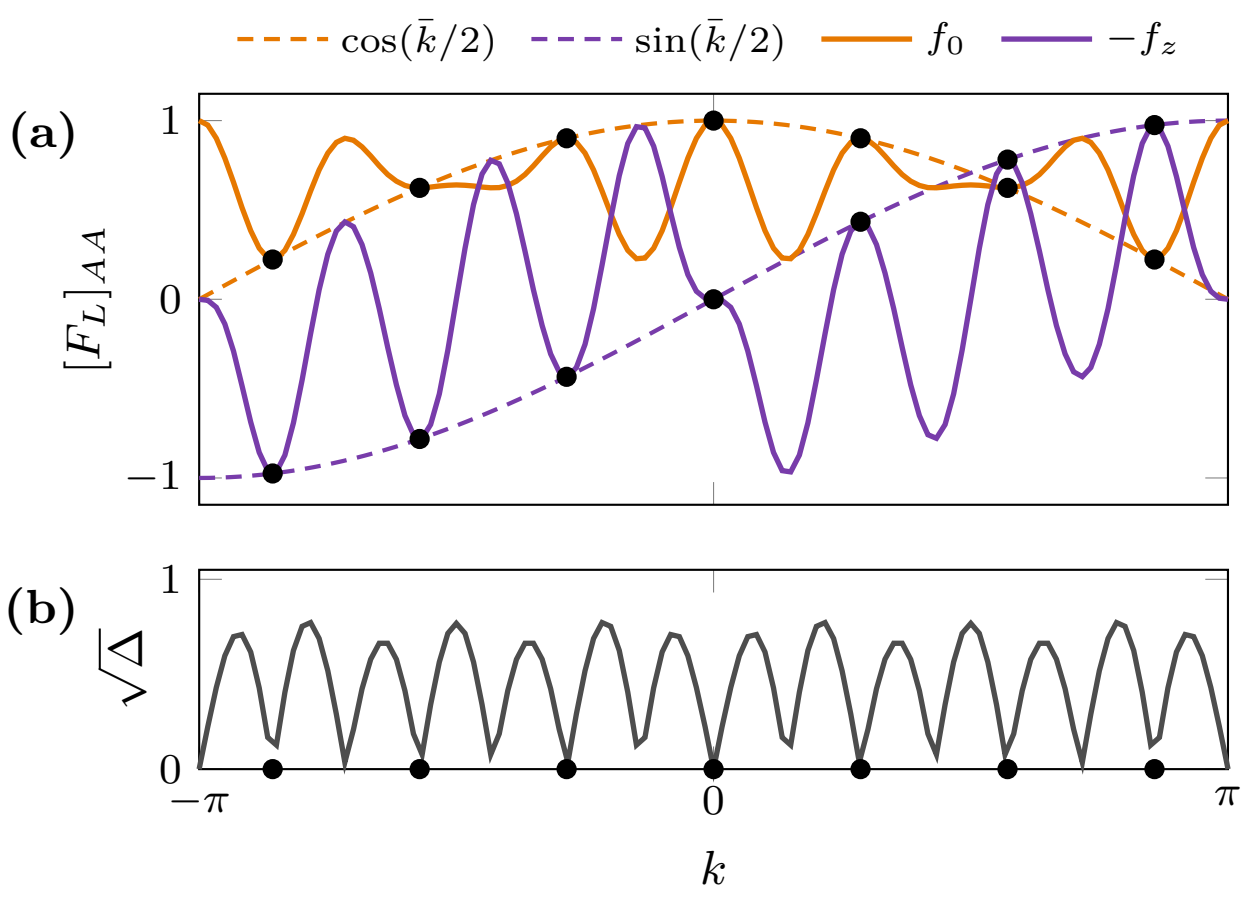
\begin{figure}[t]
\centering
\pgfplotstableread{
k f0 mfz sd
-3.142 1.000 0.000 0.000
-3.089 0.975 -0.006 0.224
-3.037 0.901 -0.045 0.430
-2.985 0.790 -0.138 0.597
-2.932 0.655 -0.290 0.698
-2.880 0.513 -0.484 0.709
-2.827 0.384 -0.687 0.617
-2.775 0.286 -0.857 0.428
-2.723 0.231 -0.959 0.167
-2.670 0.227 -0.966 0.127
-2.618 0.274 -0.872 0.405
-2.566 0.363 -0.692 0.625
-2.513 0.479 -0.452 0.752
-2.461 0.606 -0.192 0.772
-2.409 0.724 0.053 0.688
-2.356 0.818 0.252 0.518
-2.304 0.878 0.382 0.290
-2.252 0.901 0.433 0.036
-2.199 0.890 0.406 0.210
-2.147 0.853 0.306 0.423
-2.094 0.802 0.147 0.579
-2.042 0.747 -0.052 0.663
-1.990 0.698 -0.270 0.663
-1.937 0.661 -0.479 0.578
-1.885 0.637 -0.651 0.413
-1.833 0.625 -0.758 0.186
-1.780 0.624 -0.778 0.077
-1.728 0.628 -0.701 0.340
-1.675 0.634 -0.530 0.564
-1.623 0.638 -0.285 0.715
-1.571 0.640 0.000 0.768
-1.518 0.638 0.285 0.715
-1.466 0.634 0.530 0.564
-1.414 0.628 0.701 0.340
-1.361 0.624 0.778 0.077
-1.309 0.625 0.758 0.186
-1.257 0.637 0.651 0.413
-1.204 0.661 0.479 0.578
-1.152 0.698 0.270 0.663
-1.100 0.747 0.052 0.663
-1.047 0.802 -0.147 0.579
-0.995 0.853 -0.306 0.423
-0.943 0.890 -0.406 0.210
-0.890 0.901 -0.433 0.036
-0.838 0.878 -0.382 0.290
-0.785 0.818 -0.252 0.518
-0.733 0.724 -0.053 0.688
-0.681 0.606 0.192 0.772
-0.628 0.479 0.452 0.752
-0.576 0.363 0.692 0.625
-0.524 0.274 0.872 0.405
-0.471 0.227 0.966 0.127
-0.419 0.231 0.959 0.167
-0.366 0.286 0.857 0.428
-0.314 0.384 0.687 0.617
-0.262 0.513 0.484 0.709
-0.209 0.655 0.290 0.698
-0.157 0.790 0.138 0.597
-0.105 0.901 0.045 0.430
-0.052 0.975 0.006 0.224
0.000 1.000 -0.000 0.000
0.052 0.975 -0.006 0.224
0.105 0.901 -0.045 0.430
0.157 0.790 -0.138 0.597
0.209 0.655 -0.290 0.698
0.262 0.513 -0.484 0.709
0.314 0.384 -0.687 0.617
0.366 0.286 -0.857 0.428
0.419 0.231 -0.959 0.167
0.471 0.227 -0.966 0.127
0.524 0.274 -0.872 0.405
0.576 0.363 -0.692 0.625
0.628 0.479 -0.452 0.752
0.681 0.606 -0.192 0.772
0.733 0.724 0.053 0.688
0.785 0.818 0.252 0.518
0.838 0.878 0.382 0.290
0.890 0.901 0.433 0.036
0.943 0.890 0.406 0.210
0.995 0.853 0.306 0.423
1.047 0.802 0.147 0.579
1.100 0.747 -0.052 0.663
1.152 0.698 -0.270 0.663
1.204 0.661 -0.479 0.578
1.257 0.637 -0.651 0.413
1.309 0.625 -0.758 0.186
1.361 0.624 -0.778 0.077
1.414 0.628 -0.701 0.340
1.466 0.634 -0.530 0.564
1.518 0.638 -0.285 0.715
1.571 0.640 -0.000 0.768
1.623 0.638 0.285 0.715
1.675 0.634 0.530 0.564
1.728 0.628 0.701 0.340
1.780 0.624 0.778 0.077
1.833 0.625 0.758 0.186
1.885 0.637 0.651 0.413
1.937 0.661 0.479 0.578
1.990 0.698 0.270 0.663
2.042 0.747 0.052 0.663
2.094 0.802 -0.147 0.579
2.147 0.853 -0.306 0.423
2.199 0.890 -0.406 0.210
2.252 0.901 -0.433 0.036
2.304 0.878 -0.382 0.290
2.356 0.818 -0.252 0.518
2.409 0.724 -0.053 0.688
2.461 0.606 0.192 0.772
2.513 0.479 0.452 0.752
2.566 0.363 0.692 0.625
2.618 0.274 0.872 0.405
2.670 0.227 0.966 0.127
2.723 0.231 0.959 0.167
2.775 0.286 0.857 0.428
2.827 0.384 0.687 0.617
2.880 0.513 0.484 0.709
2.932 0.655 0.290 0.698
2.985 0.790 0.138 0.597
3.037 0.901 0.045 0.430
3.089 0.975 0.006 0.224
3.142 1.000 -0.000 0.000
}\profiledata
\begin{tikzpicture}
\begin{axis}[
  name=top, width=0.97\columnwidth, height=4.3cm,
  xmin=-3.1416, xmax=3.1416, ymin=-1.15, ymax=1.15,
  xtick={-3.1416,0,3.1416}, xticklabels={$-\pi$,$0$,$\pi$},
  ytick={-1,0,1}, xticklabel=\empty,
  ylabel={$[F_L]_{AA}$}, ylabel style={yshift=-4pt},
  tick label style={font=\footnotesize}, label style={font=\small},
  legend style={font=\scriptsize, at={(0.5,1.03)}, anchor=south, draw=none, legend columns=4, /tikz/every even column/.append style={column sep=4pt}},
  clip=false]
\addplot[densely dashed, line width=0.8pt, color=Acol, domain=-3.1416:3.1416, samples=101] {cos(deg(x)/2)};
\addlegendentry{$\cos(\bar k/2)$}
\addplot[densely dashed, line width=0.8pt, color=Bcol, domain=-3.1416:3.1416, samples=101] {sin(deg(x)/2)};
\addlegendentry{$\sin(\bar k/2)$}
\addplot[line width=1.0pt, color=Acol] table[x=k, y=f0] {\profiledata};
\addlegendentry{$f_0$}
\addplot[line width=1.0pt, color=Bcol] table[x=k, y=mfz] {\profiledata};
\addlegendentry{$-f_z$}
\addplot[only marks, mark=*, mark size=1.7pt, color=black] coordinates {(-2.6928,0.2225) (-1.7952,0.6235) (-0.8976,0.9010) (0.0000,1.0000) (0.8976,0.9010) (1.7952,0.6235) (2.6928,0.2225)};
\addplot[only marks, mark=*, mark size=1.7pt, color=black] coordinates {(-2.6928,-0.9749) (-1.7952,-0.7818) (-0.8976,-0.4339) (0.0000,0.0000) (0.8976,0.4339) (1.7952,0.7818) (2.6928,0.9749)};
\node[font=\bfseries\small, anchor=north east] at (rel axis cs:-0.1,1.0) {(a)};
\end{axis}
\begin{axis}[
  at={(top.below south west)}, anchor=above north west, yshift=-2pt,
  width=0.97\columnwidth, height=2.9cm,
  xmin=-3.1416, xmax=3.1416, ymin=0, ymax=1.05,
  xtick={-3.1416,0,3.1416}, xticklabels={$-\pi$,$0$,$\pi$},
  ytick={0,1}, xlabel={$k$}, ylabel={$\sqrt{\Delta}$}, ylabel style={yshift=-4pt},
  tick label style={font=\footnotesize}, label style={font=\small}, clip=false]
\addplot[line width=1.0pt, color=black!70] table[x=k, y=sd] {\profiledata};
\addplot[only marks, mark=*, mark size=1.7pt, color=black] coordinates {(-2.6928,0) (-1.7952,0) (-0.8976,0) (0.0000,0) (0.8976,0) (1.7952,0) (2.6928,0)};
\node[font=\bfseries\small, anchor=north east] at (rel axis cs:-0.1,1.0) {(b)};
\end{axis}
\end{tikzpicture}
\caption{Exact values at the allowed momenta, freedom in between.
The rotation $e^{\iu\bar k\tau_z/2}$ required by the construction
(Sec.~\ref{sec:results-construction}) and the pulse sequence $F_L$ of
Sec.~\ref{sec:preparation} that reproduces it at the allowed momenta,
for $L=7$ and $k=k_x$; $f_0$, $f_z$, and $\Delta$ are defined in
Sec.~\ref{sec:preparation}.
(a)~Dashed: real and imaginary parts, $\cos(\bar k/2)$ and
$\sin(\bar k/2)$, of the $AA$ entry $e^{\iu\bar k/2}$ of the required
rotation. The imaginary part jumps at $k=\pm\pi$, so the required
rotation is not periodic. Solid: real and imaginary parts, $f_0$ and
$-f_z$, of the $AA$ entry of $F_L$
[Eqs.~\eqref{eq:interpolated-components} and \eqref{eq:unitary-loop}],
which has period $\pi$. The two agree at the seven allowed momenta
(dots).
(b)~Added norm $\Delta^{1/2}$ of the two components that make $F_L$
unitary between the allowed momenta [Eqs.~\eqref{eq:completion-defect}
and \eqref{eq:completion-components}]. It vanishes at the allowed
momenta and, because of its period $\pi$, also at their shifts by
$\pi$.}
\label{fig:sampling}
\end{figure}

QSP is usually used to approximate a function at all values of the
signal. Here we need much less: a finite torus has only $L_\mu$
allowed values of $k_\mu$, and the pulse sequence must reproduce the
required rotation only there. The rotation
$e^{\iu\bar k_x\tau_z/2}$, which our construction needs
(Sec.~\ref{sec:results-construction}), is not periodic in $k_x$: its
entries $e^{\pm\iu\bar k_x/2}$ change sign when $k_x$ crosses $\pm\pi$.
Every pulse sequence is periodic in $k_x$, so no sequence equals this
rotation at every momentum. Pulse sequences also obey
$F(-k_x)=F(k_x)^*$ [Eq.~\eqref{eq:conj-symmetry}], so they are real
matrices at $k_x=\pi$, the same point as $-\pi$, whereas the required
rotation there, $e^{-\iu\pi\tau_z/2}=-\iu\tau_z$, is not. For even
$L_x$ the momentum $k_x=\pi$ is allowed, and no pulse sequence
reproduces the rotation there. For odd $L_x$ the allowed values are
$\bar k_x=2\pi p/L_x$ with $p=-(L_x-1)/2,\ldots,(L_x-1)/2$, all
strictly between $-\pi$ and $\pi$, and Section~\ref{sec:preparation}
constructs a sequence of at most $4(L_x-1)$ signals that equals this
rotation at every allowed momentum and differs from it in between
(Fig.~\ref{fig:sampling}).

\subsection{General local circuits}
\label{sec:general-circuits}

The lower bound allows far more control than the honeycomb pulses. A
gate may be any unitary acting on one cell or on two nearest-neighbor
cells, including a many-body interaction. Gates may vary from place to
place, and the intermediate states need not be Gaussian or have the
target AZ symmetries. The input is a product of pure cell states, and
we add no ancillas, measurements, or discarded subsystems. As in any
physical fermionic system, every gate preserves fermion
parity~\cite{BravyiKitaev2002} and
every input cell has definite parity, that is, a definite eigenvalue of
\begin{equation}
 \mathcal P_{\br}=(-1)^{\sum_a c_{a,\br}^\dagger c_{a,\br}},
 \label{eq:cell-parity}
\end{equation}
where $c_{a,\br}$ annihilates physical mode $a$ in cell $\br$.

The gates in one circuit layer act on disjoint sets of fermion
operators, such as the Majoranas of different bonds in one bond pulse,
so they commute. Each circuit layer therefore enlarges the support of an operator by at most
one nearest-neighbor step, and the circuit depth $D$ counts the circuit
layers. This limit on propagation is the locality constraint behind the
lower bound. The distance between two cells is the number of such steps
on the periodic lattice of cells, that is, $|r_x|+|r_y|$ for the
shortest displacement $\br=(r_x,r_y)$ between them. Since one honeycomb
pulse is one circuit layer, our construction is a special case of this
model.

\begin{table*}[t]
\caption{The five target states. All honeycomb layers of a target
receive the same pulse sequence. $Q$ is
the flat-band Hamiltonian $Q=2P-\Id$ of Eq.~\eqref{eq:flattening},
where $P$ is the number-conserving Bloch projector for A and AII, the
Majorana projector for D and DIII, and the spin-reduced projector for
C (Sec.~\ref{app:projectors}). $Q_1$ is given in
Eq.~\eqref{eq:seed}; $\kappa_y,\kappa_z$ and $\Id_\kappa$ are Pauli
and identity matrices acting on two blocks of components
(Sec.~\ref{sec:classes}); $K$ is complex conjugation. For AII and DIII
the total Chern number is zero, and $\nu$ is the Chern number of one
block modulo two.}
\label{tab:targets}
\begin{ruledtabular}
\begin{tabular}{cccccc}
 Class & Example & Honeycomb layers &
 Target $Q$ & Antiunitary symmetry & Strong index\\
 A & Chern insulator & 2 & $Q_1$ & None & $c_1=1$\\
 D & $p+\iu p$ superconductor & 1 & $Q_1$ &
 $\mathcal C=K$ & $c_1=1$\\
 C & $d+\iu d$ superconductor & 4 &
 $\Id_\kappa\otimes Q_1$ &
 $\mathcal C=(\iu\kappa_y\otimes\Id)K$ & $c_1=2$\\
 AII & quantum spin Hall insulator & 4 &
 $\kappa_z\otimes Q_1$ &
 $\mathcal T=(\iu\kappa_y\otimes\Id)K$ & $\nu=1$\\
 DIII & helical superconductor & 2 &
 $\kappa_z\otimes Q_1$ &
 $\mathcal T=(\iu\kappa_y\otimes\Id)K,\ \mathcal C=K$ & $\nu=1$\\
\end{tabular}
\end{ruledtabular}
\end{table*}

\section{Main results}
\label{sec:results}

This section states the results; the proofs follow in
Secs.~\ref{sec:seed}--\ref{app:regularity}.

\subsection{One pulse sequence for the five strong classes}
\label{sec:results-construction}

The construction prepares one Bloch projector on the two components
$A$ and $B$, $P_1$ below, and uses it in every class. Because QSP shapes
the dependence on one momentum coordinate at a time, we choose a state
generated by rotations that each depend on only one coordinate. The
pulses start from a product state of decoupled cells. Its flat-band
Hamiltonian on the two components is the same at every momentum: it is
$Q_0=\tau_y$, or $-\tau_y$ for the honeycomb layers that start in the
complementary state (Sec.~\ref{sec:classes}).
The rotation
\begin{align}
 U_1(\bk)={}&e^{-\iu\bar k_y\tau_z/4}\,\mathscr R_z(-\pi/4)\,
 e^{+\iu\bar k_x\tau_z/2}\notag\\
 &\times\mathscr R_z(-\pi/4)\,e^{-\iu\bar k_y\tau_z/4},
 \label{eq:branch-word}
\end{align}
whose factors act from right to left and in which
$\mathscr R_z(-\pi/4)=e^{\iu\pi\tau_y/4}$ is a fixed $z$ pulse, gives
$Q_1=U_1Q_0U_1^\dagger$ (Sec.~\ref{sec:seed-rotation}), with
\begin{align}
 Q_1(\bk)&=\bn(\bk)\cdot\btau,\qquad
 P_1=\frac{\Id+Q_1}2,\notag\\
 \bn(\bk)&=
 \begin{pmatrix}
 \sin(\bar k_y/2)\cos(\bar k_y/2)(1-\cos\bar k_x)\\
 \sin^2(\bar k_y/2)+\cos^2(\bar k_y/2)\cos\bar k_x\\
 \cos(\bar k_y/2)\sin\bar k_x
 \end{pmatrix}.
 \label{eq:seed}
\end{align}
The unit vector $\bn(\bk)$ covers the unit sphere once as $\bk$ runs
over the Brillouin zone, so $P_1$ has Chern number one
(Sec.~\ref{sec:seed-chern}); we call it the Chern-one state. $U_1$ is
not periodic in momentum, but $P_1$ is (Sec.~\ref{sec:seed-rotation}).

The five strong classes use the same pulses on one, two, or four honeycomb
layers (Table~\ref{tab:targets}). The honeycomb layers combine into
different physical fermions, and this choice, together with the initial
state of each honeycomb layer, sets the symmetry class;
Sec.~\ref{sec:classes} and Fig.~\ref{fig:classes} give the details.

\begin{theorem}[Honeycomb-pulse preparation]
\label{thm:upper}
Let $L_x$ and $L_y$ be odd. Each target state of
Table~\ref{tab:targets} is prepared exactly on the $L_x\times L_y$
torus from a product state of decoupled cells given in
Sec.~\ref{sec:classes}, without ancillas, by a circuit in which every
circuit layer is one spatially uniform honeycomb bond pulse applied to
all honeycomb layers of the target. The depth satisfies
\begin{equation}
 D_{\rm up}\le 8(L_x-1)+16(L_y-1)+5.
 \label{eq:upper}
\end{equation}
\end{theorem}

The same depth suffices for any fixed number of copies prepared in
parallel, which give other values of the strong index
(Sec.~\ref{sec:classes}).
Section~\ref{sec:preparation} builds the circuit and computes the pulse
angles explicitly.

\subsection{Locality lower bound}
\label{sec:results-lower}

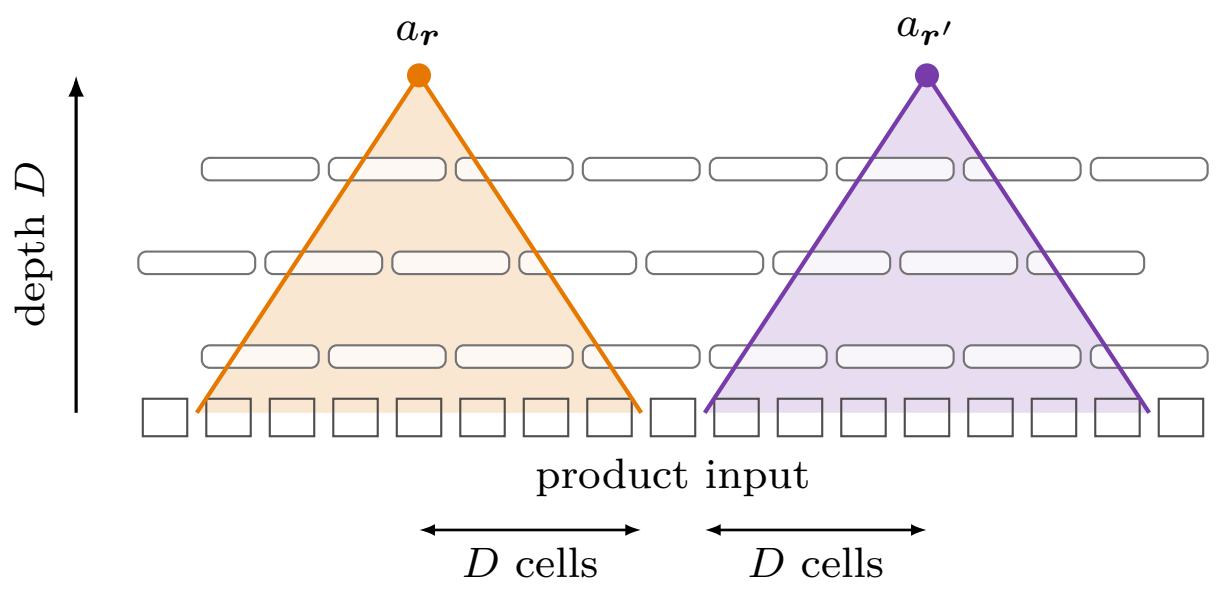
\begin{figure}[t]
\centering
\begin{tikzpicture}[x=0.42cm,y=0.62cm]
  \def\D{3}
  \def\xa{4}
  \def\xb{12}
  \fill[Acol!18] (\xa,\D+0.5) -- ({\xa-\D-0.5},-0.1) -- ({\xa+\D+0.5},-0.1) -- cycle;
  \fill[Bcol!18] (\xb,\D+0.5) -- ({\xb-\D-0.5},-0.1) -- ({\xb+\D+0.5},-0.1) -- cycle;
  \foreach \t in {1,...,\D}{
    \pgfmathtruncatemacro{\off}{mod(\t,2)}
    \foreach \c in {0,2,...,14}{
      \pgfmathsetmacro{\cc}{\c+\off}
      \ifdim \cc pt<16pt
        \draw[fill=white,fill opacity=0.75,draw=black!55,rounded corners=1.5pt,line width=0.4pt]
          ({\cc-0.42},{\t-0.62}) rectangle ({\cc+1.42},{\t-0.38});
      \fi
    }
  }
  \foreach \c in {0,...,16}{
    \draw[black!70,line width=0.4pt] ({\c-0.35},-0.35) rectangle ({\c+0.35},0.05);
  }
  \node[font=\footnotesize,anchor=north] at (8,-0.4) {product input};
  \draw[Acol,line width=0.8pt] ({\xa-\D-0.5},-0.1) -- (\xa,\D+0.5) -- ({\xa+\D+0.5},-0.1);
  \draw[Bcol,line width=0.8pt] ({\xb-\D-0.5},-0.1) -- (\xb,\D+0.5) -- ({\xb+\D+0.5},-0.1);
  \node[circle,fill=Acol,inner sep=1.6pt,label={[font=\footnotesize]above:$a_{\br}$}] at (\xa,\D+0.5) {};
  \node[circle,fill=Bcol,inner sep=1.6pt,label={[font=\footnotesize]above:$a_{\br'}$}] at (\xb,\D+0.5) {};
  \draw[-{Latex[length=4pt]},line width=0.6pt] (-1.4,-0.1) -- (-1.4,\D+0.5);
  \node[font=\footnotesize,rotate=90,anchor=south] at (-1.5,{(\D+0.4)/2}) {depth $D$};
  \draw[{Latex[length=3pt]}-{Latex[length=3pt]},line width=0.5pt] ({\xa},{-1.35}) -- ({\xa+\D+0.5},{-1.35});
  \node[font=\footnotesize,anchor=north] at ({\xa+1.75},-1.35) {$D$ cells};
  \draw[{Latex[length=3pt]}-{Latex[length=3pt]},line width=0.5pt] ({\xb-\D-0.5},{-1.35}) -- ({\xb},{-1.35});
  \node[font=\footnotesize,anchor=north] at ({\xb-1.75},-1.35) {$D$ cells};
\end{tikzpicture}
\caption{Light-cone argument for the lower bound. Time runs upward
through $D$ circuit layers of local gates acting on a product input
(bottom). The fermion operators $a_{\br}$ and $a_{\br'}$ at cells $\br$ and $\br'$ (top) have
backward light cones of half-width $D$. If the cell distance between
$\br$ and $\br'$ exceeds $2D$, the cones are disjoint and the
two-point correlation of $a_{\br}$ and $a_{\br'}$ vanishes exactly
(Lemma~\ref{lem:car-lightcone}), for interacting and
position-dependent gates alike.}
\label{fig:lightcone}
\end{figure}

The lower bound applies to every circuit of
Sec.~\ref{sec:general-circuits}; its physical origin is the finite
speed at which local gates spread correlations
(Fig.~\ref{fig:lightcone}). Set $L_x=L_y=L$, where $L$ may be even or
odd. Let $P_*$ be the Bloch projector of the target, defined at every
momentum of the Brillouin zone, on a one-particle space whose
dimension does not depend on $L$. The bound depends on how fast $P_*$
changes with momentum. We measure this by constants $M_x,M_y$ such that
\begin{align}
 \|P_*(k_x,k_y)-P_*(k_x',k_y)\|
 &\le M_x d_{\Torus}(k_x,k_x'),\notag\\
 \|P_*(k_x,k_y)-P_*(k_x,k_y')\|
 &\le M_y d_{\Torus}(k_y,k_y'),
 \label{eq:regularity}
\end{align}
where $d_{\Torus}$ is the shortest angular distance on the circle and
$\|\cdot\|$ denotes the operator norm. The constants are properties of
the target $P_*$ alone; for a fixed target they do not depend on the
system size $L$. All targets of Table~\ref{tab:targets} satisfy
Eq.~\eqref{eq:regularity} with $M_x=M_y=1/2$ (Sec.~\ref{sec:seed}).

\begin{theorem}[Locality lower bound]
\label{thm:lower}
Suppose $P_*$ has a nonzero two-dimensional stable strong index in
class A, D, C, AII, or DIII and satisfies
Eq.~\eqref{eq:regularity}. Every circuit in the model of
Sec.~\ref{sec:general-circuits} that exactly prepares on the
$L\times L$ torus the Gaussian state with Bloch projector $P_*$ at the
allowed momenta has depth
\begin{equation}
 D\ge \frac{L}{4\pi}-\frac{M_x+M_y}{4}.
 \label{eq:lower}
\end{equation}
\end{theorem}

Because the argument compares strong indices rather than Chern
numbers, the theorem also covers AII and DIII, whose total Chern number
vanishes. For ground states of gapped Hamiltonians the constants follow
from physical quantities. If $P_*$ is the negative-energy projector of
a Bloch Hamiltonian $H(\bk)=\sum_{\br}h_{\br}e^{-\iu\bk\cdot\br}$, with
hopping matrix $h_{\br}$ for the displacement $\br=(r_x,r_y)$ and
energy gap $E_g$, then $M_\mu\le v_\mu/E_g$, where
$v_\mu=\sum_{\br}|r_\mu|\,\|h_{\br}\|$ bounds the velocity
$\partial_{k_\mu}H$ (Sec.~\ref{app:regularity}); Theorem~\ref{thm:lower}
then gives $D\ge L/(4\pi)-(v_x+v_y)/(4E_g)$.
Section~\ref{sec:lower} proves Theorem~\ref{thm:lower}.

Combining the two theorems gives the depth of exact preparation of our
targets up to a constant factor.

\begin{corollary}[Optimal exact depth]
\label{cor:optimal}
For each target state of Table~\ref{tab:targets}, and for any fixed
number of copies whose total strong index is nonzero
(Sec.~\ref{sec:classes}), the minimum depth $D_{\rm min}(L)$ of a
circuit in the model of Sec.~\ref{sec:general-circuits} that prepares
it exactly on an odd $L\times L$ torus obeys
\begin{equation}
 \frac{L}{4\pi}-\frac14
 \le D_{\rm min}(L)\le 24(L-1)+5.
 \label{eq:optimal}
\end{equation}
\end{corollary}

Thus $D_{\rm min}(L)=\Theta(L)$: both bounds grow linearly with $L$,
with coefficients independent of $L$. The upper bound uses spatially
uniform free-fermion pulses, while the lower bound holds for
interacting, nonuniform circuits.

\subsection{Why linear depth: resolving the allowed momenta}
\label{sec:results-match}

The two theorems meet at one scale, the spacing $2\pi/L$ of the allowed
momenta. Write a Bloch projector as a Fourier series,
$P(\bk)=\sum_{\br}G(\br)e^{-\iu\bk\cdot\br}$, where the matrix
$G(\br)$ collects the two-point correlations between cells separated
by $\br$. We call the largest cell distance $|r_x|+|r_y|$ with
$G(\br)\ne0$ the range of $P$. The entries of a Bloch projector of
finite range are trigonometric polynomials, that is, finite Fourier
sums, whose degree in each momentum coordinate is at most the range.
A depth-$D$ circuit creates correlations only up to distance $2D$, so
for $L>8D$ the state it prepares defines a Bloch projector
$\widetilde P$ of range at most $2D$ (Lemmas~\ref{lem:car-lightcone}
and \ref{lem:alias-free}). By Bernstein's inequality, the momentum
derivatives of $\widetilde P$ are then at most $2D$ in operator norm
(Sec.~\ref{sec:lower-comparison}), so $\widetilde P$ varies
appreciably only over momentum intervals of order $1/D$. The
projector $\widetilde P$ equals the target $P_*$ at every allowed
momentum, and every momentum lies within $\pi/L$ of an allowed
momentum in each coordinate. Over this distance $\widetilde P$ changes
by at most $4\pi D/L$ and $P_*$ by at most $\pi(M_x+M_y)/L$
[Eq.~\eqref{eq:regularity}], so $\widetilde P$ stays within
$(\pi/L)(4D+M_x+M_y)$ of $P_*$ at every momentum. On the other hand, a
Bloch projector of finite range carries no strong topology, however
large its range~\cite{Read2017}, whereas $P_*$ does. By
Lemma~\ref{lem:az-homotopy}, $\widetilde P$ must therefore differ from
$P_*$ by at least one, in operator norm, at some momentum between the
allowed ones. The two statements are compatible only if
$(\pi/L)(4D+M_x+M_y)\ge1$, which is the bound of
Eq.~\eqref{eq:lower}: the depth $D$ must grow linearly with $L$.

The construction shows the other side of this scale. Its circuit is a
product of pulse matrices that are smooth and periodic in $\bk$, so it
defines a Bloch projector $P_{\rm circ}$ at every momentum, not only at
the allowed ones. Applied to the constant occupied Bloch state of the
initial state, the circuit gives an occupied Bloch state that is smooth
and periodic over the whole zone, so $P_{\rm circ}$ has Chern number
zero by the argument after Eq.~\eqref{eq:chern}~\cite{Kohmoto1985}
[Eq.~\eqref{eq:circuit-chern-zero}]. Yet $P_{\rm circ}$ equals $P_1$ at
every allowed momentum, so the circuit prepares the Chern-one state
exactly. There is no contradiction: on a finite torus only the values
at the allowed momenta are physical, and $P_{\rm circ}$ differs from
$P_1$ by at least one, in operator norm, at some momentum between them
(Lemma~\ref{lem:az-homotopy}). By the previous paragraph this requires
a range of order $L$, and the range of $P_{\rm circ}$ is at most
$8(L_x-1)+16(L_y-1)$ (Sec.~\ref{sec:circuit-extension}). In the
finite-size sense the prepared state does carry Chern number one: the
discrete Chern number of Fukui, Hatsugai, and Suzuki (FHS), computed
from the values at the allowed momenta alone, equals one for every odd
$L\ge5$ (App.~\ref{app:fhs-targets}). The circuit thus changes the FHS
number from zero to one, although unitary evolution that preserves
translation invariance cannot change the continuum Chern
number~\cite{DAlessioRigol2015,Caio2015}. Ge and Rigol found an
analogous finite-size effect: under a uniform periodic drive, the Bott
index of a finite system can change~\cite{GeRigol2017}.

\section{The Chern-one state and the five targets}
\label{sec:seed}

This section proves the properties of the targets that
Theorems~\ref{thm:upper} and \ref{thm:lower} use.

\subsection{From three one-coordinate rotations to a Chern-one projector}
\label{sec:seed-rotation}

Recall that the initial flat-band Hamiltonian is $Q_0=\tau_y$. We write
a $2\times2$ flat-band Hamiltonian as $Q=\bn\cdot\btau$ with a real
unit vector $\bn$; for $Q_0$, $\bn=(0,1,0)$. We apply $U_1$ of
Eq.~\eqref{eq:branch-word} to it. For a real angle $\vartheta$,
conjugation by $e^{-\iu\vartheta\tau_z}$ rotates $\bn$ by the angle
$2\vartheta$ about the $z$ axis. Conjugation by
$\mathscr R_z(-\pi/4)=e^{\iu\pi\tau_y/4}$ maps a vector with components
$(n_x,n_y,n_z)$ to $(-n_z,n_y,n_x)$. Applying
Eq.~\eqref{eq:branch-word} from right to left to $(0,1,0)$ gives
\begin{align}
 (0,1,0)&\longmapsto
 (-\sin(\bar k_y/2),\cos(\bar k_y/2),0)\notag\\
 &\longmapsto(0,\cos(\bar k_y/2),-\sin(\bar k_y/2))\notag\\
 &\longmapsto
 \begin{pmatrix}
 \cos(\bar k_y/2)\sin\bar k_x\\
 \cos(\bar k_y/2)\cos\bar k_x\\
 -\sin(\bar k_y/2)
 \end{pmatrix}\notag\\
 &\longmapsto
 \begin{pmatrix}
 \sin(\bar k_y/2)\\
 \cos(\bar k_y/2)\cos\bar k_x\\
 \cos(\bar k_y/2)\sin\bar k_x
 \end{pmatrix}\notag\\
 &\longmapsto\bn(\bk).
 \label{eq:seed-rotation-chain}
\end{align}
In particular, with $P_0=(\Id+\tau_y)/2$,
\begin{equation}
 U_1Q_0U_1^\dagger=Q_1,\qquad
 U_1P_0U_1^\dagger=P_1.
\end{equation}
Every step of Eq.~\eqref{eq:seed-rotation-chain} is a rotation, so
$\bn$ is a unit vector.

Although $U_1$ is not periodic, $P_1$ is. The $k_x$ factor
$e^{\iu\bar k_x\tau_z/2}$ changes sign when $k_x$ crosses $\pm\pi$, and
this sign cancels in $U_1Q_0U_1^\dagger$. At $\bar k_y=\pm\pi$, the
first and third components of $\bn$ vanish and the second is one,
independently of $k_x$. This common value makes $\bn$ continuous across
the line $k_y=\pm\pi$. Hence $\bn$ defines a continuous flat-band
Hamiltonian $Q_1$ on the full torus. It is smooth for
$\bar k_y\in(-\pi,\pi)$, with bounded derivatives on both sides of the
line $k_y=\pm\pi$.

\subsection{Chern number and the constants \texorpdfstring{$M_x,M_y$}{Mx, My}}
\label{sec:seed-chern}

The strong indices in Table~\ref{tab:targets} rest on the Chern number
one of $P_1$, and Theorem~\ref{thm:lower} with
Corollary~\ref{cor:optimal} needs the
constants $M_x,M_y$ of Eq.~\eqref{eq:regularity}. Both follow from the
derivatives of $\bn$. For $\bar k_y\in(-\pi,\pi)$, where
$\bar k_\mu=k_\mu$,
\begin{align}
 \partial_{k_x}\bn&=
 \begin{pmatrix}
 \sin(k_y/2)\cos(k_y/2)\sin k_x\\
 -\cos^2(k_y/2)\sin k_x\\
 \cos(k_y/2)\cos k_x
 \end{pmatrix},\notag\\
 \partial_{k_y}\bn&=
 \begin{pmatrix}
 \tfrac12[\cos^2(k_y/2)-\sin^2(k_y/2)](1-\cos k_x)\\
 \sin(k_y/2)\cos(k_y/2)(1-\cos k_x)\\
 -\tfrac12\sin(k_y/2)\sin k_x
 \end{pmatrix}.
 \label{eq:seed-derivatives}
\end{align}
For $P=(\Id+\bn\cdot\btau)/2$, use
$(\bm a\cdot\btau)(\bm b\cdot\btau)
=(\bm a\cdot\bm b)\Id+\iu(\bm a\times\bm b)\cdot\btau$.
Thus
\begin{align}
 [\partial_{k_x}P,\partial_{k_y}P]
 &=\frac{\iu}{2}(\partial_{k_x}\bn\times\partial_{k_y}\bn)\cdot\btau,\notag\\
 -\iu\Tr\!\left(P[\partial_{k_x}P,\partial_{k_y}P]\right)
 &=\frac12\bn\cdot(\partial_{k_x}\bn\times\partial_{k_y}\bn).
 \label{eq:pauli-curvature}
\end{align}
The first equality uses the Pauli commutator. The second uses
$\Tr\tau_a=0$ and $\Tr(\tau_a\tau_b)=2\delta_{ab}$.
Substituting Eq.~\eqref{eq:seed-derivatives} into the cross product
gives
\begin{align}
 \partial_{k_x}\bn\times\partial_{k_y}\bn
 &=\frac{\cos(k_y/2)(1-\cos k_x)}2\,\bn,\notag\\
 \bn\cdot(\partial_{k_x}\bn\times\partial_{k_y}\bn)
 &=\frac12\cos(k_y/2)(1-\cos k_x)
 \label{eq:determinant-density}\\
 &=\sin^2(k_x/2)\cos(k_y/2).
 \label{eq:seed-density}
\end{align}
The Berry curvature is therefore nonnegative, and
\begin{align}
 c_1(P_1)
 &=\frac1{4\pi}
 \int_{-\pi}^{\pi}\sin^2(k_x/2)\,dk_x
 \int_{-\pi}^{\pi}\cos(k_y/2)\,dk_y\notag\\
 &=\frac1{4\pi}
 \left[\frac{k_x}{2}-\frac{\sin k_x}{2}\right]_{-\pi}^{\pi}
 \left[2\sin(k_y/2)\right]_{-\pi}^{\pi}\notag\\
 &=\frac1{4\pi}\,\pi\,4=1.
 \label{eq:seed-chern-calculation}
\end{align}
$P_1$ is continuous everywhere and smooth except on the line
$k_y=\pm\pi$, which has zero area, so Eq.~\eqref{eq:chern} gives its
Chern number.

Some targets also contain the complementary projector $\Id-P$, which
describes the state with occupied and empty states exchanged. Since
$\partial_{k_\mu}(\Id-P)=-\partial_{k_\mu}P$ and the trace of a
commutator vanishes,
\begin{align*}
 &\Tr\!\left((\Id-P)
 [\partial_{k_x}(\Id-P),\partial_{k_y}(\Id-P)]\right)\\
 &\qquad=-\Tr\!\left(P[\partial_{k_x}P,\partial_{k_y}P]\right),
\end{align*}
so Eq.~\eqref{eq:chern} gives $c_1(\Id-P)=-c_1(P)$, and
$c_1(\Id-P_1)=-1$.

The constants $M_x,M_y$ follow from the derivative norms,
\begin{align}
 |\partial_{k_x}\bn|^2&=\cos^2(k_y/2)\le1,\notag\\
 |\partial_{k_y}\bn|^2
 &=\frac14\left[(1-\cos k_x)^2
 +\sin^2(k_y/2)\sin^2 k_x\right]\notag\\
 &\le\frac12(1-\cos k_x)\le1.
 \label{eq:seed-vector-regularity}
\end{align}
For any real vector $\bm v$, $\|\bm v\cdot\btau\|=|\bm v|$.
Consequently $\|\partial_{k_\mu}P_1\|=|\partial_{k_\mu}\bn|/2\le1/2$
for $\bar k_y\in(-\pi,\pi)$. Integrating along a path in which only
$k_\mu$ changes gives
$\|P_1(\bk')-P_1(\bk)\|\le\frac12d_{\Torus}(k_\mu,k_\mu')$; since $P_1$
is continuous across the line $k_y=\pm\pi$, the same bound holds for
paths that cross it. Therefore $P_1$ satisfies
Eq.~\eqref{eq:regularity} with
\begin{equation}
 M_x=M_y=\frac12.
 \label{eq:seed-lipschitz}
\end{equation}

\subsection{Projector conventions}
\label{app:projectors}

The honeycomb layers provide Majorana fermions. Classes D and DIII
use them directly; classes A and AII, and each spin species in class C,
use complex fermions formed from pairs of honeycomb layers.

\emph{Majorana description.} For $2n$ Majorana operators $\gamma_a$,
with $\{\gamma_a,\gamma_b\}=2\delta_{ab}$, the covariance
\begin{equation}
 \Gamma_{ab}=\frac{\iu}{2}\langle[\gamma_a,\gamma_b]\rangle
 \label{eq:covariance}
\end{equation}
is real and antisymmetric, and $\Gamma_{ab}=\langle\iu\gamma_a\gamma_b\rangle$
for $a\ne b$. A pair of Majoranas with a definite value
$\iu\gamma_1\gamma_2=s=\pm1$ has the covariance block
\begin{equation*}
 s\begin{pmatrix}0&1\\-1&0\end{pmatrix},
\end{equation*}
which squares to $-\Id$. A pure Gaussian state is a product of such
pairs after a real orthogonal transformation of its Majoranas.
Consequently $\Gamma^2=-\Id$, and therefore
\begin{equation}
 P=\frac{\Id+\iu\Gamma}{2},\qquad Q=\iu\Gamma
 \label{eq:majorana-projector}
\end{equation}
are a Hermitian projector of rank $n$ and a Hermitian matrix with
$Q^2=\Id$, both in dimension $2n$. In momentum space, the Fourier
transform $\Gamma(\bk)$ of a real covariance obeys
$\Gamma(\bk)^*=\Gamma(-\bk)$, so every Majorana Bloch projector
satisfies $Q(\bk)^*=-Q(-\bk)$: particle-hole symmetry with
$\mathcal C=K$ is built into this description.

\emph{Number-conserving states.} Two honeycomb layers form the complex
fermions $c_a=(\gamma_{a,1}+\iu\gamma_{a,2})/2$ of
Sec.~\ref{sec:honeycomb-pulses}. If the two layers are uncorrelated and
have the same covariance $\Gamma$, as they do when they start in the
same state and receive the same pulses, the anticommutation relations
give
\begin{equation}
 \langle c_b^\dagger c_a\rangle=\frac{\delta_{ab}+\iu\Gamma_{ab}}{2},
 \qquad \langle c_bc_a\rangle=0.
 \label{eq:pair-projector}
\end{equation}
The state of the pair then conserves particle number, and its Bloch
projector $P$ of Sec.~\ref{sec:topology-background} equals the
Majorana projector of Eq.~\eqref{eq:majorana-projector} of either
layer.

\emph{Spin-reduced description (class C).} For a superconductor with
spin-rotation symmetry and modes $c_{a\uparrow},c_{a\downarrow}$ in
each cell, the spin-reduced Nambu vector keeps the spin-up particles
and the spin-down holes,
\begin{equation}
 \Psi_{\bk}=
 \begin{pmatrix}
 c_{\bk,A\uparrow}\\c_{\bk,B\uparrow}\\
 c_{-\bk,A\downarrow}^{\dagger}\\c_{-\bk,B\downarrow}^{\dagger}
 \end{pmatrix},
 \label{eq:c-reduced-vector}
\end{equation}
and its Bloch projector is
$(P_C)_{ab}(\bk)=\langle\Psi_{\bk,b}^{\dagger}\Psi_{\bk,a}\rangle$.
The full Nambu description, which contains every mode both as a
particle and as a hole, would double each Chern number. We use the
spin-reduced description throughout~\cite{Read2017,Ryu2010}.

\subsection{The five targets, class by class}
\label{sec:classes}

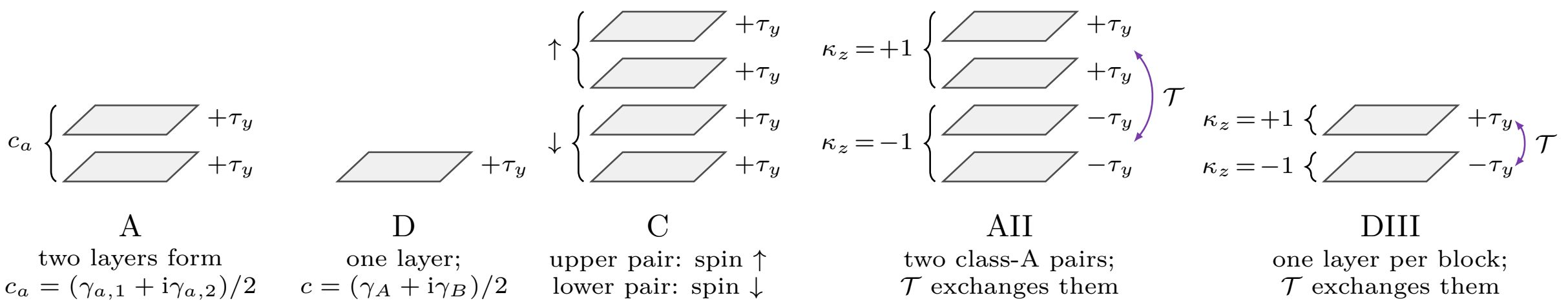
\begin{figure*}[t]
\centering
\begin{tikzpicture}[x=1cm,y=1cm]
  \def\sheet#1#2#3{
    \filldraw[fill=gray!12,draw=black!70,line width=0.5pt]
      (#1,#2) -- ++(1.05,0) -- ++(0.32,0.3) -- ++(-1.05,0) -- cycle;
    \node[font=\scriptsize,anchor=west,inner sep=1pt] at ($(#1,#2)+(1.42,0.15)$) {$#3\tau_y$};}
  \def\grp#1#2#3#4{
    \draw[decorate,decoration={brace,amplitude=3pt},line width=0.5pt]
      (#1,#2) -- (#1,#3) node[midway,left=3pt,font=\scriptsize,align=right] {#4};}
  \def\gap{0.48}
  \begin{scope}[shift={(0.6,0)}]
    \sheet{0}{0}{+}\sheet{0}{\gap}{+}
    \grp{-0.08}{0.0}{0.78}{$c_a$}
    \node[font=\small] at (0.68,-0.45) {A};
    \node[font=\scriptsize,align=center] at (0.68,-0.95) {two layers form\\ $c_a=(\gamma_{a,1}+\iu\gamma_{a,2})/2$};
  \end{scope}
  \begin{scope}[shift={(3.4,0)}]
    \sheet{0}{0}{+}
    \node[font=\small] at (0.68,-0.45) {D};
    \node[font=\scriptsize,align=center] at (0.68,-0.95) {one layer;\\ $c=(\gamma_A+\iu\gamma_B)/2$};
  \end{scope}
  \begin{scope}[shift={(6.0,0)}]
    \sheet{0}{0}{+}\sheet{0}{\gap}{+}\sheet{0}{2*\gap}{+}\sheet{0}{3*\gap}{+}
    \grp{-0.08}{0.0}{0.78}{$\downarrow$}
    \grp{-0.08}{0.96}{1.74}{$\uparrow$}
    \node[font=\small] at (0.68,-0.45) {C};
    \node[font=\scriptsize,align=center] at (0.68,-0.95) {upper pair: spin $\uparrow$\\ lower pair: spin $\downarrow$};
  \end{scope}
  \begin{scope}[shift={(9.6,0)}]
    \sheet{0}{0}{-}\sheet{0}{\gap}{-}\sheet{0}{2*\gap}{+}\sheet{0}{3*\gap}{+}
    \grp{-0.08}{0.0}{0.78}{$\kappa_z\!=\!-1$}
    \grp{-0.08}{0.96}{1.74}{$\kappa_z\!=\!+1$}
    \draw[{Latex[length=3pt]}-{Latex[length=3pt]},line width=0.5pt,Bcol]
      (1.95,0.4) to[bend right=40] node[right,font=\scriptsize,black] {$\mathcal T$} (1.95,1.35);
    \node[font=\small] at (0.68,-0.45) {AII};
    \node[font=\scriptsize,align=center] at (0.68,-0.95) {two class-A pairs;\\ $\mathcal T$ exchanges them};
  \end{scope}
  \begin{scope}[shift={(13.5,0)}]
    \sheet{0}{0}{-}\sheet{0}{\gap}{+}
    \grp{-0.08}{0.0}{0.3}{$\kappa_z\!=\!-1$}
    \grp{-0.08}{0.48}{0.78}{$\kappa_z\!=\!+1$}
    \draw[{Latex[length=3pt]}-{Latex[length=3pt]},line width=0.5pt,Bcol]
      (1.95,0.15) to[bend right=40] node[right,font=\scriptsize,black] {$\mathcal T$} (1.95,0.63);
    \node[font=\small] at (0.68,-0.45) {DIII};
    \node[font=\scriptsize,align=center] at (0.68,-0.95) {one layer per block;\\ $\mathcal T$ exchanges them};
  \end{scope}
\end{tikzpicture}
\caption{Honeycomb layers and physical fermions of the five targets.
Each slab is one honeycomb layer; all honeycomb layers receive the same
pulses and are never coupled. The label on a honeycomb layer is the
flat-band Hamiltonian of its initial state on the $A,B$ components:
$+\tau_y$ ($\iu\gamma_{A,\br}\gamma_{B,\br}=-1$ on every $z$ bond) or
$-\tau_y$ ($+1$). Braces group honeycomb layers into the physical
fermions or the blocks $\kappa_z=\pm1$ of Sec.~\ref{sec:classes}, and
arrows marked $\mathcal T$ join the blocks exchanged by time reversal.
In C, the upper pair forms the spin-up and the lower pair the
spin-down fermions; in the spin-reduced description, the block
$\kappa_z=+1$ holds the spin-up particles at $\bk$ and $\kappa_z=-1$
the spin-down holes at $-\bk$.}
\label{fig:classes}
\end{figure*}

All five targets are built from $P_1$. They differ in the number of
honeycomb layers, in how the Majoranas of the honeycomb layers combine
into physical fermions, and in the initial state of each honeycomb
layer (Fig.~\ref{fig:classes}). The initial state is a product state of
the matter fermions. With $u_{ij}=1$, the matter-fermion Hamiltonian of
the $z$ bonds is $H_z=\iu J_z^{\rm K}\sum_{\br}\gamma_{A,\br}\gamma_{B,\br}$
[Eq.~\eqref{eq:kitaev-bond-matter}]. For $J_z^{\rm K}>0$ its unique
ground state has $\iu\gamma_{A,\br}\gamma_{B,\br}=-1$ on every $z$ bond.
Its covariance [Eq.~\eqref{eq:covariance}] has
$\Gamma_{AB}=\langle\iu\gamma_A\gamma_B\rangle=-1=-\Gamma_{BA}$ in every
cell, so its flat-band Hamiltonian on the components $A$ and $B$ is
$\iu\Gamma=\tau_y$ [Eq.~\eqref{eq:majorana-projector}], with projector
$P_0=(\Id+\tau_y)/2$. We call it the $+\tau_y$ state of a honeycomb
layer. The ground state for $J_z^{\rm K}<0$, with
$\iu\gamma_{A,\br}\gamma_{B,\br}=+1$ on every $z$ bond, is the
$-\tau_y$ state, with flat-band Hamiltonian $-\tau_y$ and projector
$\Id-P_0$. By Eq.~\eqref{eq:pair-projector}, a pair of honeycomb layers
in the same state has the same projector for its complex fermions. For classes with two blocks of components,
$\kappa_x,\kappa_y,\kappa_z$ denote Pauli matrices and $\Id_\kappa$ the
identity acting on the block index; the blocks are specified for each
class below. In classes C and AII, spin labels the fermion species
formed by a pair of honeycomb layers, not the Kitaev spins.

\emph{Class A (example: Chern insulator).} Two honeycomb layers form
the complex fermions
$c_{a,\br}=(\gamma_{a,1,\br}+\iu\gamma_{a,2,\br})/2$ of
Sec.~\ref{sec:honeycomb-pulses}, $a=A,B$, and the pulses act on them as
hopping pulses [Eq.~\eqref{eq:hopping-majorana-copies}]. Both
honeycomb layers start in the $+\tau_y$ state, so the complex fermions
have the Bloch projector $P_0=(\Id+\tau_y)/2=\psi_0\psi_0^\dagger$
[Eq.~\eqref{eq:pair-projector}], with the occupied Bloch state
$\psi_0=(1,\iu)^T/\sqrt2$ at every momentum, and $Q_0=\tau_y$. In real
space, each cell holds one fermion in the state
\begin{equation}
 |\Omega\rangle
 =\frac{c_A^\dagger+\iu c_B^\dagger}{\sqrt2}|0\rangle.
 \label{eq:atomic-A}
\end{equation}
The target is $P_1$ of Eq.~\eqref{eq:seed}, with no
antiunitary symmetry imposed: a Chern insulator in which the fermions
fill one band with $c_1(P_1)=1$, so the Hall conductance is $e^2/h$.

\emph{Class D (example: $p+\iu p$ superconductor).} One honeycomb
layer suffices, read in the Majorana description. It starts in the
$+\tau_y$ state, so $Q_0=\tau_y$. The target is $Q_1$, read as a Majorana flat-band Hamiltonian. As
such it must obey $Q(\bk)^*=-Q(-\bk)$ (Sec.~\ref{app:projectors}), and
it does because of the parities of the components of $\bn$ under
inversion, $n_x(-\bk)=-n_x(\bk)$, $n_y(-\bk)=n_y(\bk)$, and
$n_z(-\bk)=-n_z(\bk)$. Together with $\tau_x^*=\tau_x$,
$\tau_z^*=\tau_z$, and $\tau_y^*=-\tau_y$, they give
\begin{equation}
 Q_1(-\bk)=-Q_1(\bk)^*,
 \label{eq:seed-phs}
\end{equation}
that is, $P_1(\bk)^*=\Id-P_1(-\bk)$: the second relation of
Eq.~\eqref{eq:az-projectors} with $U_C=\Id$, so $\mathcal C=K$ and
$\mathcal C^2=+1$. The index is $c_1(P_1)=1$. In terms of the complex
fermion $c=(\gamma_A+\iu\gamma_B)/2$, whose vacuum is the initial
state, the even component $n_y$ of $Q_1$ conserves the number of $c$
fermions, while $n_x$ and $n_z$, odd in $\bk$, are pairing amplitudes:
the target is a paired state of spinless fermions with Chern number
one, as in a $p+\iu p$ superconductor.

\emph{Class C (example: $d+\iu d$ superconductor).} Four honeycomb
layers are used: the upper pair forms spin-up fermions
$c_{a\uparrow}=(\gamma_{a,1}+\iu\gamma_{a,2})/2$, and the lower pair
forms spin-down fermions $c_{a\downarrow}$ in the same way. The same
pulse on all four honeycomb layers is a hopping pulse that acts
identically on both spins; its generator is a sum of identical spin-up
and spin-down terms and commutes with $SU(2)$ spin rotations. All four
honeycomb layers start in the $+\tau_y$ state, so, as in class A, each
spin fills $\psi_0$, and
\begin{equation*}
 |\Omega\rangle
 =\frac{(c_{A\uparrow}^\dagger+\iu c_{B\uparrow}^\dagger)
 (c_{A\downarrow}^\dagger+\iu c_{B\downarrow}^\dagger)}{2}|0\rangle,
\end{equation*}
a spin singlet with two fermions per cell. Both spins
have the number-conserving Bloch projector $P$ ($P_0$ initially, $P_1$
finally), and there is no pairing. Our class-C representative is thus
a spin-degenerate Chern state without pairing. Viewed as a
superconductor, that is, without imposing charge conservation, it has
spin-rotation symmetry and no time-reversal symmetry, so it lies in
class C, the class of $d+\iu d$ superconductors. In the spin-reduced vector of
Eq.~\eqref{eq:c-reduced-vector}, the hole block follows from the
canonical anticommutation relations:
\begin{align*}
 \langle c_{-\bk,b\downarrow}c_{-\bk,a\downarrow}^\dagger\rangle
 &=\delta_{ab}-\langle
 c_{-\bk,a\downarrow}^\dagger c_{-\bk,b\downarrow}\rangle\\
 &=[\Id-P(-\bk)^T]_{ab}.
\end{align*}
Hermiticity and Eq.~\eqref{eq:seed-phs} give
$P_1(-\bk)^T=P_1(-\bk)^*=\Id-P_1(\bk)$, hence
\begin{align}
 P_C(\bk)
 &=P_1(\bk)\oplus[\Id-P_1(-\bk)^T]\notag\\
 &=\Id_\kappa\otimes P_1(\bk),
 \label{eq:c-reduced-projector}
\end{align}
where $\kappa_z=+1$ labels the particle block and $\kappa_z=-1$ the
hole block. The same calculation with $P_0$ gives the initial
flat-band Hamiltonian $\Id_\kappa\otimes Q_0=\Id_\kappa\otimes\tau_y$.
With $U_C=\iu\kappa_y\otimes\Id$,
\begin{align}
 U_C Q_C(\bk)^*U_C^\dagger
 &=\Id_\kappa\otimes Q_1(\bk)^*\notag\\
 &=-Q_C(-\bk),\qquad U_CU_C^*=-\Id,
 \label{eq:c-symmetry-check}
\end{align}
where $Q_C=2P_C-\Id=\Id_\kappa\otimes Q_1$. This is the class-C
relation, with $\mathcal C^2=-1$. The occupied projector is
$P_1\oplus P_1$, with $c_1(P_C)=2$, the smallest nonzero value in
$2\Z$.

\emph{Class AII (example: quantum spin Hall insulator).} Four
honeycomb layers form two class-A pairs: the upper pair gives
$c_{a\uparrow}$ (block $\kappa_z=+1$), and the lower pair gives
$c_{a\downarrow}$ (block $\kappa_z=-1$). Time reversal of spin-$1/2$
fermions, $c_\uparrow\mapsto c_\downarrow$ and
$c_\downarrow\mapsto-c_\uparrow$, has $U_T=\iu\kappa_y\otimes\Id$, which
is real and obeys $U_TU_T^*=-\Id$. Because time reversal is
antiunitary, it maps the $+\tau_y$ state of one pair to the $-\tau_y$
state of the other. A time-reversal-invariant initial state with the
upper pair in the $+\tau_y$ state therefore has the lower pair in the
$-\tau_y$ state, whose projector $\Id-P_0$ is that of the one-fermion
cell state $(c_A^\dagger-\iu c_B^\dagger)|0\rangle/\sqrt2$. The initial
cell state
\begin{equation*}
 |\Omega\rangle
 =\frac{(c_{A\uparrow}^\dagger+\iu c_{B\uparrow}^\dagger)
 (c_{A\downarrow}^\dagger-\iu c_{B\downarrow}^\dagger)}{2}|0\rangle
\end{equation*}
has two fermions per cell, and the initial flat-band Hamiltonian is
$\tau_y\oplus(-\tau_y)=\kappa_z\otimes Q_0$. The target is
\begin{equation}
 Q_{\rm TR}=\kappa_z\otimes Q_1=Q_1\oplus(-Q_1),
 \label{eq:tr-double}
\end{equation}
with occupied projector $P_1\oplus(\Id-P_1)$. Since
$\kappa_y\kappa_z\kappa_y=-\kappa_z$,
\begin{align}
 U_T Q_{\rm TR}(\bk)^*U_T^\dagger
 &=(-\kappa_z)\otimes Q_1(\bk)^*\notag\\
 &=\kappa_z\otimes Q_1(-\bk)
 =Q_{\rm TR}(-\bk).
 \label{eq:tr-symmetry-check}
\end{align}
Spin-up fermions fill a band with Chern number one and spin-down
fermions a band with Chern number minus one
(Sec.~\ref{sec:seed-chern}), so the total Chern number vanishes and the
strong $\Z_2$ index is $\nu=c_1(P_1)\bmod2=1$: the target is a quantum
spin Hall insulator made of two time-reversed Chern insulators, with
one Kramers pair of helical edge modes.

\emph{Class DIII (example: helical superconductor).} Two honeycomb
layers, labeled $+$ (block $\kappa_z=+1$, upper in
Fig.~\ref{fig:classes}) and $-$ (block $\kappa_z=-1$, lower), are read
in the Majorana description. Time reversal is the antiunitary operator
$\widehat T$ on Fock space with
\begin{equation}
 \widehat T\gamma_{a,+}\widehat T^{-1}=\gamma_{a,-},\qquad
 \widehat T\gamma_{a,-}\widehat T^{-1}=-\gamma_{a,+},\qquad a=A,B,
 \label{eq:diii-physical-tr}
\end{equation}
so $U_T=\iu\kappa_y\otimes\Id$ and $U_TU_T^*=-\Id$. Because
$\widehat T$ reverses the sign of $\iu$, it maps
$\iu\gamma_{A,+}\gamma_{B,+}$ to $-\iu\gamma_{A,-}\gamma_{B,-}$ and
$\iu\gamma_{A,-}\gamma_{B,-}$ to $-\iu\gamma_{A,+}\gamma_{B,+}$: as in
AII, it exchanges the $+\tau_y$ state of one layer with the $-\tau_y$
state of the other. A time-reversal-invariant initial state with the
upper layer in the $+\tau_y$ state therefore has the lower one in the
$-\tau_y$ state, and the initial flat-band Hamiltonian is
$\kappa_z\otimes Q_0$. The target is $\kappa_z\otimes Q_1$ of
Eq.~\eqref{eq:tr-double}, and Eq.~\eqref{eq:tr-symmetry-check} holds as
in AII. As a Majorana flat-band Hamiltonian, the target must obey
$Q(\bk)^*=-Q(-\bk)$, and it does by Eq.~\eqref{eq:seed-phs}:
$Q_{\rm TR}(\bk)^*=\kappa_z\otimes Q_1(\bk)^*=-Q_{\rm TR}(-\bk)$, which is
particle-hole symmetry with $\mathcal C=K$. Thus $\mathcal T^2=-1$ and
$\mathcal C^2=+1$, as in class DIII, and $\nu=1$ as in AII. In terms of
$c_+=(\gamma_{A,+}+\iu\gamma_{B,+})/2$ and
$c_-=(\gamma_{A,-}-\iu\gamma_{B,-})/2$, whose common vacuum is the
initial state, $\widehat T$ is the spin-$1/2$ time reversal
$c_+\mapsto c_-$, $c_-\mapsto-c_+$, and the blocks $\kappa_z=\pm1$, with
Chern numbers $\pm1$, are the spin-up $p+\iu p$ and spin-down
$p-\iu p$ parts of a helical superconductor.

The number of physical modes per cell, one complex mode in D, two in
A and DIII, and four in C and AII, does not depend on the system size,
as Theorem~\ref{thm:lower} requires. Every target is
a direct sum of copies of $P_1$ and $\Id-P_1$. The operator norm of a
block-diagonal matrix is the largest norm of its blocks, and
$[\Id-P_1(\bk)]-[\Id-P_1(\bk')]=-[P_1(\bk)-P_1(\bk')]$, so all targets
satisfy Eq.~\eqref{eq:regularity} with $M_x=M_y=1/2$.

Copies give other indices. A block started in the $-\tau_y$ state ends
in $-Q_1$, that is, in $\Id-P_1$, with Chern number $-1$
[Eq.~\eqref{eq:complement-preparation}]. Hence $|c_1|$ parallel copies
of the A or D target, or $|c_1|/2$ copies of the C target, all started
in the $-\tau_y$ state when $c_1<0$, give any fixed Chern number
$c_1\ne0$ in A and D and any fixed even $c_1\ne0$ in C. The strong index
is additive over copies, and the copies satisfy
Eq.~\eqref{eq:regularity} with the same constants.
Corollary~\ref{cor:optimal} covers every set of copies whose total
strong index is nonzero; for example, an odd number of AII or DIII
copies has $\nu=1$.

\section{Exact preparation with honeycomb pulses}
\label{sec:preparation}

We now prove Theorem~\ref{thm:upper}. The momentum-dependent factors
of the rotation $U_1$ of Eq.~\eqref{eq:branch-word} have the form
$e^{-\iu\vartheta(k_\mu)\tau_z}$, each depending on one momentum
coordinate; we call them one-coordinate rotations. They are not
periodic, but on the finite torus they are needed only at the allowed
momenta (Sec.~\ref{sec:qsp-background}).

A pulse sequence with $d$ signals, Eq.~\eqref{eq:qsp-background}, is a
$2\times2$ matrix function $F(k)$ of one coordinate $k=k_\mu$. Each
signal contains only $e^{\pm\iu k}$, so every entry of $F$ is a
trigonometric polynomial $\sum_{n=-d}^{d}a_ne^{\iu nk}$
(Sec.~\ref{sec:results-match}), whose degree is the largest $|n|$ with
$a_n\ne0$. With $\zeta=e^{\iu k}$, it is a Laurent polynomial in
$\zeta$. The Fourier coefficients are real-space amplitudes, so $d$ signals move a
fermion by at most $d$ cells.

Every pulse matrix also obeys a conjugation symmetry. By
Eq.~\eqref{eq:bond-generators}, $g_\mu(-k)=-g_\mu(k)^*$ and
$g_z^*=-g_z$, so
$\mathscr R_\mu(\theta;-k)=\exp[\iu\theta g_\mu(k)^*]
=\mathscr R_\mu(\theta;k)^*$ and $\mathscr R_z(\theta)^*=\mathscr R_z(\theta)$
for every real angle $\theta$. Every pulse sequence therefore satisfies
\begin{equation}
 F(-k)=F(k)^*.
 \label{eq:conj-symmetry}
\end{equation}
Writing $F(k)=\sum_nF_ne^{\iu nk}$, Eq.~\eqref{eq:conj-symmetry} states
that the Fourier coefficient matrices $F_n$ are real $2\times2$
matrices. The problem thus reduces to the following one-dimensional
statement.

\begin{proposition}[One-coordinate pulse sequence]
\label{prop:profile}
Let $L=2m+1$, $k_p=2\pi p/L$ with $p=-m,\ldots,m$, and let the
prescribed real angles $\vartheta_p$ obey
$\vartheta_{-p}=-\vartheta_p$ and $\vartheta_0=0$.
There is a $2\times2$ matrix function $F_L(k)$, unitary with
determinant one at every $k$, whose entries are trigonometric
polynomials of degree at most $4(L-1)$, with
\begin{equation*}
 F_L(k_p)=e^{-\iu\vartheta_p\tau_z},\qquad
 F_L(-k)=F_L(k)^*.
\end{equation*}
It is realized exactly by at most $8(L-1)+1$ circuit layers, each a
spatially uniform $z$ pulse or $x$ (or $y$) pulse acting on the $A$
and $B$ components.
\end{proposition}

The conditions on the angles make the prescribed matrices compatible
with Eq.~\eqref{eq:conj-symmetry}. Since $\tau_z$ is real,
$(e^{-\iu\vartheta_p\tau_z})^*=e^{-\iu\vartheta_{-p}\tau_z}$ holds when
$\vartheta_{-p}=-\vartheta_p$, and at $k_0=0$ the matrix must be real,
which $\vartheta_0=0$ ensures. We prove the proposition in four
steps. Positive interpolation
(Sec.~\ref{app:completion}) fixes the values at the allowed momenta
while keeping the norm of the rotation at most one. A Fej\'er--Riesz
completion (Sec.~\ref{sec:completion}) restores exact unitarity
between the allowed momenta without changing those values. A product
decomposition (Sec.~\ref{app:factorization}) factors the resulting
matrix function into signals and control pulses with explicit angles,
and Sec.~\ref{app:gates} counts the bond pulses. Interpolation, the
Fej\'er--Riesz theorem, and the product decomposition are standard
tools~\cite{FejerRiesz,Haah2019}; we give the short proofs because we
need versions with real coefficients and only even powers.
Sections~\ref{sec:assembly} and \ref{sec:physical-doubling} then
assemble the circuit for all five strong classes, and
Sec.~\ref{sec:circuit-extension} computes the Chern number of the
circuit's own Bloch projector.

\subsection{Positive interpolation}
\label{app:completion}

Consider one coordinate of odd length $L=2m+1$, with allowed momenta
$k_p=2\pi p/L$, $p=-m,\ldots,m$, and real angles $\vartheta_p$ that
satisfy $\vartheta_{-p}=-\vartheta_p$ and $\vartheta_0=0$. We first
build two real trigonometric polynomials $f_0,f_z$ such that
$f_0+\iu f_z$ equals $e^{\iu\vartheta_p}$ at each $k_p$ and never leaves
the closed unit disk. Define the weights
\begin{equation}
 \lambda_p(k)=\frac1L\sum_{n=-m}^{m}e^{2\iu n(k-k_p)},\qquad
 \rho_p(k)=|\lambda_p(k)|^2.
 \label{eq:cardinal}
\end{equation}
At an allowed momentum $k_{p'}$,
\begin{equation}
 \lambda_p(k_{p'})=\frac1L\sum_{n=-m}^{m}
 \exp\!\left(\frac{4\pi\iu n(p'-p)}L\right)=\delta_{pp'}.
 \label{eq:cardinal-proof}
\end{equation}
For $p'=p$ every term is one. For $p'\ne p$, the terms are the $L$
consecutive powers $n=-m,\ldots,m$ of $e^{4\pi\iu(p'-p)/L}$. This is an
$L$th root of unity different from one, because $L$ is odd and
$0<|p'-p|<L$, so $L$ does not divide $2(p'-p)$; the sum of $L$
consecutive powers of such a root vanishes. Expanding the squared
modulus gives
\begin{equation}
 \sum_p\rho_p(k)
 =\frac1{L^2}\sum_{n,n'=-m}^{m}e^{2\iu(n-n')k}
 \sum_p e^{-2\iu(n-n')k_p}=1.
 \label{eq:partition-proof}
\end{equation}
The inner sum is $L\delta_{nn'}$ by the same argument, with $n-n'$ in
place of $p'-p$, since $|n-n'|\le L-1$. The weights are nonnegative,
$\rho_p\ge0$. Each $\rho_p$ contains only the powers $e^{2\iu\ell k}$
with $|\ell|\le L-1$, so it has degree at most $2(L-1)$ and only even
powers of $\zeta=e^{\iu k}$. In particular $\rho_p(k+\pi)=\rho_p(k)$,
and $\rho_{-p}(-k)=\rho_p(k)$ because $\lambda_{-p}(-k)=\lambda_p(k)^*$.

Interpolate the cosine and sine components with these weights:
\begin{equation}
 f_0(k)=\sum_p\rho_p(k)\cos\vartheta_p,\qquad
 f_z(k)=\sum_p\rho_p(k)\sin\vartheta_p.
 \label{eq:interpolated-components}
\end{equation}
By Eq.~\eqref{eq:cardinal-proof}, $f_0(k_p)=\cos\vartheta_p$ and
$f_z(k_p)=\sin\vartheta_p$. The relations $\rho_{-p}(-k)=\rho_p(k)$ and
$\vartheta_{-p}=-\vartheta_p$ give
\begin{equation}
 f_0(-k)=f_0(k),\qquad f_z(-k)=-f_z(k).
 \label{eq:component-parities}
\end{equation}
The two components stay in the closed unit disk:
\begin{align}
 |f_0(k)+\iu f_z(k)|
 &=\left|\sum_p\rho_p(k)e^{\iu\vartheta_p}\right|\notag\\
 &\le\sum_p\rho_p(k)|e^{\iu\vartheta_p}|=1,
 \label{eq:convex-profile-bound}
\end{align}
where the inequality uses $\rho_p\ge0$ and the last equality uses
Eq.~\eqref{eq:partition-proof}. Therefore the unitarity defect
\begin{equation}
 \Delta(k)=1-f_0(k)^2-f_z(k)^2\ge0,\qquad \Delta(k_p)=0.
 \label{eq:completion-defect}
\end{equation}
It is real valued and even in $k$, contains only even powers of
$\zeta$, and has degree at most $4(L-1)$. Its Fourier coefficients are
therefore real and symmetric:
\begin{equation}
 \Delta(k)=\sum_{n=-2(L-1)}^{2(L-1)}\Delta_ne^{2\iu nk},
 \qquad \Delta_n=\Delta_{-n}\in\mathbb R.
 \label{eq:defect-fourier}
\end{equation}

\subsection{Unitary completion by Fej\'er--Riesz factorization}
\label{sec:completion}

The components $(f_0,f_z)$ have the required values at the allowed
momenta, but between them the defect $\Delta$ is in general positive.
We add two components $f_x,f_y$ that vanish at the allowed momenta and
supply the missing norm, $f_x^2+f_y^2=\Delta$
[Fig.~\ref{fig:sampling}(b)]. This requires writing
$\Delta$ as a squared modulus of a polynomial with real coefficients.
The Fej\'er--Riesz theorem~\cite{FejerRiesz} provides such a factor;
we write out its standard construction because we need the real
coefficients.

If $\Delta$ vanishes identically, set $f_x=f_y=0$. Otherwise let
$d_\Delta\le2(L-1)$ be the largest $n$ with $\Delta_n\ne0$, and write
$w=e^{2\iu k}$, so that $\Delta=\sum_{n=-d_\Delta}^{d_\Delta}\Delta_nw^n$.
The polynomial $w^{d_\Delta}\Delta$ has degree $2d_\Delta$, real
coefficients, and nonzero constant term $\Delta_{-d_\Delta}$, so
\begin{equation}
 w^{d_\Delta}\Delta=\Delta_{d_\Delta}\prod_{i=1}^{2d_\Delta}(w-w_i)
 \label{eq:defect-roots}
\end{equation}
with nonzero roots $w_i$. Three properties fix the structure of the
roots. First, the coefficients are real, so the nonreal roots come in
complex-conjugate pairs with equal multiplicities. Second,
$\Delta_n=\Delta_{-n}$ gives
$w^{2d_\Delta}\,[w^{-d_\Delta}\Delta(1/w)]=w^{d_\Delta}\Delta(w)$, so
with every root $w_i$ also $1/w_i$, and hence $1/w_i^*$, is a root with
the same multiplicity. Third, a root on the unit circle $|w|=1$ has
even multiplicity, because $\Delta(k)\ge0$ is real: near a zero of odd
order it would change sign as a function of $k$. Let $S$ contain every
root with $|w_i|<1$ with its full multiplicity, and every root with
$|w_i|=1$ with half its multiplicity. By the first two properties, the roots
off the circle pair as $w_i$ and $1/w_i^*$, one inside and one outside,
so $S$ has $d_\Delta$ elements. By the first property, $S$ is closed
under complex conjugation, so
\begin{equation}
 \tilde\eta(w)=\left(\frac{|\Delta_{d_\Delta}|}
 {\prod_{i\in S,\,|w_i|<1}|w_i|}\right)^{1/2}
 \prod_{i\in S}(w-w_i)
 \label{eq:fr-factor}
\end{equation}
has real coefficients. For $|w|=1$ and $w_i\ne0$,
$|w-1/w_i^*|=|w-w_i|/|w_i|$. Applying this to every root outside the
circle in Eq.~\eqref{eq:defect-roots} gives, on the unit circle,
\begin{equation*}
 \Delta=|w^{d_\Delta}\Delta|
 =|\Delta_{d_\Delta}|\prod_{i=1}^{2d_\Delta}|w-w_i|
 =|\tilde\eta(w)|^2.
\end{equation*}
Finally put $\eta(\zeta)=\tilde\eta(\zeta^2)$. It is a polynomial in
$\zeta=e^{\iu k}$ of degree $2d_\Delta\le4(L-1)$, with real
coefficients and only even powers, and
\begin{equation}
 \Delta(k)=|\eta(e^{\iu k})|^2.
 \label{eq:real-spectral-factor}
\end{equation}

Real coefficients give $\eta(e^{-\iu k})=\eta(e^{\iu k})^*$. Define
the real and imaginary parts
\begin{align}
 f_y(k)&=\operatorname{Re}\eta(e^{\iu k})
 =\frac{\eta(\zeta)+\eta(\zeta^{-1})}{2},\notag\\
 f_x(k)&=\operatorname{Im}\eta(e^{\iu k})
 =\frac{\eta(\zeta)-\eta(\zeta^{-1})}{2\iu}.
 \label{eq:completion-components}
\end{align}
They are real-valued trigonometric polynomials of degree at most
$4(L-1)$ with only even powers of $\zeta$, they satisfy
$f_x^2+f_y^2=\Delta$, and $f_y$ is even and $f_x$ odd in $k$. With
these components, define
\begin{equation}
 F_L(k)=f_0(k)\Id-\iu\bigl(f_x(k)\tau_x+f_y(k)\tau_y+f_z(k)\tau_z\bigr).
 \label{eq:unitary-loop}
\end{equation}
Since the four components are real, Pauli multiplication verifies exact
unitarity at every momentum:
\begin{equation}
 F_L^\dagger F_L=(f_0^2+f_x^2+f_y^2+f_z^2)\Id=\Id,
 \qquad \det F_L=1.
 \label{eq:completion-unitarity}
\end{equation}
At an allowed momentum, $\Delta(k_p)=0$ and
Eq.~\eqref{eq:real-spectral-factor} give $\eta(e^{\iu k_p})=0$, so
$f_x(k_p)=f_y(k_p)=0$, while $f_0(k_p)=\cos\vartheta_p$ and
$f_z(k_p)=\sin\vartheta_p$. Hence
$F_L(k_p)=\cos\vartheta_p\,\Id-\iu\sin\vartheta_p\,\tau_z
=e^{-\iu\vartheta_p\tau_z}$.

The parities of the four components, with $\tau_x^*=\tau_x$,
$\tau_y^*=-\tau_y$, and $\tau_z^*=\tau_z$, give
\begin{align}
 F_L(-k)&=f_0\Id+\iu f_x\tau_x-\iu f_y\tau_y+\iu f_z\tau_z\notag\\
 &=F_L(k)^*,
 \label{eq:FL-conjugation}
\end{align}
where the components on the right are evaluated at $k$. Thus $F_L$
obeys the conjugation symmetry~\eqref{eq:conj-symmetry}: its Fourier
coefficient matrices are real, although its values, such as
$F_L(k_p)=e^{-\iu\vartheta_p\tau_z}$, are complex matrices. In summary,
\begin{align}
 F_L(k_p)&=e^{-\iu\vartheta_p\tau_z},\notag\\
 F_L(-k)&=F_L(k)^*,\qquad \deg F_L\le4(L-1),
 \label{eq:profile-properties}
\end{align}
and all exponents of $\zeta$ in $F_L$ are even.

Because $F_L$ contains only even powers of $\zeta$, it has period
$\pi$. This gives a second reason, besides the one in
Sec.~\ref{sec:qsp-background}, why the construction uses odd $L$: for
odd $L$ each point $k_p+\pi$ lies halfway between two allowed momenta, whereas on an
even grid $k_p+\pi$ is itself allowed, and the prescribed matrices at
$k_p$ and $k_p+\pi$ need not agree.

\subsection{Factorization into signals and control pulses}
\label{app:factorization}

The completed $F_L$ is a unitary matrix function of degree at most
$4(L-1)$. To turn it into pulses, we remove one signal at a time,
lowering the degree by one at each step. Consider
\begin{equation}
 F(\zeta)=\sum_{n=-d}^{d}F_n\zeta^n,
 \label{eq:factorization-input}
\end{equation}
unitary with determinant one on the unit circle, with real coefficient
matrices $F_n$, exact degree $d$ (so $F_d$ or $F_{-d}$ is nonzero), and
only exponents $n$ with $n\equiv d\pmod 2$. By
Eq.~\eqref{eq:profile-properties}, $F_L$ has these properties for some
even $d\le4(L-1)$. For a matrix or scalar Laurent polynomial, such as
$F(\zeta)=\sum_nF_n\zeta^n$, define
\begin{equation}
 F^\sharp(\zeta)=\sum_nF_n^\dagger\zeta^{-n},
 \label{eq:sharp}
\end{equation}
where $F_n^\dagger=F_n^*$ for a scalar. On the unit circle
$F^\sharp=F^\dagger$. Two Laurent polynomials that agree on the circle
are equal, so unitarity gives $FF^\sharp=F^\sharp F=\Id$ as Laurent
polynomials.

For $d>0$, both extreme coefficients $F_d$ and $F_{-d}$ are nonzero. To
see this, recall that a unitary $2\times2$ matrix with determinant one
has the form $\bigl(\begin{smallmatrix}a&b\\-b^*&a^*\end{smallmatrix}\bigr)$,
so
\begin{equation}
 F=\begin{pmatrix}a&b\\-b^\sharp&a^\sharp\end{pmatrix}
 \label{eq:su2-endpoint-pairing}
\end{equation}
with scalar Laurent polynomials $a=\sum_na_n\zeta^n$ and
$b=\sum_nb_n\zeta^n$. The coefficient of $\zeta^{\pm d}$ in $a^\sharp$
is $a_{\mp d}^*$, and similarly for $b^\sharp$. Hence
\begin{equation}
 F_d=\begin{pmatrix}a_d&b_d\\-b_{-d}^*&a_{-d}^*\end{pmatrix},\qquad
 F_{-d}=\begin{pmatrix}a_{-d}&b_{-d}\\-b_d^*&a_d^*\end{pmatrix}.
 \label{eq:extreme-coefficients}
\end{equation}
Both vanish exactly when $a_{\pm d}=b_{\pm d}=0$. Since the degree is
$d$, neither vanishes. The coefficient of $\zeta^{2d}$ in $FF^\sharp=\Id$
gives
\begin{equation}
 F_dF_{-d}^\dagger=0.
 \label{eq:endpoint-orthogonality}
\end{equation}
Since $F_{-d}\ne0$, the matrix $F_d$ is not invertible, so $F_d$ has
rank one, and likewise $F_{-d}$; Eq.~\eqref{eq:endpoint-orthogonality}
states that their rows are orthogonal. Thus
\begin{equation}
 \Pi_d=\frac{F_d^\dagger F_d}{\Tr(F_d^\dagger F_d)}
 \label{eq:endpoint-projector}
\end{equation}
is a real rank-one orthogonal projector onto the row space of $F_d$,
and
\begin{equation}
 F_d(\Id-\Pi_d)=0,\qquad F_{-d}\Pi_d=0,
 \label{eq:endpoint-cancellation}
\end{equation}
the second because $F_{-d}F_d^\dagger=(F_dF_{-d}^\dagger)^\dagger=0$.

To lower the degree, we use, for a real rank-one orthogonal projector
$\Pi$, the elementary factor
\begin{equation}
 E_\Pi(\zeta)=\zeta\Pi+\zeta^{-1}(\Id-\Pi).
 \label{eq:primitive}
\end{equation}
On the unit circle $E_\Pi$ is unitary with eigenvalues $\zeta$ and
$\zeta^{-1}$, so $\det E_\Pi=1$, and
$E_\Pi^\sharp=\zeta^{-1}\Pi+\zeta(\Id-\Pi)$ is its inverse. Multiplying
$F$ on the right by $E_{\Pi_d}^\sharp$ gives
\begin{align}
 F^{(1)}(\zeta)&=F(\zeta)E_{\Pi_d}^\sharp(\zeta),\notag\\
 F^{(1)}_n&=F_{n+1}\Pi_d+F_{n-1}(\Id-\Pi_d).
 \label{eq:degree-recursion}
\end{align}
The product could contain the powers $\zeta^{d+1}$ and $\zeta^{-d-1}$,
but their coefficients vanish by Eq.~\eqref{eq:endpoint-cancellation}:
\begin{align*}
 F^{(1)}_{d+1}&=F_d(\Id-\Pi_d)=0,\\
 F^{(1)}_{-d-1}&=F_{-d}\Pi_d=0.
\end{align*}
All exponents of $F^{(1)}$ are congruent to $d-1$ modulo two, so the
powers $\zeta^{\pm d}$ are absent as well. The next powers have the
coefficients $F^{(1)}_{d-1}=F_d\Pi_d+F_{d-2}(\Id-\Pi_d)$ and
$F^{(1)}_{-d+1}=F_{-d+2}\Pi_d+F_{-d}(\Id-\Pi_d)$. Multiplying the first
by $\Pi_d$ and the second by $\Id-\Pi_d$ gives
\begin{align}
 F^{(1)}_{d-1}\Pi_d&=F_d\ne0,\notag\\
 F^{(1)}_{-d+1}(\Id-\Pi_d)&=F_{-d}\ne0.
 \label{eq:recursion-exact-degree}
\end{align}
Thus $F^{(1)}$ has exact degree $d-1$, exponents of the opposite
parity, real coefficients, and is unitary with determinant one on the
circle.

Let $\Pi_s$ denote the projector of Eq.~\eqref{eq:endpoint-projector}
constructed when the degree equals $s$. Iterating down to degree zero
and reversing the steps gives
\begin{equation}
 F(\zeta)=E_0E_{\Pi_1}(\zeta)\cdots E_{\Pi_d}(\zeta),
 \label{eq:primitive-product}
\end{equation}
where $E_0$ is a constant real unitary matrix with determinant one.
This is the product decomposition of Ref.~\cite{Haah2019}, in the form
with real coefficients and exponents of one parity, and the steps above give a
finite algorithm for the angles. For $d=0$,
Eq.~\eqref{eq:primitive-product} is the constant $F=E_0$ itself.

Each factor $E_{\Pi_s}$ is one signal between two control pulses. This
step uses that $\Pi_s$ is real; it is not the general Euler
decomposition of a unitary. A real rank-one projector on the two
components projects onto a real unit vector $(-\sin\alpha_s,\cos\alpha_s)^T$ for
some real angle $\alpha_s$. Since
$\mathscr R_z(\alpha)=e^{-\iu\alpha\tau_y}$ is the real rotation matrix
that maps $(0,1)^T$ to $(-\sin\alpha,\cos\alpha)^T$,
\begin{equation}
 \Pi_s=\mathscr R_z(\alpha_s)\Pi_B \mathscr R_z(-\alpha_s),
 \qquad \Pi_B=\frac{\Id-\tau_z}{2},
 \label{eq:real-projector-rotation}
\end{equation}
where $\Pi_B$ projects onto the $B$ component. A complex projector
would need an extra phase rotation, which the $z$ pulses do not
provide. In the basis of the $A$ and $B$ components, in which $\tau_z$
is diagonal,
\begin{equation}
 E_{\Pi_B}(e^{\iu k})
 =\diag(e^{-\iu k},e^{\iu k})=e^{-\iu k\tau_z},
 \label{eq:fixed-signal}
\end{equation}
which is the signal $W_\mu(k)$ of Eq.~\eqref{eq:two-layer-signal}. Since $E_{R\Pi R^\dagger}=RE_\Pi R^\dagger$
for any rotation $R$, each factor equals
$E_{\Pi_s}(e^{\iu k})=\mathscr R_z(\alpha_s)e^{-\iu k\tau_z}\mathscr R_z(-\alpha_s)$.
A constant real unitary matrix with determinant one is a rotation,
$E_0=\mathscr R_z(\beta)$. Substitution into
Eq.~\eqref{eq:primitive-product} and multiplication of adjacent
rotations give
\begin{equation}
 F_L(k)=\mathscr R_z(\varphi_0)
 \prod_{s=1}^{d}\left[e^{-\iu k\tau_z}\mathscr R_z(\varphi_s)\right],
 \qquad d\le4(L-1),
 \label{eq:qsp}
\end{equation}
with explicit angles
\begin{align}
 \varphi_0&=\beta+\alpha_1,\notag\\
 \varphi_s&=\alpha_{s+1}-\alpha_s,\qquad 1\le s<d,\notag\\
 \varphi_d&=-\alpha_d.
 \label{eq:qsp-angles}
\end{align}
For $d=0$, the product is empty and $\varphi_0=\beta$. Equation~\eqref{eq:qsp}
is the QSP sequence of Eq.~\eqref{eq:qsp-background}, with signal
$W_\mu=e^{-\iu k\tau_z}$; the rightmost factor acts first.

\subsection{Realization by bond pulses}
\label{app:gates}

Equation~\eqref{eq:qsp} is a pulse sequence. Each signal is the
two-pulse product $W_\mu=\mathscr R_z(-\pi/2)\mathscr R_\mu(\pi/2;k_\mu)$
of Eq.~\eqref{eq:two-layer-signal}, and each control is a $z$ pulse
$\mathscr R_z(\varphi_s)$. The $z$ pulse $\mathscr R_z(-\pi/2)$ that
ends each signal merges with the control pulse on its left, so
\begin{align}
 F_L={}&\mathscr R_z(\varphi_0-\pi/2)\mathscr R_\mu(\pi/2)
 \mathscr R_z(\varphi_1-\pi/2)\mathscr R_\mu(\pi/2)\notag\\
 &\cdots\mathscr R_z(\varphi_{d-1}-\pi/2)\mathscr R_\mu(\pi/2)
 \mathscr R_z(\varphi_d),
 \label{eq:merged-sequence}
\end{align}
with the momentum argument of $\mathscr R_\mu$ suppressed. This leaves
$d$ intercell bond pulses and $d+1$ $z$ pulses, so the depth of one
one-coordinate rotation is
\begin{equation}
 D_{\rm rot}\le2d+1\le8(L-1)+1.
 \label{eq:profile-cost}
\end{equation}
This completes the proof of Proposition~\ref{prop:profile}.

\subsection{Assembling the circuit and counting depth}
\label{sec:assembly}

Apply Proposition~\ref{prop:profile} in the coordinate $k_x$ with the
angles $\vartheta_p=-\bar k_p/2$, and in $k_y$ with
$\vartheta_q=\bar k_q/4$, where $\bar k_p=2\pi p/L_x$ and
$\bar k_q=2\pi q/L_y$, with $p,q$ running from $-(L_\mu-1)/2$ to
$(L_\mu-1)/2$, are the allowed values in $(-\pi,\pi)$. These values are
symmetric about zero and include zero, so both lists satisfy
$\vartheta_{-p}=-\vartheta_p$, $\vartheta_{-q}=-\vartheta_q$, and
$\vartheta_0=0$. Writing the resulting matrix functions as $F_{x,L_x}$
and $F_{y,L_y}$, the finite circuit is
\begin{equation}
 U_{L_x,L_y}=F_{y,L_y}\,\mathscr R_z(-\pi/4)\,F_{x,L_x}\,
 \mathscr R_z(-\pi/4)\,F_{y,L_y}.
 \label{eq:finite-word}
\end{equation}
At every allowed momentum it equals $U_1$ of
Eq.~\eqref{eq:branch-word}, so
\begin{equation*}
 U_{L_x,L_y}(\bk)\tau_y U_{L_x,L_y}(\bk)^\dagger=Q_1(\bk),
 \qquad \bk\in\mathcal K_{L_x,L_y}.
\end{equation*}
Thus it prepares the Chern-one state exactly. Using the cost of
Eq.~\eqref{eq:profile-cost} for its three one-coordinate rotations gives
\begin{align}
 D_{\rm up}&\le
 [8(L_x-1)+1]+2[8(L_y-1)+1]+2\notag\\
 &=8(L_x-1)+16(L_y-1)+5.
 \label{eq:depth-count}
\end{align}
The two extra circuit layers are the fixed pulses $\mathscr R_z(-\pi/4)$.

\subsection{Several honeycomb layers: the five strong classes and copies}
\label{sec:physical-doubling}

Section~\ref{sec:assembly} prepares $P_1$ from $P_0=(\Id+\tau_y)/2$.
This is the target of classes A and D, because the hopping pulse on two
honeycomb layers and the Majorana pulse on one honeycomb layer act on
the $A$ and $B$ components with the same matrix
(Sec.~\ref{sec:honeycomb-pulses}). In the other classes, the same pulse
is applied to all honeycomb layers at once. Pulses on different
honeycomb layers act on disjoint Majoranas and commute, so one pulse on
all honeycomb layers is still one circuit layer, and the depth remains
that of Eq.~\eqref{eq:depth-count}. Below we write $U=U_{L_x,L_y}$ for
the circuit of Eq.~\eqref{eq:finite-word}.

On the blocks of Sec.~\ref{sec:classes}, a pulse with one-particle
matrix $V(\bk)$ acts as $\Id_\kappa\otimes V(\bk)$. In AII and DIII
this is immediate, because both blocks are honeycomb layers that
receive the same pulse. In C, the hole components $c_{-\bk,a\downarrow}^\dagger$
transform with $V(-\bk)^*$, and Eq.~\eqref{eq:conj-symmetry} gives
$V(\bk)\oplus V(-\bk)^*=\Id_\kappa\otimes V(\bk)$. The full circuit is
therefore
\begin{equation}
 \mathcal U(\bk)=\Id_\kappa\otimes U(\bk),
 \label{eq:doubled-word}
\end{equation}
and at the allowed momenta
\begin{align}
 \mathcal U(\kappa_z\otimes\tau_y)
 \mathcal U^\dagger
 &=\kappa_z\otimes Q_1,\notag\\
 \mathcal U(\Id_\kappa\otimes\tau_y)
 \mathcal U^\dagger
 &=\Id_\kappa\otimes Q_1.
 \label{eq:doubled-preparation}
\end{align}
A block started in the $-\tau_y$ state of Sec.~\ref{sec:classes} ends
in
\begin{equation}
 U(-\tau_y)U^\dagger=-Q_1
 \quad\text{on }\mathcal K_{L_x,L_y}.
 \label{eq:complement-preparation}
\end{equation}
Copies of the targets, including those started in the $-\tau_y$ state,
act on disjoint honeycomb layers and receive the same pulses, so they
leave the depth unchanged. This completes the proof of
Theorem~\ref{thm:upper}.

\subsection{The circuit's own Bloch projector has Chern number zero}
\label{sec:circuit-extension}

Section~\ref{sec:results-match} uses the following computation. The
finite circuit is periodic in momentum, whereas $U_1$ is not. The
circuit therefore defines a Bloch projector at every momentum,
\begin{equation}
 P_{\rm circ}(\bk)=
 U(\bk)P_0U(\bk)^\dagger.
 \label{eq:circuit-extension}
\end{equation}
With the occupied Bloch state $\psi_0=(1,\iu)^T/\sqrt2$ of $P_0$
(Sec.~\ref{sec:classes}),
\begin{equation}
 \psi_{\rm circ}(\bk)
 =U(\bk)\psi_0
 \label{eq:global-circuit-frame}
\end{equation}
is a smooth, periodic, normalized occupied Bloch state of
$P_{\rm circ}$.
Its Berry connection
$\mathcal A=-\iu\psi_{\rm circ}^\dagger d\psi_{\rm circ}$ is defined on
the whole torus, so
\begin{equation}
 c_1(P_{\rm circ})
 =\frac1{2\pi}\int_{\Torus^2}d\mathcal A=0.
 \label{eq:circuit-chern-zero}
\end{equation}
Yet $P_{\rm circ}=P_1$ at every allowed momentum. The entries of
$P_{\rm circ}$ are trigonometric polynomials of degree at most
$8(L_x-1)$ in $k_x$ and $16(L_y-1)$ in $k_y$, so its range is at most
$8(L_x-1)+16(L_y-1)$, linear in the system size. Because the Chern
numbers differ, Lemma~\ref{lem:az-homotopy} gives
$\|P_{\rm circ}-P_1\|\ge1$ at some momentum; for $L_x=L_y=L$,
$P_{\rm circ}$ there departs by at least $1-\pi/L$ from its value at
the nearest allowed momentum, where it equals $P_1$, because $P_1$
changes by at most $\pi/L$ over that distance
[Eq.~\eqref{eq:seed-lipschitz}].

The circuits of the other classes behave in the same way. They give
$\kappa_z\otimes Q_{\rm circ}$ and $\Id_\kappa\otimes Q_{\rm circ}$,
with $Q_{\rm circ}=2P_{\rm circ}-\Id$, whose entries are trigonometric
polynomials. These matrices obey the relations of
Eq.~\eqref{eq:az-projectors} at every momentum, because the circuit
respects the antiunitary symmetries of the targets. For the matrix
$\iu\kappa_y\otimes\Id$ of time reversal in AII and DIII and of
particle-hole symmetry in C, Eq.~\eqref{eq:conj-symmetry} gives
\begin{align}
 (\iu\kappa_y\otimes\Id)\,\mathcal U(\bk)^*\,
 (\iu\kappa_y\otimes\Id)^\dagger
 &=(\iu\kappa_y)(\iu\kappa_y)^\dagger\otimes U(\bk)^*\notag\\
 &=\Id_\kappa\otimes U(-\bk)=\mathcal U(-\bk),
 \label{eq:doubled-unitary-covariance}
\end{align}
and for $\mathcal C=K$ in classes D and DIII,
$\mathcal U(\bk)^*=\mathcal U(-\bk)$ directly. Their blocks are
copies of $P_{\rm circ}$ and $\Id-P_{\rm circ}$, with Chern number zero
by Eq.~\eqref{eq:circuit-chern-zero}, so their strong indices vanish,
although they equal the targets at every allowed momentum.

\section{Proof of the locality lower bound}
\label{sec:lower}

We now prove Theorem~\ref{thm:lower}. Fix $L$, even or odd, and
suppose that a depth-$D$ circuit of Sec.~\ref{sec:general-circuits}
prepares the target exactly on the $L\times L$ torus. The target is
given by its Bloch projector $P_*$, defined at every momentum of the
Brillouin zone. The prepared state has a Bloch projector $P$ defined
only at the allowed momenta, where it equals $P_*$ by assumption. From
the correlations of the prepared state we construct a second projector
$\widetilde P$ on the whole zone and compare it with $P_*$. The proof
is by contradiction: if $D$ were below the bound of
Eq.~\eqref{eq:lower}, $\widetilde P$ would be close to $P_*$ and yet
have a different strong index. Gaussianity, translation invariance,
and the AZ symmetries enter only through the final state, not through
the circuit.

The argument has four steps (Fig.~\ref{fig:lightcone}).
\begin{enumerate}
\item A depth-$D$ circuit applied to a product input creates no
two-point correlations between cells
more than $2D$ apart (Lemma~\ref{lem:car-lightcone},
Sec.~\ref{sec:lower-lightcone}).
\item If $L>8D$, these correlations define a Bloch projector
$\widetilde P$ of range at most $2D$ on the whole Brillouin zone, which
agrees with $P_*$ at every allowed momentum and obeys the same AZ
relations (Lemma~\ref{lem:alias-free}, Sec.~\ref{sec:lower-lift}).
\item $\widetilde P$ changes by at most $2D$ per unit momentum, and
$P_*$ by at most $M_x$ or $M_y$; since every momentum lies within
$\pi/L$ of an allowed momentum in each coordinate, the two stay close
everywhere (Lemma~\ref{lem:projector-comparison},
Sec.~\ref{sec:lower-comparison}).
\item Close projectors have the same strong index
(Lemma~\ref{lem:az-homotopy}, Sec.~\ref{sec:lower-homotopy}), whereas a
Bloch projector of finite range carries no strong topology, however
large its range (Lemma~\ref{lem:read}, Sec.~\ref{sec:lower-read}). Since
the strong index of $P_*$ is nonzero, the two cannot be close, which
gives the bound (Sec.~\ref{sec:lower-completion}).
\end{enumerate}

\subsection{Shallow circuits have strictly short-range correlations}
\label{sec:lower-lightcone}

The first step is the light-cone argument of
Fig.~\ref{fig:lightcone}.

\begin{lemma}[Short-range correlations]
\label{lem:car-lightcone}
In the depth-$D$ circuit model of Sec.~\ref{sec:general-circuits}, let
$a_{\br}$ and $a_{\br'}$ be fermion operators in cells $\br$ and
$\br'$, that is, odd polynomials in the annihilation and creation (or
Majorana) operators of these cells. If the cell distance between $\br$
and $\br'$ exceeds $2D$, the two-point correlation function of the
final state vanishes:
\begin{equation}
 \langle a_{\br}a_{\br'}\rangle=0.
 \label{eq:correlation-range}
\end{equation}
Intermediate states and the input need not be Gaussian.
\end{lemma}

\begin{proof}
Let $U$ be the circuit on the lattice $\Lambda_L\equiv\Lambda_{L,L}$,
and write $a_{\br}(D)=U^\dagger a_{\br}U$ for the operator after the
$D$ circuit layers, and similarly $a_{\br'}(D)$. Let
$\mathcal B_D(\br)$ denote the set of cells within distance $D$ of
$\br$ on the periodic lattice. Because every gate preserves parity,
$a_{\br}(D)$ remains odd. Because every circuit layer is a product of
gates on disjoint sets of operators, each acting on one cell or on two
neighboring cells, the support of $a_{\br}(D)$ lies in
$\mathcal B_D(\br)$.

Suppose the distance between $\br$ and $\br'$ exceeds $2D$. Then
$\mathcal B=\mathcal B_D(\br)$ and $\mathcal B'=\mathcal B_D(\br')$ are
disjoint sets of whole cells that contain the supports of $a_{\br}(D)$
and $a_{\br'}(D)$. Denote the input by $|\Omega\rangle$; the final
state is $U|\Omega\rangle$, so its two-point function is
$\langle\Omega|a_{\br}(D)a_{\br'}(D)|\Omega\rangle$. Since each input
cell has definite parity, $|\Omega\rangle$ is an eigenstate of the
region parity
$\mathcal P_{\mathcal B}=\prod_{\bm s\in\mathcal B}\mathcal P_{\bm s}$,
the product of the cell parities of Eq.~\eqref{eq:cell-parity}, and
$\mathcal P_{\mathcal B}^2=\Id$. Oddness and disjoint support give
\begin{equation}
 \mathcal P_{\mathcal B}\,a_{\br}(D)\,\mathcal P_{\mathcal B}=-a_{\br}(D),\qquad
 [\mathcal P_{\mathcal B},a_{\br'}(D)]=0.
 \label{eq:local-parity-relations}
\end{equation}
Thus
\begin{align}
 \langle\Omega|a_{\br}(D)a_{\br'}(D)|\Omega\rangle
 &=\langle\Omega|\mathcal P_{\mathcal B}a_{\br}(D)a_{\br'}(D)\mathcal P_{\mathcal B}|\Omega\rangle\notag\\
 &=-\langle\Omega|a_{\br}(D)a_{\br'}(D)|\Omega\rangle=0.
 \label{eq:parity-range-proof}
\end{align}
Fermionic annihilation, creation, and Majorana operators are odd, so
every two-point function between cells more than $2D$ apart has this
form and vanishes.
\end{proof}

\subsection{Short-range correlations define a Bloch projector of range \texorpdfstring{$2D$}{2D}}
\label{sec:lower-lift}

The second step turns the correlations of the final state into a Bloch
projector on the whole Brillouin zone, with range at most $2D$ in the
sense of Sec.~\ref{sec:results-match}. The condition $L>8D$ below
ensures that distinct short displacements stay distinct on the torus,
and that products of two such projectors are still fixed by their
values at the allowed momenta.

\begin{lemma}[Bloch projector of range $2D$]
\label{lem:alias-free}
Suppose the exact final state of a depth-$D$ circuit is a
translation-invariant Gaussian state with Bloch projector $P(\bk)$ at
the allowed momenta $\bk\in\mathcal K_{L,L}$, and
\begin{equation}
 L>8D.
 \label{eq:no-alias}
\end{equation}
Then there is a Hermitian orthogonal projector $\widetilde P(\bk)$,
defined at every $\bk\in\Torus^2$ and of range at most $2D$, that equals
$P(\bk)$ at every allowed momentum. If $P$ obeys the relations of
Eq.~\eqref{eq:az-projectors} at the allowed momenta, $\widetilde P$
obeys them at every momentum.
\end{lemma}

\begin{proof}
The final state is translation invariant, even if its preparing
circuit is not. Its correlations between cells separated by $\br$ form
the matrix
\begin{equation}
 G(\br)=\frac1{L^2}\sum_{\bk\in\mathcal K_{L,L}}
 P(\bk)e^{\iu\bk\cdot\br},\qquad \br\in\Lambda_L.
 \label{eq:finite-kernel}
\end{equation}
For $\br\ne0$, each entry of $G(\br)$ is a two-point correlation
function of fermion operators in two cells separated by $\br$, for
example $\langle c_{b,\bm s}^\dagger c_{a,\bm s+\br}\rangle$ in a
number-conserving state, and Majorana or pairing correlations in the
descriptions of Sec.~\ref{app:projectors}. Lemma~\ref{lem:car-lightcone}
therefore gives
\begin{equation}
 G(\br)=0\quad\text{when the cell distance of $\br$ exceeds $2D$.}
 \label{eq:kernel-diamond}
\end{equation}
Because $L>8D$, each displacement on the torus of cell distance at
most $2D$ corresponds to exactly one integer vector $\br\in\Z^2$ with
$|r_x|+|r_y|\le2D$: two such vectors would differ by a nonzero vector
of $L\Z^2$, of length $|r_x|+|r_y|\ge L>4D$, whereas their difference
has length at most $4D$. Using these integer vectors, define for every
$\bk\in\Torus^2$
\begin{equation}
 \widetilde P(\bk)=\sum_{|r_x|+|r_y|\le2D}G(\br)e^{-\iu\bk\cdot\br}.
 \label{eq:laurent-lift}
\end{equation}
It has range at most $2D$, so its entries are trigonometric
polynomials of degree at most $2D$ in each coordinate. At an allowed
momentum, finite Fourier inversion gives
$P(\bk)=\sum_{\br\in\Lambda_L}G(\br)e^{-\iu\bk\cdot\br}$, in which only
the terms of Eq.~\eqref{eq:laurent-lift} are nonzero, so
$\widetilde P=P$ there. The relation $G(-\br)=G(\br)^\dagger$ makes
$\widetilde P$ Hermitian at all momenta.

To extend the projector property, we use a uniqueness statement. A
scalar trigonometric polynomial $f(k)=\sum_{n=-N}^{N}f_ne^{\iu nk}$ of
degree at most $N$ that vanishes at the $L$ allowed values of $k$ is
identically zero when $L>2N$: multiplying by $e^{\iu Nk}$ gives a
polynomial of degree at most $2N$ in $e^{\iu k}$ with $L>2N$ distinct
roots. In two coordinates, fix any allowed $k_y$. Vanishing at all $L$
values of $k_x$ forces every $k_x$ Fourier coefficient to vanish at
that $k_y$. Each such coefficient is itself a trigonometric polynomial
in $k_y$ of degree at most $N$ that vanishes at all $L$ values of
$k_y$, so it is identically zero. This argument applies to every matrix
entry.

The defect $\widetilde P^2-\widetilde P$ has degree at most $4D$ in each
coordinate and vanishes at all allowed momenta. Since $L>8D$, it
vanishes identically:
\begin{equation}
 \widetilde P^2=\widetilde P
 \quad\text{on the full torus}.
 \label{eq:global-idempotency}
\end{equation}
The defects of the relations in Eq.~\eqref{eq:az-projectors},
such as $U_T\widetilde P(\bk)^*U_T^\dagger-\widetilde P(-\bk)$, have
degree at most $2D$ and vanish at the allowed momenta, because
$\bk\mapsto-\bk$ maps the grid to itself. The same argument extends
these relations to the full torus.
\end{proof}

\subsection{The projector of range \texorpdfstring{$2D$}{2D} stays close to the target}
\label{sec:lower-comparison}

The third step uses the equality at the allowed momenta to control
$\widetilde P$ between them. For a matrix-valued function $A(\bk)$ on
the torus we write $\|A\|_{\max}=\max_{\bk}\|A(\bk)\|$, the largest
operator norm over all momenta.

\begin{lemma}[Closeness to the target]
\label{lem:projector-comparison}
Assume Eq.~\eqref{eq:no-alias}, let $P_*$ satisfy
Eq.~\eqref{eq:regularity} with constants $M_x,M_y$, and let
$\widetilde P$ be the projector of Lemma~\ref{lem:alias-free}, equal to
$P_*$ at every allowed momentum. Then
\begin{equation}
 \|\widetilde P-P_*\|_{\max}
 \le\frac{\pi}{L}(4D+M_x+M_y).
 \label{eq:comparison}
\end{equation}
The bound does not depend on the dimension of the one-particle space.
\end{lemma}

\begin{proof}
For a scalar trigonometric polynomial $f$ of degree at most $2D$,
Bernstein's inequality gives
\begin{equation}
 \|f'\|_{\max}\le 2D\|f\|_{\max}.
 \label{eq:scalar-bernstein}
\end{equation}
We use the trigonometric form of this standard
inequality~\cite{QueffelecZarouf2019}. For fixed $k_y$ and any unit
vectors $u,v$, apply it to $u^\dagger\widetilde P(k_x,k_y)v$. Taking
the maximum over the fixed vectors and momenta gives
\begin{align}
 \|\partial_{k_x}\widetilde P\|_{\max}
 &=\max_{k_y}\max_{k_x}\max_{\|u\|=\|v\|=1}
 |\partial_{k_x}(u^\dagger\widetilde P v)|\notag\\
 &\le 2D\max_{k_y}\max_{\|u\|=\|v\|=1}
 \max_{k_x}|u^\dagger\widetilde P v|\notag\\
 &=2D\|\widetilde P\|_{\max}= 2D.
 \label{eq:matrix-bernstein-x}
\end{align}
All maxima range over independent variables, so their order can be
interchanged. The scalar Bernstein inequality acts at fixed
$k_y,u,v$, and the maximum over unit vectors is the operator norm. The
same estimate holds in $k_y$. The last equality holds because
$\widetilde P$ is an orthogonal projector at every momentum, by
Lemma~\ref{lem:alias-free}.

Given any momentum $\bk$, choose a nearest allowed momentum
$\bk_0=(2\pi p/L,2\pi q/L)$. Since neighboring allowed values of each
coordinate are $2\pi/L$ apart, $d_{\Torus}(k_x,2\pi p/L)\le\pi/L$ and
$d_{\Torus}(k_y,2\pi q/L)\le\pi/L$. Moving one coordinate at a time and
using equality at the allowed momenta yields
\begin{align}
 \|\widetilde P(\bk)-P_*(\bk)\|
 &\le\|\widetilde P(\bk)-\widetilde P(\bk_0)\|
 +\|P_*(\bk_0)-P_*(\bk)\|\notag\\
 &\le (2D+M_x)\,d_{\Torus}(k_x,2\pi p/L)\notag\\
 &\quad+(2D+M_y)\,d_{\Torus}(k_y,2\pi q/L)\notag\\
 &\le\frac{\pi}{L}(4D+M_x+M_y).
 \label{eq:nearest-node-proof}
\end{align}
This proves Eq.~\eqref{eq:comparison}.
\end{proof}

\subsection{Close projectors have the same topology}
\label{sec:lower-homotopy}

The fourth step turns this closeness into equality of strong indices.

\begin{lemma}[Same topology when close]
\label{lem:az-homotopy}
Let $P_*$ and $\widetilde P$ be continuous orthogonal projectors on
$\Torus^2$ that satisfy the relations of Eq.~\eqref{eq:az-projectors}
with the same matrices $U_T,U_C$, whichever are present. If
$\|\widetilde P-P_*\|_{\max}<1$, they are homotopic within that
symmetry class, and their stable strong indices agree.
\end{lemma}

\begin{proof}
Define
\begin{equation}
 Q_*=2P_*-\Id,\qquad \widetilde Q=2\widetilde P-\Id,
 \qquad H_t=(1-t)Q_*+t\widetilde Q,
 \label{eq:homotopy-H}
\end{equation}
where $0\le t\le1$. Suppose $H_t v=0$ for a unit vector $v$ at some
momentum. Then $(1-t)Q_*v=-t\widetilde Qv$. Both $Q_*$ and
$\widetilde Q$ are Hermitian with square one, so they preserve the norm,
and $t=1/2$. At that value, $Q_*v=-\widetilde Qv$, which implies
\begin{equation}
 \|(Q_*-\widetilde Q)v\|=\|2Q_*v\|=2.
 \label{eq:homotopy-gap-proof}
\end{equation}
This contradicts $\|Q_*-\widetilde Q\|=2\|P_*-\widetilde P\|<2$. Thus
$H_t$ is invertible at every momentum and every $t$.

Continuous spectral flattening gives
\begin{equation}
 Q_t=H_t(H_t^2)^{-1/2},\qquad P_t=\frac{\Id+Q_t}{2}.
 \label{eq:equivariant-homotopy}
\end{equation}
Both ends of the interpolation satisfy Eq.~\eqref{eq:az-projectors} with
the same matrices $U_T,U_C$. The interpolation therefore satisfies,
whenever the corresponding symmetry is present,
\begin{align*}
 U_T H_t(\bk)^*U_T^\dagger&=H_t(-\bk),\\
 U_C H_t(\bk)^*U_C^\dagger&=-H_t(-\bk).
\end{align*}
The minus sign disappears in $H_t^2$, so functional calculus gives
\begin{align*}
 U_T Q_t(\bk)^*U_T^\dagger&=Q_t(-\bk),\\
 U_C Q_t(\bk)^*U_C^\dagger&=-Q_t(-\bk).
\end{align*}
Returning to projectors gives
$U_TP_t(\bk)^*U_T^\dagger=P_t(-\bk)$ and
$U_CP_t(\bk)^*U_C^\dagger=\Id-P_t(-\bk)$. Thus $P_t$ is a continuous
homotopy inside the target AZ class. It does not assume that the
physical preparing circuit follows this homotopy.
\end{proof}

\subsection{Bloch projectors of finite range carry no strong topology}
\label{sec:lower-read}

The last ingredient is a result of Read~\cite{Read2017}. A
vector-valued trigonometric polynomial
$v(\bk)=\sum_{\br}v_{\br}e^{-\iu\bk\cdot\br}$, with finitely many nonzero
vectors $v_{\br}$, is the Fourier transform of a one-particle state
supported on finitely many cells, that is, a compactly supported
Wannier-type state. Read showed that such states cannot span an
occupied subspace with nonzero strong topology, in any of the ten classes, when
the spanning set carries the symmetry structure of the class. For Chern
insulators, related results for strictly local flat-band Hamiltonians
and compactly supported Wannier functions appear in
Refs.~\cite{ChenMazaheriSeidelTang2013,DubailRead2013}.

A Bloch projector $P$ of finite range, whose entries are trigonometric
polynomials, supplies such a spanning set. With fixed real basis vectors $e_a$ of the
one-particle space, its columns
$P(\bk)e_a$ are vector-valued trigonometric polynomials, and they span
the occupied subspace, because every $v$ in the range of $P(\bk)$ obeys
\begin{equation}
 v=P(\bk)v=
 \sum_a(e_a^\dagger v)P(\bk)e_a.
 \label{eq:polynomial-generators}
\end{equation}
The columns of $\Id-P$ span the empty subspace in the same way. Time
reversal and particle-hole symmetry map a vector $v(-\bk)$ to
$U_Tv(-\bk)^*$ and $U_Cv(-\bk)^*$ at $\bk$, and the relations of
Eq.~\eqref{eq:az-projectors} at $-\bk$ give
\begin{align}
 U_T[P(-\bk)e_a]^*&=P(\bk)U_Te_a,\notag\\
 U_C[P(-\bk)e_a]^*&=[\Id-P(\bk)]U_Ce_a.
 \label{eq:polynomial-symmetry-generators}
\end{align}
Time reversal thus maps the columns of $P$ into the occupied subspace
and particle-hole symmetry maps them into the empty subspace, so the
spanning sets carry the symmetry structure of the class. We use Read's
result in the following form.

\begin{lemma}[Read~\cite{Read2017}]
\label{lem:read}
A Bloch projector on $\Torus^2$ of finite range that obeys the relations
of Eq.~\eqref{eq:az-projectors} of class A, D, C, AII, or DIII has
zero stable strong index, however large its range.
\end{lemma}

\subsection{Completing the proof}
\label{sec:lower-completion}

\begin{proof}[Proof of Theorem~\ref{thm:lower}]
Suppose a circuit of depth $D$ prepares the target exactly. If
$L\le8D$, Eq.~\eqref{eq:lower} holds trivially, so we assume $L>8D$.
Then Lemma~\ref{lem:alias-free} gives a projector
$\widetilde P$ of range at most $2D$ that equals $P_*$ at every allowed
momentum and obeys the AZ relations of the target. By
Lemma~\ref{lem:read}, its strong index is zero. If
$(\pi/L)(4D+M_x+M_y)<1$, Lemma~\ref{lem:projector-comparison} gives
$\|\widetilde P-P_*\|_{\max}<1$, and Lemma~\ref{lem:az-homotopy} then
gives $\widetilde P$ the nonzero strong index of $P_*$, a
contradiction. Hence $(\pi/L)(4D+M_x+M_y)\ge1$, which is
Eq.~\eqref{eq:lower}.
\end{proof}

\begin{proof}[Proof of Corollary~\ref{cor:optimal}]
All targets of Table~\ref{tab:targets} and their copies satisfy
Eq.~\eqref{eq:regularity} with $M_x=M_y=1/2$ (Sec.~\ref{sec:classes}).
The targets, and the copies whose total strong index is nonzero, also
satisfy the index condition of Theorem~\ref{thm:lower}. Inserting
$M_x=M_y=1/2$ into Theorem~\ref{thm:lower} gives the lower bound in
Eq.~\eqref{eq:optimal}. Theorem~\ref{thm:upper} with
$L_x=L_y=L$ gives the upper bound $8(L-1)+16(L-1)+5=24(L-1)+5$.
\end{proof}

\section{Which targets vary slowly in momentum}
\label{app:regularity}

Theorem~\ref{thm:lower} needs constants $M_x,M_y$ for the target. For
a continuously differentiable $P_*$, the maxima of
$\|\partial_{k_\mu}P_*\|$ over the Brillouin zone can serve as $M_\mu$,
and $P_1$ has $M_x=M_y=1/2$ (Sec.~\ref{sec:seed-chern}). For the ground
state of a gapped Hamiltonian, the constants follow from the gap and
the hopping.

Let the Bloch Hamiltonian have the expansion
\begin{equation}
 H(\bk)=\sum_{\br}h_{\br}e^{-\iu\bk\cdot\br},
 \label{eq:general-hopping}
\end{equation}
where $h_{\br}$ is the hopping matrix for the displacement
$\br=(r_x,r_y)$ and $r_\mu$ is its component along $\mu=x,y$. Assume a
fixed finite band dimension, Hermiticity, and a gap $E_g>0$ around zero
energy:
\begin{equation}
 \spec H(\bk)\cap(-E_g/2,E_g/2)=\varnothing
 \quad\text{for all }\bk.
 \label{eq:uniform-half-gap}
\end{equation}
Assume also that the hopping bounds
\begin{equation}
 v_\mu=\sum_{\br}|r_\mu|\,\|h_{\br}\|,
 \qquad\mu=x,y,
 \label{eq:weighted-hopping}
\end{equation}
are finite. Then
$\sum_{\br}\|h_{\br}\|\le\|h_0\|+v_x+v_y$, so the series for $H$ and
its derivative converge absolutely. The derivative, the velocity
operator, satisfies
\begin{align*}
 \partial_{k_\mu}H(\bk)
 &=-\sum_{\br}\iu r_\mu h_{\br}e^{-\iu\bk\cdot\br},\\
 \|\partial_{k_\mu}H\|_{\max}
 &\le\sum_{\br}|r_\mu|\,\|h_{\br}\|=v_\mu.
\end{align*}
Hoppings that vanish beyond some distance are a special case.

Let $P_*$ be the negative-energy projector. Differentiate
$P_*^2=P_*$:
\begin{equation*}
 (\partial_{k_\mu}P_*)P_*
 +P_*(\partial_{k_\mu}P_*)=\partial_{k_\mu}P_*.
\end{equation*}
Multiplying on both sides by $P_*$ shows that the block of the
derivative within the occupied subspace is zero. Multiplying on both
sides by $\Id-P_*$ similarly makes the block within the empty subspace
zero. Thus the derivative only connects the occupied and empty
subspaces. At fixed $\bk$, put
\begin{equation}
 X_\mu=(\Id-P_*)(\partial_{k_\mu}P_*)P_*.
 \label{eq:projector-derivative-block}
\end{equation}
Then $\partial_{k_\mu}P_*=X_\mu+X_\mu^\dagger$. In a basis adapted to
the occupied and empty subspaces, $X_\mu$ maps the occupied subspace to
the empty subspace, and
\begin{align*}
 \partial_{k_\mu}P_*
 &=\begin{pmatrix}0&X_\mu^\dagger\\X_\mu&0\end{pmatrix},\\
 (\partial_{k_\mu}P_*)^2
 &=\begin{pmatrix}X_\mu^\dagger X_\mu&0\\0&X_\mu X_\mu^\dagger\end{pmatrix}.
\end{align*}
The two diagonal blocks in the square have the same nonzero
eigenvalues. Hence $\|\partial_{k_\mu}P_*\|^2=\|X_\mu\|^2$.

To bound $X_\mu$, let $H_+$ and $H_-$ be the restrictions of $H$ to the
empty and occupied subspaces. Equation~\eqref{eq:uniform-half-gap}
gives $H_+\ge E_g/2$ and $-H_-\ge E_g/2$. Differentiating
$[H,P_*]=0$ gives
\begin{equation*}
 [\partial_{k_\mu}H,P_*]
 +[H,\partial_{k_\mu}P_*]=0.
\end{equation*}
Taking the block from the occupied to the empty subspace yields the
Sylvester equation
\begin{equation}
 H_+X_\mu-X_\mu H_-=-Y_\mu,\qquad
 Y_\mu=(\Id-P_*)(\partial_{k_\mu}H)P_*.
 \label{eq:sylvester}
\end{equation}
In eigenbases of $H_+$ and $H_-$, the homogeneous equation multiplies
each entry by a positive energy difference of at least $E_g$. It
therefore has only the zero solution. The inhomogeneous solution is
\begin{equation}
 X_\mu=-\int_0^\infty e^{-tH_+}Y_\mu e^{tH_-}\,dt.
 \label{eq:sylvester-integral}
\end{equation}
Indeed, both exponentials have norm at most $e^{-E_gt/2}$, so the
integral converges and the integrand vanishes as $t\to\infty$. Since
$-H_+(e^{-tH_+}Y_\mu e^{tH_-})+(e^{-tH_+}Y_\mu e^{tH_-})H_-$ is the
$t$ derivative of the integrand, substitution gives
\begin{align*}
 H_+X_\mu-X_\mu H_-
 &=\int_0^\infty\frac{d}{dt}
   \left(e^{-tH_+}Y_\mu e^{tH_-}\right)dt\\
 &=\left[e^{-tH_+}Y_\mu e^{tH_-}\right]_0^\infty=-Y_\mu,
\end{align*}
and uniqueness identifies this solution with the derivative block.
Finally, projection on the occupied and empty subspaces gives
$\|Y_\mu\|\le\|\partial_{k_\mu}H\|$. Therefore
\begin{align}
 \|\partial_{k_\mu}P_*\|
 &=\|X_\mu\|\le\int_0^\infty e^{-E_gt}\|Y_\mu\|\,dt\notag\\
 &\le\frac{\|\partial_{k_\mu}H\|}{E_g}
 \le\frac{v_\mu}{E_g}.
 \label{eq:gap-regularity-bound}
\end{align}
As in first-order perturbation theory, only mixing between the
occupied and empty subspaces changes the projector. The gap suppresses
this mixing, while $v_\mu$ bounds the velocity operator
$\partial_{k_\mu}H$.

Integrating the derivative estimate along coordinate paths gives
Eq.~\eqref{eq:regularity} with $M_\mu=v_\mu/E_g$. Only the gap between
the occupied and empty subspaces enters; degeneracies within each
subspace are allowed.

\section{Discussion}
\label{sec:discussion}

We have shown that uniform Kitaev honeycomb bond pulses prepare
representative states of all five strong classes exactly on odd tori,
in depth linear in the lattice size. Every local fermionic circuit that
prepares these targets exactly needs linear depth, even if it is
interacting, spatially nonuniform, or symmetry breaking. Both results
are set by the spacing $2\pi/L$ of the allowed momenta. The
construction resolves this spacing with two components that move in
opposite directions and interfere; the lower bound shows that, because
local gates spread correlations at a finite speed, resolving it needs a
depth of order $L$.

The construction shows how QSP can serve as a physical
state-preparation method: one looks for the signal among the
operations that a system already provides. In the honeycomb model, an
$x$ (or $y$) pulse followed by a $z$ pulse supplies the signal exactly,
and a $z$ pulse of tunable angle supplies the control; the control
angles of Sec.~\ref{sec:preparation} set the momentum dependence of the
final rotation. No measurements or Trotter steps are used. The physical
requirements are only selective, uniform control of the three bond
types.

The pulse-sequence step applies beyond these five states.
Proposition~\ref{prop:profile} turns any list of one-coordinate
rotation angles with $\vartheta_{-p}=-\vartheta_p$ and
$\vartheta_0=0$ into a sequence of the same bond pulses. A Gaussian target that
factors into a short product of such rotations could therefore be
prepared in the same way. Which physically relevant ground states have
this structure, and how many rotations they need, remains open. The
five strong classes already show that one control scheme reaches very
different topological indices.

Because only the values at the allowed momenta are physical, the
discrete Chern number computed from them~\cite{FHS2005} is the natural
finite-size index. It agrees with the continuous index when the grid
is fine compared with the momentum variation of the projector
(Lemma~\ref{lem:fhs-mesh}), and it gives a second, weaker lower bound
that uses only the data at the allowed momenta
(Theorem~\ref{thm:fhs-lower} in App.~\ref{app:fhs-lower}).

Several questions remain open. Our lower bound concerns exact
preparation; extending it to a fixed nonzero error requires a robust
comparison between the final state and the target. The construction
uses odd $L$, while the lower bound holds for every $L$; even $L$
needs a different construction. Our results concern the matter
fermions of the Kitaev model at fixed bond variables $u_{ij}=1$.
Preparing the corresponding spin states from a product spin state,
which requires choosing the flux sector and projecting onto the
physical spin space~\cite{PedrocchiChesiLoss2011}, is a separate problem that we leave open. Open
boundaries and robustness to pulse errors and decoherence remain to be
analyzed. Interacting targets lie outside the free-fermion setting of
the construction~\cite{FidkowskiKitaev2010}, and higher-dimensional analogs are another
direction.

\begin{acknowledgments}
X.Y. acknowledges the Peterson postdoctoral fellowship from the
Department of Computing and Mathematical Sciences at Caltech and the
Simons Targeted Grant in Mathematics and Physical Sciences on
Moir\'e Materials Magic, Award No.~896630. X.G. acknowledges funding
from the U.S. Department of Energy, Quantum Systems Accelerator, and
from NSF PFC grant No.~PHYS 2317149.

\emph{AI disclosure}---We acknowledge the use of GPT (OpenAI) and
Claude Opus (Anthropic) for assistance with technical derivations,
literature search, figure preparation, and writing and proofreading of
the manuscript. The authors take full responsibility for the accuracy
and integrity of the manuscript and its conclusions.
\end{acknowledgments}

\section*{Data Availability}

The curves in Fig.~\ref{fig:sampling} are evaluated numerically from
the explicit formulas of Secs.~\ref{app:completion} and
\ref{sec:completion}; no other data were generated or analyzed. All
constructions and proofs supporting the results are contained in this
article.

\appendix

\section{Discrete Chern numbers of the target states}
\label{app:fhs-targets}

\begin{figure}[t]
\centering
\begin{tikzpicture}[x=0.82cm,y=0.82cm]
  \def\Lg{5}
  \draw[black!45,dashed,line width=0.5pt] (-0.5,-0.5) rectangle (4.5,4.5);
  \fill[Acol!18] (-0.5,4) rectangle (4.5,4.5);
  \fill[Acol!18] (-0.5,-0.5) rectangle (4.5,0);
  \foreach \p in {0,...,4}{\draw[black!60,line width=0.5pt] (\p,0) -- (\p,4);}
  \foreach \q in {0,...,4}{\draw[black!60,line width=0.5pt] (0,\q) -- (4,\q);}
  \foreach \q in {0,...,4}{
    \draw[Bcol,line width=0.6pt,densely dashed] (4,\q) -- (4.5,\q);
    \draw[Bcol,line width=0.6pt,densely dashed] (-0.5,\q) -- (0,\q);}
  \foreach \p in {0,...,4}{
    \draw[Bcol,line width=0.6pt,densely dashed] (\p,4) -- (\p,4.5);
    \draw[Bcol,line width=0.6pt,densely dashed] (\p,-0.5) -- (\p,0);}
  \foreach \p in {0,...,4}{\foreach \q in {0,...,4}{\fill (\p,\q) circle (1.6pt);}}
  \begin{scope}[line width=1.2pt,xcol,-{Latex[length=4pt]}]
    \draw (1.08,1) -- (1.92,1);
    \draw (2,1.08) -- (2,1.92);
    \draw (1.92,2) -- (1.08,2);
    \draw (1,1.92) -- (1,1.08);
  \end{scope}
  \node[font=\scriptsize,below] at (1.5,1) {$\mathcal L_x(p,q)$};
  \node[font=\scriptsize,right] at (2,1.5) {$\mathcal L_y(p{+}1,q)$};
  \node[font=\scriptsize,above] at (1.5,2) {$\mathcal L_x(p,q{+}1)^{-1}$};
  \node[font=\scriptsize,left] at (0.9,1.5) {$\mathcal L_y(p,q)^{-1}$};
  \node[font=\scriptsize,below] at (-0.5,-0.55) {$-\pi$};
  \node[font=\scriptsize,below] at (4.5,-0.55) {$\pi$};
  \node[font=\small,below] at (2,-0.75) {$k_x$};
  \node[font=\scriptsize,left] at (-0.55,-0.5) {$-\pi$};
  \node[font=\scriptsize,left] at (-0.55,4.5) {$\pi$};
  \node[font=\small,left] at (-0.85,2) {$k_y$};
  \node[font=\scriptsize,right,align=left] at (4.6,4.25) {top row};
  \node[font=\scriptsize,right,align=left,Bcol] at (4.6,2.6) {wrapping\\ edges};
\end{tikzpicture}
\caption{Momentum grid of App.~\ref{app:fhs-targets}, drawn for $L=5$.
Dots are the allowed momenta $(k_{x,p},k_{y,q})$ of
Eq.~\eqref{eq:fhs-sorted-grid}, with $p$ increasing to the right and
$q$ upward; the dashed square is the boundary $k_x=\pm\pi$,
$k_y=\pm\pi$ of the Brillouin zone. The four blue
arrows mark one plaquette $(p,q)$, traversed counterclockwise through
the links $\mathcal L_x(p,q)$, $\mathcal L_y(p+1,q)$,
$\mathcal L_x(p,q+1)^{-1}$, and $\mathcal L_y(p,q)^{-1}$ of
Eq.~\eqref{eq:fhs-definition}. The purple dashed stubs at the zone
boundary are the wrapping edges: they leave through one side and
re-enter through the opposite side, joining $p=L-1$ to $p=0$ across
$k_x=\pm\pi$ and $q=L-1$ to $q=0$ across $k_y=\pm\pi$. The shaded
strip is the top row of plaquettes, $q=L-1$, whose vertical edges
cross $k_y=\pm\pi$; it appears in two halves, above $q=L-1$ and below
$q=0$, because $k_y=\pi$ and $k_y=-\pi$ are identified.}
\label{fig:fhs-grid}
\end{figure}
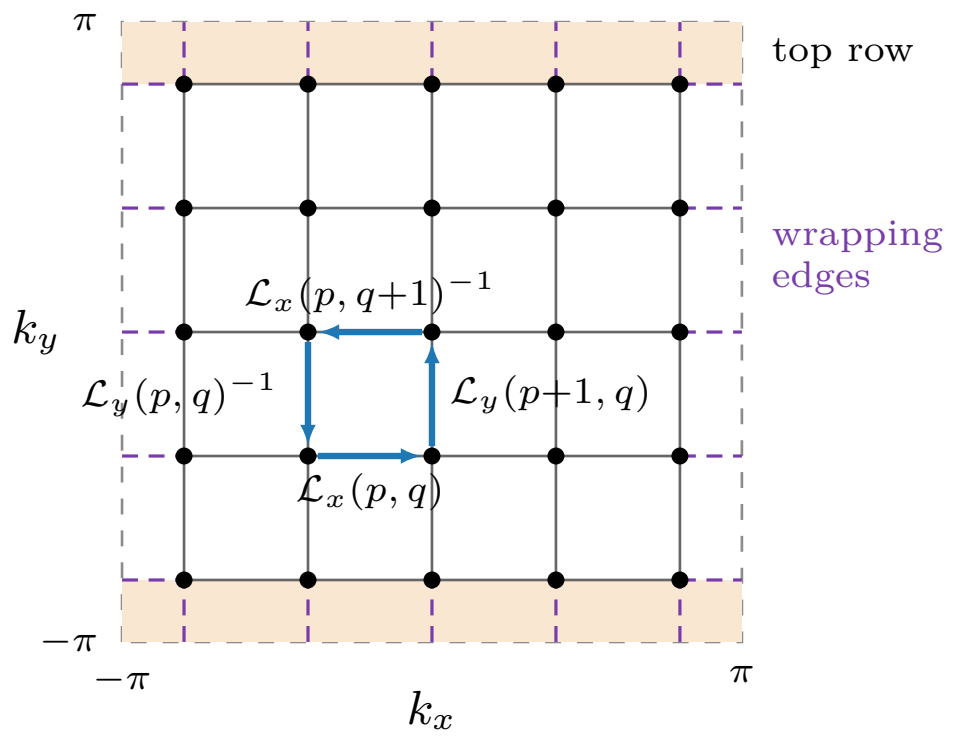

This appendix computes the discrete Chern number of Fukui, Hatsugai,
and Suzuki (FHS)~\cite{FHS2005} for the targets and shows that it
agrees with their continuous indices for every odd $L\ge5$. The FHS
number uses only the values of the projector at the allowed momenta.
Figure~\ref{fig:fhs-grid} shows the grid, the links, the wrapping
edges, and the top row of plaquettes used below.

\subsection{Definition and sign convention}

The FHS number is defined for every $L$ on the grid of
Eq.~\eqref{eq:grid}, with labels $p,q\in\Z_L$ taken modulo $L$, so the
label $L$ means $0$. For a projector of rank one at each allowed
momentum, choose any unit vector $u_{p,q}$ spanning its range at
$(2\pi p/L,2\pi q/L)$. The links are
\begin{equation}
 \mathcal L_x(p,q)=
 \frac{u_{p,q}^\dagger u_{p+1,q}}
 {|u_{p,q}^\dagger u_{p+1,q}|},\qquad
 \mathcal L_y(p,q)=
 \frac{u_{p,q}^\dagger u_{p,q+1}}
 {|u_{p,q}^\dagger u_{p,q+1}|},
 \label{eq:fhs-links}
\end{equation}
defined when the overlaps are nonzero. The plaquette $(p,q)$ is the
square with corners at the labels $(p,q)$, $(p+1,q)$, $(p+1,q+1)$, and
$(p,q+1)$. With
$\operatorname{Arg}\in(-\pi,\pi]$, its phase and the FHS number are
\begin{align}
 \Phi_{p,q}&=\operatorname{Arg}\!\bigl[
 \mathcal L_x(p,q)\mathcal L_y(p+1,q)\notag\\
 &\qquad{}\times
 \mathcal L_x(p,q+1)^{-1}\mathcal L_y(p,q)^{-1}\bigr],\notag\\
 C_{\mathrm{FHS},L}&=\frac{1}{2\pi}\sum_{p,q}\Phi_{p,q}.
 \label{eq:fhs-definition}
\end{align}
This is the gauge-invariant integer of Fukui, Hatsugai, and
Suzuki~\cite{FHS2005}. For a projector of higher rank, the columns of
$u_{p,q}$ form an orthonormal basis of its range, and each overlap in
Eq.~\eqref{eq:fhs-links} is replaced by its determinant before
normalizing. The continuous counterpart is Eq.~\eqref{eq:chern}, for
which $c_1(P_1)=1$ [Eq.~\eqref{eq:seed-chern-calculation}].

For the explicit calculations below, $L$ is odd, and we label the
allowed values of each coordinate in increasing order in $(-\pi,\pi)$:
\begin{equation}
 k_{x,p}=-\pi+(2p+1)\delta,\qquad k_{y,q}=-\pi+(2q+1)\delta,
 \label{eq:fhs-sorted-grid}
\end{equation}
with $\delta=\pi/L$ and $p,q=0,\ldots,L-1$. These are the momenta
$2\pi p'/L$ with $p'=p+(L+1)/2$ modulo $L$, and similarly for $q$. This
cyclic relabeling keeps neighboring momenta neighbors, so it permutes
the links and plaquettes and leaves $C_{\mathrm{FHS},L}$ unchanged. In
these labels, the horizontal edges from $p=L-1$ to $p=0$ cross the line
$k_x=\pm\pi$, and the vertical edges from $q=L-1$ to $q=0$ cross the
line $k_y=\pm\pi$; we call both kinds wrapping edges. The plaquettes
with $q=L-1$, whose vertical edges are wrapping edges, form the top
row.

\subsection{Exact overlaps, including the wrapping edges}

Let $L\ge3$ be odd. In the constant eigenbasis $|\pm y\rangle$ of
$\tau_y$ (Sec.~\ref{sec:qsp-background}), the Chern-one state has the
normalized occupied Bloch state
\begin{align}
 u(\bk)={}&|{+y}\rangle\!\left[\cos\frac{k_x}2
 -\iu\sin\frac{k_y}2\sin\frac{k_x}2\right]\notag\\
       &+|{-y}\rangle\cos\frac{k_y}2\sin\frac{k_x}2
 \label{eq:fhs-seed-spinor}
\end{align}
for $k_x,k_y\in[-\pi,\pi)$; direct multiplication gives
$uu^\dagger=P_1$. We take $u_{p,q}=u(k_{x,p},k_{y,q})$.

\emph{Horizontal edges.} For $p\le L-2$, the overlap is
\begin{equation}
 u_{p,q}^\dagger u_{p+1,q}=\cos\delta-\iu\sin\frac{k_{y,q}}2\sin\delta.
 \label{eq:fhs-horizontal}
\end{equation}
Its real part is $\cos\delta>0$, so
$\mathcal L_x(p,q)=e^{\iu\chi_x(k_{y,q})}$ with
$\chi_x(k_y)=-\arctan[\tan\delta\,\sin(k_y/2)]\in(-\pi/2,\pi/2)$. On
the wrapping edge $p=L-1$, the vector at $p=0$ is
$u_{0,q}=-u(k_{x,L-1}+2\delta,k_{y,q})$, because
$u(k_x+2\pi,k_y)=-u(k_x,k_y)$. The overlap is therefore minus that of
Eq.~\eqref{eq:fhs-horizontal}, and
$\mathcal L_x(L-1,q)=-e^{\iu\chi_x(k_{y,q})}$. Each plaquette with
$p=L-1$ contains two such links, $\mathcal L_x(L-1,q)$ and
$\mathcal L_x(L-1,q+1)^{-1}$, so the two signs cancel, including in the
corner plaquette.

\emph{Vertical edges.} For $q\le L-2$, the overlap at
$k_x=k_{x,p}$ is
\begin{align}
 u_{p,q}^\dagger u_{p,q+1}={}&\cos^2\frac{k_x}2+\cos\delta\sin^2\frac{k_x}2\notag\\
 &+\iu\left(\sin\frac{k_{y,q}}2-\sin\frac{k_{y,q+1}}2\right)
 \cos\frac{k_x}2\sin\frac{k_x}2.
 \label{eq:fhs-vertical}
\end{align}
Its real part is at least $\cos\delta>0$, so
$\mathcal L_y(p,q)=e^{\iu\chi_{y,q}(k_{x,p})}$, where
$\chi_{y,q}(k_x)\in(-\pi/2,\pi/2)$ is the phase of the right side of
Eq.~\eqref{eq:fhs-vertical}; it is $2\pi$ periodic in $k_x$.

\emph{Wrapping vertical edges.} For $q=L-1$, the edge joins
$k_y=\pi-\delta$ to $k_y=-\pi+\delta$. Put $a_\delta=\cos(\delta/2)$
and $b_\delta=(1-\cos\delta)/2$, so that $a_\delta^2=(1+\cos\delta)/2$
and $a_\delta^2+b_\delta=1$. The overlap at $k_x=k_{x,p}$ is
\begin{equation}
 u_{p,L-1}^\dagger u_{p,0}=b_\delta+a_\delta^2\cos k_x+\iu a_\delta\sin k_x.
 \label{eq:fhs-seam}
\end{equation}
As $k_x$ runs from $-\pi$ to $\pi$, the right side traces an ellipse
centered at $b_\delta$ on the real axis, with semi-axes $a_\delta^2$
and $a_\delta$. Since $a_\delta^2>b_\delta$, it encloses the origin
once counterclockwise and never vanishes: it equals $1$ at $k_x=0$ and
$-\cos\delta$ at $k_x=\pi$, and its imaginary part vanishes only at
these two points. Let $\chi_{y,L-1}(k_x)$ be the continuous phase of
Eq.~\eqref{eq:fhs-seam} with $\chi_{y,L-1}(0)=0$. Then
$\chi_{y,L-1}(k_x+2\pi)=\chi_{y,L-1}(k_x)+2\pi$, and
$\mathcal L_y(p,L-1)=e^{\iu\chi_{y,L-1}(k_{x,p})}$. Together,
Eqs.~\eqref{eq:fhs-horizontal}--\eqref{eq:fhs-seam} show that every
link of $P_1$ is defined for all odd $L\ge3$.

\subsection{Plaquette phases for odd \texorpdfstring{$L\ge5$}{L>=5}}

Each plaquette product is the exponential of $\iu$ times the signed
sum of its four edge phases. If this edge-phase sum lies in
$(-\pi,\pi)$, it equals the plaquette phase $\Phi_{p,q}$ of
Eq.~\eqref{eq:fhs-definition}; otherwise $\operatorname{Arg}$ would
shift it by a multiple of $2\pi$. We show that for $L\ge5$ every such
sum lies in $(-\pi/2,\pi/2)$, so we denote the sums by $\Phi_{p,q}$
from the start.

Two horizontal phases are needed explicitly. Since
$\sin(k_{y,0}/2)=-\cos(\delta/2)$ and $\sin(k_{y,L-1}/2)=\cos(\delta/2)$,
\begin{equation}
 \chi_x(k_{y,0})=-\chi_x(k_{y,L-1})=t_\delta
 =\arctan(a_\delta\tan\delta).
 \label{eq:fhs-endpoint-phase}
\end{equation}
With the links of the previous subsection, the edge-phase sums of the
regular plaquettes ($q\le L-2$) and of the top row ($q=L-1$) are
\begin{align}
 \Phi_{p,q}={}&\chi_x(k_{y,q})-\chi_x(k_{y,q+1})\notag\\
 &+\chi_{y,q}(k_{x,p}+2\delta)-\chi_{y,q}(k_{x,p}),\notag\\
 \Phi_{p,L-1}={}&\chi_{y,L-1}(k_{x,p}+2\delta)\notag\\
 &-\chi_{y,L-1}(k_{x,p})-2t_\delta.
 \label{eq:fhs-raw-phases}
\end{align}
In the regular rows, the signs of the wrapping horizontal edges cancel,
and $\chi_{y,q}$ is $2\pi$ periodic, so $k_{x,p+1}$ may be replaced by
$k_{x,p}+2\delta$ also for $p=L-1$. In the top row, the horizontal
edges at $q=L-1$ and $q=0$ contribute $-t_\delta$ each. There the
replacement of $k_{x,p+1}$ by $k_{x,p}+2\delta$ is a choice, because
$\chi_{y,L-1}$ is not periodic: it grows by $2\pi$ when $k_x$ grows by
$2\pi$. For the corner plaquette $p=L-1$, the right edge lies at
$k_{x,0}=k_{x,L-1}+2\delta-2\pi$. Its link is a single complex number,
but its phase in the edge-phase sum can be taken as
$\chi_{y,L-1}(k_{x,0})$ or as
$\chi_{y,L-1}(k_{x,L-1}+2\delta)=\chi_{y,L-1}(k_{x,0})+2\pi$. We take the
second, so that the corner has the same edge-phase sum,
Eq.~\eqref{eq:fhs-raw-phases}, as the other plaquettes of the row. The
plaquette product does not depend on this choice; the bounds below
show that our sum lies in $(-\pi,\pi)$, whereas the other would lie
$2\pi$ lower. The extra $2\pi$ is the winding of the wrapping link, and
it supplies the Chern number in the row sum below.

For $L\ge5$, $\delta\le\pi/5$ and $\cos\delta>4/5$. From the formula
for $\chi_x$, $|\chi_x'(k_y)|\le\tan\delta/2$, so
$|\chi_x(k_{y,q})-\chi_x(k_{y,q+1})|\le\delta\tan\delta\le\delta^2/\cos\delta$.
Also $|\sin(k_{y,q}/2)-\sin(k_{y,q+1}/2)|\le\delta$, because
$k_{y,q+1}-k_{y,q}=2\delta$ and $\sin(k_y/2)$ changes at a rate of at
most $1/2$; with $|\cos(k_x/2)\sin(k_x/2)|\le1/2$, the imaginary part of
Eq.~\eqref{eq:fhs-vertical} has magnitude at most $\delta/2$, while its
real part is at least $\cos\delta$; hence
$|\chi_{y,q}(k_x)|\le\delta/(2\cos\delta)$. Therefore
\begin{equation}
 |\Phi_{p,q}|\le\frac{\delta^2+\delta}{\cos\delta}
 <\frac{\pi^2}{20}+\frac\pi4<\frac\pi2,\qquad q\le L-2.
 \label{eq:fhs-regular-margin}
\end{equation}
For the top row, direct differentiation of the phase of
Eq.~\eqref{eq:fhs-seam} yields
\begin{equation}
 \chi_{y,L-1}'(k_x)=
 \frac{a_\delta(a_\delta^2+b_\delta\cos k_x)}
 {|b_\delta+a_\delta^2\cos k_x+\iu a_\delta\sin k_x|^2}>0,
 \label{eq:fhs-seam-derivative}
\end{equation}
where positivity uses $a_\delta^2>b_\delta$. The identity
\begin{equation}
 \begin{aligned}
 &|b_\delta+a_\delta^2\cos k_x+\iu a_\delta\sin k_x|^2
 -\cos\delta\,(a_\delta^2+b_\delta\cos k_x)\\
 &\quad=b_\delta(1+\cos k_x)[1+a_\delta^2(1-\cos k_x)]
 \ge0
 \end{aligned}
 \label{eq:fhs-seam-identity}
\end{equation}
implies $0<\chi_{y,L-1}'(k_x)\le a_\delta/\cos\delta$. Thus
\begin{align}
 0<\chi_{y,L-1}(k_x+2\delta)-\chi_{y,L-1}(k_x)
 &\le\frac{2\delta a_\delta}{\cos\delta}<\frac\pi2,\notag\\
 0<2t_\delta&\le2\delta<\frac\pi2.
 \label{eq:fhs-seam-margin}
\end{align}
The difference of these two quantities, which is $\Phi_{p,L-1}$, lies
strictly between $-\pi/2$ and $\pi/2$.

Summing the regular plaquettes of row $q$ over $p$ cancels the
$\chi_{y,q}$ terms, because $\chi_{y,q}$ is $2\pi$ periodic. The
remaining terms, $L[\chi_x(k_{y,q})-\chi_x(k_{y,q+1})]$, add up over
$q=0,\ldots,L-2$ to $L[\chi_x(k_{y,0})-\chi_x(k_{y,L-1})]=2Lt_\delta$.
In the top row, the $\chi_{y,L-1}$ terms add up to
$\chi_{y,L-1}(k_{x,0}+2\pi)-\chi_{y,L-1}(k_{x,0})=2\pi$, so the top row
contributes $2\pi-2Lt_\delta$. Consequently
\begin{align}
 \sum_{p,q}\Phi_{p,q}&=2\pi,\notag\\
 C_{\mathrm{FHS},L}(P_1)&=1=c_1(P_1),
 \quad L\ge5\ \text{odd}.
 \label{eq:fhs-seed-all-L}
\end{align}

The same calculation covers the other targets. In class C, the
spin-reduced projector $\Id_\kappa\otimes P_1$ has the orthonormal
occupied Bloch states $u\oplus0$ and $0\oplus u$, the columns of $u_{p,q}$
at each allowed momentum. Each edge overlap is the corresponding overlap
of $P_1$ times the $2\times2$ identity, so each determinant link is the
square of the link of $P_1$, and each edge-phase sum doubles. Since
every sum above lies in $(-\pi/2,\pi/2)$, the doubled sums lie in
$(-\pi,\pi)$ and are the plaquette phases; hence
$C_{\mathrm{FHS},L}=2=c_1(P_C)$ for odd $L\ge5$. In AII and DIII, the
block $\kappa_z=+1$ of the target is $P_1$ itself, so its FHS number is
one and its parity equals the strong index $\nu=1$. The block
$\kappa_z=-1$ is $\Id-P_1$. Writing $u=u_1|{+y}\rangle+u_2|{-y}\rangle$,
its range is spanned by $u_2^*|{+y}\rangle-u_1^*|{-y}\rangle$, whose
overlaps are the complex conjugates of those of $u$. Its FHS number is
therefore $-1$, and the total is zero, as is the total Chern number.

\section{When the discrete and continuous Chern numbers agree}
\label{app:fhs-mesh}

This appendix gives a condition under which the FHS number on the
$L\times L$ grid equals the continuous Chern number.
Appendix~\ref{app:fhs-lower} uses it for the projector of range $2D$
produced by a hypothetical shallow circuit. We are not aware of a
reference that states this quantitative condition, so we prove it.

\begin{lemma}[Rank-one mesh agreement]
\label{lem:fhs-mesh}
Let $P(\bk)$ be a smooth rank-one orthogonal projector on $\Torus^2$,
acting on a finite-dimensional space of any dimension, with
$\|\partial_{k_x}P(\bk)\|\le M$ and $\|\partial_{k_y}P(\bk)\|\le M$ for
all $\bk$. If $2\pi M/L\le1/2$, then on the $L\times L$ grid of
Eq.~\eqref{eq:grid} every link is nonzero, every plaquette phase
satisfies $|\Phi_{p,q}|\le4\arctan(1/2)<\pi$, and
\begin{equation}
 C_{\mathrm{FHS},L}(P)=c_1(P).
 \label{eq:fhs-mesh-agreement}
\end{equation}
\end{lemma}

\begin{proof}
\emph{One grid square.} Regard $P$ as a $2\pi$-periodic function on the
plane. Each plaquette, including those that contain wrapping edges,
then covers an ordinary square of side $2\pi/L$ in the plane. Let
$\bk_c$ be its center and let $e$ be a unit vector in the range of
$P(\bk_c)$. Every $\bk$ in the square has
$|k_x-k_{c,x}|\le\pi/L$ and $|k_y-k_{c,y}|\le\pi/L$, so integrating the
derivative bounds along coordinate paths gives
\begin{equation}
 \|P(\bk)-P(\bk_c)\|\le M(|k_x-k_{c,x}|+|k_y-k_{c,y}|)
 \le \frac{2\pi M}{L}\le\frac12.
 \label{eq:fhs-cell-distance}
\end{equation}
Put $s(\bk)=e^\dagger P(\bk)e$. Since $P(\bk_c)e=e$ and $\Id-P(\bk)$ is
a projector,
\begin{align*}
 1-s&=e^\dagger[\Id-P(\bk)]e=\|[\Id-P(\bk)]e\|^2\\
 &=\|[P(\bk_c)-P(\bk)]e\|^2\le\frac14,
\end{align*}
so $s\ge3/4$. Hence $P(\bk)e\ne0$ spans the range of $P(\bk)$, and
$u_c(\bk)=P(\bk)e/\sqrt s$ is a smooth unit vector in that range,
because $\|P(\bk)e\|^2=e^\dagger P(\bk)e=s$. Decompose
$P(\bk)e=s(e+\xi)$ with $\xi=P(\bk)e/s-e$, which satisfies
$e^\dagger\xi=s/s-1=0$. Then $1=\|u_c\|^2=s(1+\|\xi\|^2)$, so
\begin{equation}
 u_c=\frac{e+\xi}{\sqrt{1+\|\xi\|^2}},\qquad
 \|\xi\|^2=\frac{1-s}{s}\le\frac13.
 \label{eq:fhs-cell-frame}
\end{equation}
For two points $\bm a,\bm b$ of the square, with $\xi_a=\xi(\bm a)$ and
$\xi_b=\xi(\bm b)$, the relation $e^\dagger\xi=0$ gives
\begin{equation}
 u_c(\bm a)^\dagger u_c(\bm b)
 =\frac{1+\xi_a^\dagger\xi_b}
 {\sqrt{(1+\|\xi_a\|^2)(1+\|\xi_b\|^2)}}.
 \label{eq:fhs-frame-overlap}
\end{equation}
By the Cauchy--Schwarz inequality, $|\xi_a^\dagger\xi_b|\le1/3$, so the
numerator has real part at least $2/3$ and imaginary part at most $1/3$
in magnitude, while the denominator is at most $4/3$. Thus
$|u_c(\bm a)^\dagger u_c(\bm b)|\ge1/2$, and the phase of this overlap
has magnitude at most $\arctan(1/2)$. The plaquette product is gauge
invariant, so it may be computed with the vectors $u_c$. Its four edge
phases then add up to at most $4\arctan(1/2)<\pi$ in magnitude, and
this sum is the plaquette phase. In particular every link is nonzero.

\emph{Refinement.} Next we show that $C_{\mathrm{FHS},L}(P)=C_{\mathrm{FHS},2L}(P)$.
For $0\le t\le1$, consider the grid with the $2L$ momentum values
\begin{align}
 k^{(t)}_{2p}&=\frac{2\pi p}{L},\notag\\
 k^{(t)}_{2p+1}&=\frac{2\pi p}{L}+\frac{\pi t}{L},
 \qquad p=0,\ldots,L-1,
 \label{eq:fhs-moving-grid}
\end{align}
in each of the two coordinates $k_x$ and $k_y$. At $t=0$ each point
with an odd label coincides with the point before it; at $t=1$ the
grid is the ordinary $2L\times2L$ grid. Every plaquette of this moving grid lies inside one
square of the $L\times L$ grid, so the first part of the proof applies
throughout: every link is nonzero, and every plaquette phase, computed
with the vector $u_c$ of that square, has magnitude below $\pi$ and is a
continuous function of $t$. The sum of all plaquette phases is $2\pi$
times an integer, because every link appears in two plaquettes with
opposite orientations, so the product of all plaquette products is one.
Being continuous in $t$ and quantized, the sum does not depend on $t$.
At $t=0$, a plaquette between coinciding points has zero phase, since
each of its links is either a link from a point to itself or appears
twice with opposite orientations. The remaining plaquettes, whose lower
left corner has odd labels in both coordinates, are the plaquettes of
the $L\times L$ grid. Hence $C_{\mathrm{FHS},L}(P)=C_{\mathrm{FHS},2L}(P)$.
Repeating the argument gives
\begin{equation}
 C_{\mathrm{FHS},L}(P)=C_{\mathrm{FHS},2^\ell L}(P),\qquad \ell=1,2,\ldots.
 \label{eq:fhs-refinement}
\end{equation}

\emph{Continuum limit.} On the grid with spacing
$\varepsilon_\ell=2\pi/(2^\ell L)$, use a smooth occupied Bloch state
$u$, defined locally, with Berry connection $\mathcal A_\mu=-\iu u^\dagger\partial_{k_\mu}u$
(Sec.~\ref{sec:topology-background}). The phase of
an edge of length $\varepsilon_\ell$ in direction $\mu$ is
$\varepsilon_\ell\mathcal A_\mu+\varepsilon_\ell^2\partial_{k_\mu}\mathcal A_\mu/2
+O(\varepsilon_\ell^3)$, so a plaquette has phase
\begin{align}
 \Phi&=\varepsilon_\ell^2(\partial_{k_x}\mathcal A_y-\partial_{k_y}\mathcal A_x)
 +O(\varepsilon_\ell^3),\notag\\
 \partial_{k_x}\mathcal A_y-\partial_{k_y}\mathcal A_x
 &=-\iu\Tr\{P[\partial_{k_x}P,\partial_{k_y}P]\}.
 \label{eq:fhs-fine-limit}
\end{align}
Cover the torus by finitely many patches on each of which such a
vector $u$ exists, and by smaller patches whose closures lie inside
them. For small $\varepsilon_\ell$, every plaquette lies in one of
the larger patches, and all
remainders have a common bound. There are $4\pi^2/\varepsilon_\ell^2$
plaquettes, so the total remainder is $O(\varepsilon_\ell)$, and the
sum of the leading terms is a Riemann sum for $2\pi c_1(P)$. Since the
left side of Eq.~\eqref{eq:fhs-refinement} does not depend on $\ell$,
the limit $\ell\to\infty$ proves Eq.~\eqref{eq:fhs-mesh-agreement}.
\end{proof}

We state the lemma for rank one, which is all
Appendix~\ref{app:fhs-lower} needs, because every target has a rank-one
block. For a projector of constant rank $r$, the same argument works
with determinant links and an orthonormal frame in place of $u_c$; each
edge phase is then a sum of $r$ small phases, so the condition on
$2\pi M/L$ must be tightened as $r$ grows.

\section{A lower bound from the discrete FHS data}
\label{app:fhs-lower}

Theorem~\ref{thm:lower} needs a continuous target $P_*$ and its
constants $M_x,M_y$. A weaker bound follows from the values at the
allowed momenta alone, when the final state contains a rank-one block
with nonzero FHS number. We use a block rather than the whole
projector because in AII and DIII the total FHS number vanishes, like
the total Chern number; only one of the two blocks related by time
reversal has a nonzero FHS number.

Fix a set of modes in each cell, the same at every momentum, for
example the spin-up particle components in class C. In the
decomposition into these modes and the rest, write the Bloch projector
of the final state at an allowed momentum as
\begin{equation}
 P=\begin{pmatrix}B&X\\X^\dagger&B'\end{pmatrix}.
 \label{eq:fhs-block-matrix}
\end{equation}
We say that the final state has a rank-one block if $B(\bk)$ is a
rank-one orthogonal projector at every allowed momentum. Then $P^2=P$
gives, in the upper-left block, $B^2+XX^\dagger=B$, so
$XX^\dagger=B-B^2=0$ and $X=0$: the block is decoupled from the other
modes.

\begin{theorem}[FHS lower bound]
\label{thm:fhs-lower}
Consider the circuit model of Sec.~\ref{sec:general-circuits} on an
$L\times L$ torus. Suppose its exact final state is a
translation-invariant pure Gaussian state with a rank-one block $B$
whose FHS number is nonzero, $C_{\mathrm{FHS},L}(B)\ne0$. Then
\begin{equation}
 D>\frac{L}{8\pi}.
 \label{eq:fhs-direct-depth}
\end{equation}
Intermediate gates need not preserve Gaussianity, translation
invariance, AZ symmetries, or the block.
\end{theorem}

\begin{proof}
Assume instead $D\le L/(8\pi)$. Then $8D\le L/\pi<L$, so
Lemma~\ref{lem:alias-free} applies: the correlations of the final state
define a Bloch projector $\widetilde P$ of range at most $2D$ on the
whole torus, equal to the projector $P$ of the final state at every
allowed momentum. Write $\widetilde P$ in the block form of
Eq.~\eqref{eq:fhs-block-matrix}, with blocks $\widetilde B$ and
$\widetilde X$. The block $\widetilde X$ has degree at most $2D$ in
each coordinate and vanishes at all allowed momenta, because $X=0$
there. Since $L>4D$, the uniqueness statement in the proof of
Lemma~\ref{lem:alias-free} gives $\widetilde X=0$ at every momentum.
Hence $\widetilde B$ is an orthogonal projector at every momentum, with
trigonometric-polynomial entries. Its rank is its trace, which is
continuous and integer valued on the connected torus and equals one at
the allowed momenta, so $\widetilde B$ has rank one everywhere.

The Bernstein estimate of Eq.~\eqref{eq:matrix-bernstein-x} gives
$\|\partial_{k_\mu}\widetilde B\|_{\max}\le2D$, and
\begin{equation}
 2D\,\frac{2\pi}{L}=\frac{4\pi D}{L}\le\frac12.
 \label{eq:fhs-depth-mesh}
\end{equation}
Lemma~\ref{lem:fhs-mesh} with $M=2D$ therefore gives
$C_{\mathrm{FHS},L}(\widetilde B)=c_1(\widetilde B)$, and
Lemma~\ref{lem:read} in class A gives $c_1(\widetilde B)=0$. Since
$\widetilde B=B$ at every allowed momentum,
\begin{equation}
 C_{\mathrm{FHS},L}(B)
 =C_{\mathrm{FHS},L}(\widetilde B)
 =c_1(\widetilde B)=0,
 \label{eq:fhs-direct-contradiction}
\end{equation}
contrary to the assumption.
\end{proof}

For the five targets and every odd $L\ge5$, a rank-one block with FHS
number one exists: $P_1$ itself in A and D, the spin-up particle block
($\kappa_z=+1$) of the spin-reduced projector in C, and the block
$\kappa_z=+1$ in AII and DIII (App.~\ref{app:fhs-targets}).
Theorem~\ref{thm:fhs-lower} therefore applies to all five targets,
including those whose total Chern number vanishes. Its coefficient
$1/(8\pi)$ is smaller than the coefficient $1/(4\pi)$ of
Theorem~\ref{thm:lower}, but it uses only the data at the allowed
momenta: it needs neither a continuous target $P_*$ nor the constants
$M_x,M_y$.

\bibliography{references}

\end{document}